\UseRawInputEncoding
\documentclass[times,final,longtitle]{elsarticle}
\usepackage{amssymb}
\usepackage{amsmath}
\usepackage{color}
\usepackage{graphics}
\usepackage{epsfig,epstopdf}
\usepackage{graphicx}
\usepackage{epsfig}
\usepackage{float}
\usepackage[normalem]{ulem}
\usepackage{epsf}

\usepackage{algorithm}                                                                     
\usepackage{algpseudocode}  

\usepackage{listings}

\newcommand{\refeq}[1]{(\ref{#1})}

\newcommand\E{{\bf e}}

\newcommand\G{{\bf g}}
\newcommand\M{{\bf m}}
\newcommand\N{{\bf n}}

\newcommand\U{{\bf u}}

\newcommand\X{{\bf x}}

\newcommand\DD{{\bf D}}

\newcommand\dert{\partial_t}
\newcommand\derx{\partial_x}

\newcommand\derz{\partial_z}

\newcommand\romandt[1]{\frac{{\rm d} #1}{{\rm d}t}}

\newcommand\grad{\nabla}

\newcommand\eps{\epsilon}
\newcommand\ii{{\rm i}}
\newcommand\om{{\omega}}

\newcommand\be{\begin{equation}}
\newcommand\nd{\end{equation}}
\newcommand\bed{\begin{displaymath}}
\newcommand\ndd{\end{displaymath}}

\newcommand\ba{\begin{array}}
\newcommand\ea{\end{array}}

\newcommand\Order{{\cal O}}
\newcommand\cijk{C_{i,j,k}}

\newcommand\bea{\begin{eqnarray}}
\newcommand\nda{\end{eqnarray}}

\renewcommand\Re{{\rm Re}\,}

\newcommand\We{{\rm We}\,}

\newcommand\ubar{{\bar u}}

\def\ubar{{\bf u}}

\def\grad{{\nabla}}

\journal{Computers and Fluids}

\newcommand {\red}[1]{{{#1}}}
\newcommand {\blue}[1]{{{#1}}}

\begin{document}


\begin{frontmatter}



\title{\red{A mass-momentum consistent}, Volume-of-Fluid method for incompressible flow on staggered grids}

\author[b]{T. Arrufat}
\author[f]{M. Crialesi-Esposito}
\author[b]{D. Fuster}
\author[a]{Y. Ling}
\author[b,d]{L. Malan}
\author[b]{S. Pal}
\author[e]{R. Scardovelli}
\author[label1]{G. Tryggvason}
\author[b]{S. Zaleski}
\address[b]{Sorbonne Universit\'e et CNRS, \\Institut Jean Le Rond d'Alembert, UMR 7190, Paris, France}
\address[f]{CMT-Motores T\'ermicos, Universitat Polit\'ecnica de Val\'encia, Camino de Vera, s/n, Edificio 6D, Valencia, Spain}
\address[d]{InCFD, Dept. of Mechanical Engineering, University of Cape Town, South Africa}
\address[e]{ DIN - Lab. di Montecuccolino, Universit\`a di Bologna, I-40136 Bologna, Italy}
\address[label1]{Mechanical Engineering, Johns Hopkins University, Baltimore, USA}
\address[a]{Dept. of Mechanical Engineering, Baylor University, Waco, TX, USA}

\begin{abstract}
  The computation of flows with large density contrasts is notoriously   difficult. To alleviate the difficulty we consider a
  discretization of the Navier-Stokes equation that advects mass and momentum in a consistent manner. Incompressible flow with capillary forces is modeled   and the discretization is performed on a staggered grid of Marker   and Cell type. The Volume-of-Fluid method is used to track the   interface and a Height-Function method is used to compute surface   tension. The advection of the volume fraction is performed using   either the Lagrangian-Explicit / CIAM (Calcul d'Interface Affine par Morceaux)    method or the Weymouth and Yue (WY)   Eulerian-Implicit method. The WY method conserves fluid mass to machine    accuracy provided incompressibility is satisfied.     To improve the stability of these methods   momentum fluxes are advected in a manner ``consistent'' with the   volume-fraction fluxes, that is a discontinuity of the momentum is   advected at the same speed as a discontinuity of the density. To find the density on the staggered cells on which the velocity is   centered, an auxiliary reconstruction of the density is   performed. The method is tested for a droplet without surface   tension in uniform flow, for a droplet suddenly accelerated in a   carrying gas at rest at very large density ratio without viscosity   or surface tension,\red{ for the Kelvin-Helmholtz instability, for a 3mm-diameter } falling raindrop and for an atomizing flow in air-water conditions.

\end{abstract}


\end{frontmatter}

\newcommand\division[1]{\subsection{#1}}
\newcommand\subdivision[1]{\subsubsection{#1}}
\newcommand\onlybook[1]{{}}
\newcommand\opus{article}
\newcommand\reduit[1]{#1}
\newcommand\LLL{{\cal L}}
\section{Introduction} 
Multiphase flows abound in nature, but their stable and accurate computation remains elusive in many cases.
As a case in point, many numerical methods used for two-phase incompressible flow are
strongly unstable for large density contrasts and large Reynolds numbers. 
Experience with such simulations shows that the presence of
surface tension is an aggravating factor. The large density contrasts that are of interest are 
 air/water or gas/liquid-metal, with $\rho_l/\rho_g$ of the order of $10^3$ or $10^4$.
The large density contrasts are a difficulty whether one deals with any of the three major
interface advection methods, Level-Set, Volume-of-Fluid (VOF) or Front-Tracking, or with combinations 
such as CLSVOF.  (The term density contrast is preferable to density ratio since it encompasses ratios both much larger than one and much smaller than one.)

Several methods have been used to alleviate the high-density-contrast difficulties.
It has been observed by several authors that making the momentum-advection method conservative improves the situation. 
For incompressible flow, ``momentum-conserving'' methods have been initially proposed by \cite{rudman98},
and by several other authors since \cite{bussmann2002modeling,desjardins10,raessi12,le13,Vaudor:2017ip,zuzio2020new}.
These methods have been shown to
improve the stability of the numerical results in various situations. In particular, liquid-gas flows
 with very contrasted densities, as for example in the process of atomization,
cause serious problems that are resolved by using momentum-conserving methods. 
In that case another oft-suggested solution is to increase the number of 
equations from the standard four equations to five, six or seven equations, 
by introducing new field variables in each phase. 
The addition of one more density $\rho_i$, momentum $\rho_i \U_{i}$ or energy variable
$\rho_i e_i$  increases  the number of equations. 
The authors of ref. \cite{Saurel99} used seven equations, 
those of \cite{allaire02} used five equations and six equations were used in \cite{Pelanti14}. 
The last three references also use a momentum-conserving formulation. 
Several authors, including some of those cited above, have argued that the difficulty may come from 
gas velocities of order $u_g$ being mixed with liquid densities of order $\rho_l$. 
Both the $\rho_l$ and $u_g$ scales are large and the appearance of an nonphysical $\rho_l u_g^2$ dynamic pressure scale could create numerical pressure fluctuations of the same order and nonphysical pressure spikes, 
as the one nicely illustrated in \cite{xiao2012large} in the front part of a
suddenly accelerated small droplet. One way of avoiding this nonphysical mixture of liquid and gas
quantities is to extrapolate liquid and gas pressure and velocity in the ``other'' phase, 
as in the ghost fluid method. This extrapolation strategy was used successfully in \cite{Xiao:2014vs}. 

It may be argued that a way to avoid this numerical diffusion of liquid and gas quantities is to advect 
the volume fraction and the conserved quantities that depend on it (density, momentum and energy) in a consistent manner.
In incompressible flow in which we specialize in this paper, it means that the volume fraction and momentum
or velocity must be advected in the same way. This is equivalent to request 
that the discontinuity of the Heaviside function $H$, marking the phase transition,
should be advected at the same speed as the discontinuity in momentum. 
This can be expressed by the following consistency requirement: 
if momentum is initially exactly proportional to volume fraction, 
it should remain so after advection. 
We call such a method VOF-consistent.

To satisfy this requirement the idea is to solve
the advection equation for momentum with the same numerical scheme that
is used for the VOF color function.
{This consistency property minimizes the nonphysical transfer of momentum from one phase to another due to the differences in the numerical schemes used. The consistency is especially important
when dealing with fluids with a large density contrast where a small numerical momentum transfer from the dense phase to the light phase results in large numerical errors in the velocity field which in turn creates numerical instabilities.}

In this work, we present a modification of the classical 
momentum-preserving scheme proposed by \cite{rudman98} \red{
for the case of a staggered grid and VOF method. The scheme is then not
momentum-conserving although it is built by discretization of the momentum conserving formulation
of the advection terms, so we rather call it mass-momentum consistent. }

The paper is organized as follow: the second section deals with the continuum mechanics formulation 
for incompressible flow and sharp interfaces. Section 3 describes our numerical method, starting
with an overview of already-known methods for spatial discretization, time-stepping, and VOF advection. We 
continue with the new momentum advection-VOF-consistent method. Section 4 is devoted to tests of the 
method, followed by a conclusion. 

Among the authors, Gretar Tryggvason, Ruben Scardovelli, Yue Ling and St\'ephane
Zaleski have been involved in the construction of the base of the
ParisSimulator VOF and Front-Tracking code that was used to implement
and test the ideas in this paper. ParisSimulator is itself based on a
Front-Tracking code developed by Gretar Tryggvason, Sadegh Dabiri and
Jiacai Lu.  Daniel Fuster was involved in the development of the
momentum advection method consistent with VOF advection, with help
from Tomas Arrufat, Leon Malan and Yue Stanley Ling. \red{The Kelvin-Helmholtz analysis and testing
were done by St\'ephane Zaleski.} The falling
raindrop testing and the corresponding figures were done by Tomas
Arrufat and Sagar Pal. The large density droplet and shear layer tests were done by Sagar Pal. 
The atomisation testing was done by Marco Crialesi-Esposito.


\section{Navier--Stokes equations with interfaces} 
\label{nse}

We model flows with sharp interfaces defined implicitly by a characteristic function $H(\X,t)$
defined such that fluid 1 corresponds to $H=1$ and fluid 2 to $H=0$. The viscosity $\mu$
and density $\rho$ are calculated as an average
\be
\mu = \mu_1 H + \mu_2 (1-H)\,, \qquad \rho = \rho_1 H + \rho_2 (1 - H) \,. 
\label{muH}
\nd
There is no phase change so the interface, 
almost always a smooth differentiable surface $S$, advances at the speed of the 
flow, that is $V_S=\U\cdot \N$ where $\U$ is the local fluid velocity 
and $\N$ a unit normal vector perpendicular to the interface. Equivalently 
the interface motion can be expressed in weak form
\be
\dert H + \U \cdot \grad H = 0 \,, 
\label{interfadv}
\nd
which expresses the fact that the singularity of $H$, located on $S$, moves at velocity $V_S=\U\cdot \N$. 
For incompressible flows, which we will consider in what follows, we have
\be
\nabla \cdot \U = 0 \,.
\label{divu}
\nd
The Navier--Stokes equations for incompressible, Newtonian flow with surface tension may 
conveniently be written in operator form
\be
 \dert (\rho \U) = \LLL(\rho,\U) - \grad p \label{nse1}
\nd
where 
$
\LLL =  \LLL_{\rm conv} + \LLL_{\rm diff} +   \LLL_{\rm cap} + \LLL_{\rm ext}
$
so that the operator $\LLL$ is the sum of advective, diffusive, capillary force and
external force terms. The first two terms are 
\be
\LLL_{\rm conv} = -\grad \cdot ( \rho \U  \U )\,, \qquad \LLL_{\rm diff} =  \grad \cdot \DD \,,
\nd
where $\DD$ is a stress tensor whose expression for incompressible flow is
\be
\DD = \mu \left[ \nabla\U + (\nabla\U)^T \right] \,,
\nd
where $\mu$ is computed from $H$ using (\ref{muH}). 
The capillary term is
\be
\LLL_{\rm cap} =  {\sigma \kappa \delta_S \N} \,,  \qquad \kappa = 1/R_1 + 1/R_2 \,,
\nd
where $\sigma$ is the surface tension coefficient, $\N$ is the unit normal perpendicular to the 
interface, $\kappa$ is the sum of the principal curvatures and $\delta_S$ is 
a Dirac distribution concentrated on the interface.  
We assume a constant coefficient $\sigma$. Finally
$\LLL_{\rm ext}$ represents external forces such as gravity. 


\section{Method} 

\subsection{Spatial discretization}

\begin{figure}
\begin{center}
    \includegraphics[width=0.45 \textwidth]{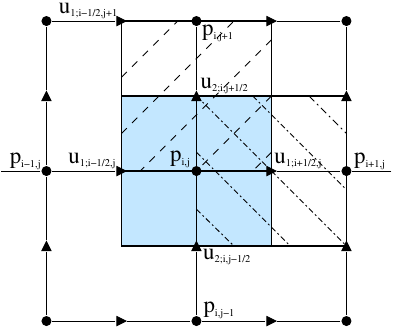}
\end{center}
\caption{Representation of the staggered spatial discretization in two dimensions. 
The pressure $p_{i,j}$ is located at the center of its control volume (light color area); 
the horizontal velocity component $u_{1;i+1/2,j}$ is stored in the middle of the
right edge of the pressure control volume and at the center of its control volume
(dash-dotted area); the vertical velocity component $u_{2;i,j+1/2}$ is stored in the 
middle of the top edge  and at the center of its control volume
(dashed area).}
\label{stag-grid}
\end{figure}

We assume a regular cubic or square grid. This can be easily
generalized to rectangular or cuboid grids, and with some efforts to
quadtree and octree grids. We also use staggered velocity and pressure
grids, as represented in two dimensions in Fig. \ref{stag-grid}. 

The control volume surrounding the pressure $p$ is used for other
scalar quantities, such as the density $\rho$ and volume fraction $C$. 
The control volumes of the velocity components, $u_1$ and $u_2$,
and momentum components are shifted respectively half cell horizontally and
vertically, with respect to the pressure control volume.
The use of staggered control volumes has the advantage of
suppressing neutral modes often observed in collocated methods but
leads to more complex discretizations (see \cite{Tryggvason11} for a
detailed discussion.) This type of staggered representation
is easily generalized to three dimensions, and
the discrete version of the continuity equation \eqref{divu}
is rather compact on such a grid 
\begin{equation}
{u_{1;i+1/2,j,k}- u_{1;i-1/2,j,k}\over \Delta x}+ 
{u_{2;i,j+1/2,k}- u_{2;i,j-1/2,k}\over \Delta y}+
{u_{3;i,j,k+1/2}- u_{3;i,j,k-1/2}\over \Delta z}=0 \,.
\label{cont2}
\end{equation}


In what follows, we shall use the subscript $f=m\pm$, with the integer index $m=1,2,3$,
to note the face of any control volume located in the positive or negative Cartesian direction 
$m$, and $\N_f$ for the unit normal vector of face $f$ pointing outwards of the control volume.
On a cubic grid the spatial step is
$\Delta x = \Delta y = \Delta z = h$ so the discrete continuity equation becomes
\be
\grad^h \cdot \U = \sum_{m=1}^3 (u_{m+} + u_{m-})/h =0 \,, 
\label{incompdisc3}
\nd
where $u_f=u_{m\pm}= \U \cdot \N_f $ is the velocity component normal to face $f$. 
The discretization of the interface location is performed using a VOF method.
VOF methods typically attempt to solve approximately equation (\ref{interfadv}) 
which involves the Heaviside function $H$, whose integral 
in the cell $\Omega$ indexed by $i,j,k$ defines the volume fraction 
$\cijk$ from the relation
\be
h^3 \,\cijk  =\int_\Omega  H \, {\rm d}\X \,.
\nd
$\cijk$ represents the fraction of the cell filled with fluid 1, 
taken to be the reference fluid. 

\subsection{Time Marching}

The volume fraction field is updated as
\be
C^{n+1} = C^{n} + \LLL_{\rm VOF}(C^{n},\U^{n}\tau/h) \,,
\label{cnp1}
\nd
where $\LLL_{VOF}$ represents the operator that updates the Volume of Fluid data
given the velocity field. Once volume fraction is updated, the
velocity field is updated in a couple of steps. A projection method is first used, 
in which a provisional velocity field $\U^*$ is computed
\be
\rho^{n+1} \U^* = \rho^n \U^{n} +  \tau \LLL^h_{\rm conv}(\rho^{n},\U^{n}) + \tau \left[\LLL^h_{\rm diff}(\mu^{n},\U^{n}) +  \LLL^h_{\rm cap}(C^{n+1}) + \LLL^h_{\rm ext}(C^{n+1})\right] \label{conspredictedvel}
\nd
It goes without saying that the above operators depend on the discretization time step
$\tau$ and spatial step $h$ as well as the fluid parameters. 
The discussion of the $\LLL^h_{\rm conv}$ operator is the main point of this paper. 
In the second step, the projection step, the pressure gradient force 
is added to yield the velocity at the new time step
\be
\U^{n+1} = \U^* - \frac{\tau }{\rho^{n+1}} \nabla^h p \,. 
\label{fotm}
\nd
The pressure is determined by the requirement that the 
velocity at the end of the time step must be divergence free
\be
\grad^{h} \cdot \ubar^{n+1}=0 \,,
\label{cont-eq1}
\nd
which leads to a Poisson-like equation for the pressure
\be
\grad^h \cdot \frac{\tau }{\rho^{n+1}} \nabla^h p =  \grad^h \cdot \U^* \,.
\label{Pois}
\nd

\subsection{Volume-Of-Fluid}
\label{vof}
\red{
In this section we detail only the necessary steps to 
illustrate the momentum advection method based on the VOF method.
To simplify the presentation we rescale space and time variables so 
that the cell volume and the time step are both equal to 1. 
Any velocity component $u$ becomes $u^\prime = u \tau / h$ and 
the $\cijk$ value is also the measure of the volume of reference fluid in cell 
$i,j,k$.
At the beginning of any simulation the volume fraction field is
initialized with the {\sc Vofi} library described in \cite{bna2015numerical} 
and \cite{bna2016vofi}. This allows a highly accurate numerical integration of 
the measure of fluid volumes. 

\subsubsection{Normal vector and plane constant determination} 

The VOF method proceeds in two steps, reconstruction and advection. 
In the first step we consider a PLIC reconstruction in each cell cut by the 
interface where the unit normal vector $\N$ is computed 
with the MYC method described in \cite{Tryggvason11}. We then consider
the colinear normal vector $\M$ whose components satisfy the relation
$|m_x| + |m_y| + |m_z|=1$. Given the
volume $V=\cijk$ in cell $i,j,k$ occupied by fluid 1 and the normal $\M$
we consider the family of planes 
\be 
\M \cdot \X = \alpha \label{mxalpha}
\nd 
By changing the value of the plane constant $\alpha$ a different
volume of fluid 1 is cut in the cell. The correct value of
$\alpha$ is determined by the resolution of a cubic equation \cite{Scardovelli00}.

\subsubsection{General split-direction advection}
\label{generalsplit}
The interface reconstruction at time $t_{n}$ is then used to obtain the  
position of the interface and the volumes $\cijk$ at the next discrete time $t_{n+1}$. 
The following discussion of momentum advection is based on 
two VOF advection methods, Lagrangian Explicit and Weymouth and Yue's 
schemes. We first describe their common features. 

After addition and subtraction of a term proportional to the velocity divergence, 
equation (\ref{interfadv}) leads to 
\be
\dert H + \nabla \cdot (\U H) = H\, \nabla \cdot \U \,.
\nd
This equation is integrated in the time step and cell volume
\be
{\cijk^{n+1} - \cijk^{n}} = - \sum_{\rm{faces}\, f} F^{(c)}_f + \int_{t_n}^{t_{n+1}}  
{\rm d}t \int_\Omega  H \,\nabla \cdot \U\,  {\rm d}\X   \,,
\label{sumf}
\nd
where the first term on the right-hand side is the sum over the cell faces $f$  
of the fluxes $F^{(c)}_f$ of $(\U H)$. Obviously the ``compression'' term on the 
right-hand side disappears for incompressible flow, however it is essential 
in split-advection methods.  In the previous equation the flux $F_f^{(c)}$ is 
\be
F_f^{(c)} = \int_{t_n}^{t_{n+1}} {\rm d}t \int_{f} u_f(\X,t) H(\X,t) \, {\rm d}\X \,,
\label{faceint}
\nd
where $u_f = \U\cdot \N_f$.
Once an approximation for the evolution of $(\U H)$ during
the time step is chosen, a four-dimensional integral remains to be
computed in equation (\ref{faceint}). The two methods we consider here
are directionally split and are also designed to preserve the
property $0 \le \cijk \le 1$ which we call $C$-bracketing. It
is important to preserve $C$-bracketing in order to avoid arbitrary addition or
removal of mass. \red{Furthermore they do not produce in the bulk of the
two fluids deviations from the correct values $0$ and $1$.}

Directional splitting results in the breakdown of equation (\ref{sumf}) 
into three equations
\be
{\cijk^{n,l+1} - \cijk^{n,l}} = - F^{(c)}_{m-} - F^{(c)}_{m+} 
+ c_m \partial_{m}^h u_m \,,
\label{sumf2}
\nd
where the superscript $l=0,1,2$ is the substep index, i.e.
$\cijk^{n,0} = \cijk^{n}$ and $\cijk^{n,3} = \cijk^{n+1}$. 
The face with subscript $m-$ is the ``left'' face in direction $m$ with 
$F^{(c)}_{m-} \ge 0$ if the flow is locally from right to left. A similar reasoning 
applies to the ``right'' face $m+$.
 We have also  approximated the compression term in 
(\ref{sumf}) by
\be
 \int_{t_n}^{t_{n+1}}  {\rm d}t \int_\Omega  H \partial_m u_m  {\rm d}\X \simeq  c_m 
 \partial_{m}^h u_m \,, 
 \label{comp}
\nd
with no implicit summation rule. 
 In the RHS of (\ref{sumf2}) and (\ref{comp}) the flux terms $F_{f}^{(c)}$ and the 
 partial derivative $\partial_{m} u_m$ must
be evaluated with the same discretized velocities. In particular,
$\partial_{m}^h u_m$ is a finite difference or finite volume approximation of the 
spatial derivative of the $m$th component of the velocity vector in direction $m$, and
the ``compression coefficient'' $c_m$ approximates the color fraction. 
Its exact expression is dependent on the advection method and 
it preserves the $C$-bracketing condition. 
Since the coefficient $c_m$ may not be the same along the three Cartesian 
directions, the sum $\sum_m c_m \partial_{m}^h u_m$ is not necessarily vanishing
even if the flow is incompressible.
After each advection substep (\ref{sumf2}), the interface is reconstructed
with the updated volumes $\cijk^{n,l+1}$, then the 
fluxes $F^{(c)}_{f}$ are computed for the next substep. 

In each substep the velocity field in direction $m$ is usually dependent
only on the spatial component $x_m$, $u_m (x_m)$. This approximation ensures that the 
fluid areas fluxing across a cell side are rectangular in two dimensions, as shown in
Figs. \ref{lagfig} and \ref{eulflux}. The multi-dimensionality of the flow
is considered in unsplit methods, where the fluxing areas are described by more 
complex polygons \cite{Cervone09}.

\subsubsection{Lagrangian Explicit advection}

The Lagrangian Explicit / CIAM method refers to a specific type
of split advection and it is most naturally explained
as a Lagrangian transport of the Heaviside function \cite{li95,Scardovelli02} . 
The velocity field is linearly interpolated between the face velocities $u_{m-}$ on
the left and $u_{m+}$ on the right, so that 
$u_m (x_m) = - u_{m-}( 1 - x_m) + u_{m+} x_m$, with the origin of
$x_m$ on the left face. 
The equation of motion $d x_m / dt = u_m (x_m)$ is integrated with a
first-order in time, explicit $u_m (x^n_m)$ approximation, to get in rescaled variables 
\be
x_m^{n+1} =  - u_{m-} +  (1 +  u_{m+} + u_{m-}) \,x^n_m \,.
\label{map}
\nd
This transformation gives the position of advected points as a function of the 
original position and compresses distances along direction $m$ by a factor 
$(1 + u_{m+} +  u_{m-})$. Points over faces and linear interface are advected
in the same way, and 
in the two-dimensional case of Fig. \ref{lagfig} the advection substep results 
in the three contributions $V_1$, $V_2$ and $V_3$ to the central cell.
The intermediate value of the color function in the central cell
will be given by the sum of these three contributions.
\begin{figure}
\begin{center}
    \includegraphics[width=\textwidth]{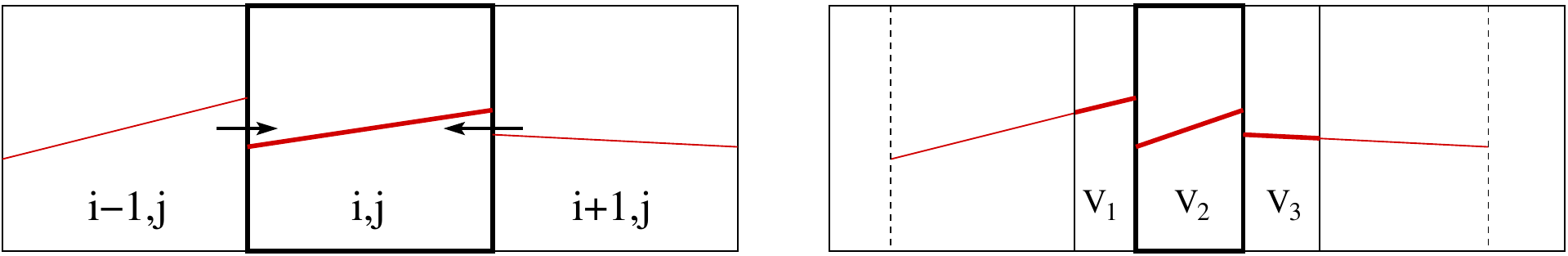}
\end{center}
\caption{Formation of the areas  $V_1,V_2$ and $V_3$ by Lagrangian advection in the 
horizontal direction: initial reconstruction with the horizontal velocities on the faces 
of the central cell (left); segments and areas $V_i$ after Lagrangian 
advection (right). In three dimensions rectangles become cuboids.}
\label{lagfig}
\end{figure}

There is a correspondence between the geometrical interpretation of the Lagrangian 
Explicit advection and the definition \eqref{faceint} of $F^{(c)}_{f}$. 
For example, for the central cell of Fig. \ref{lagfig} the flux on
the left face is from left to right, since $u_{1;i-1/2,j} > 0$.
Then with $m = 1$ and $f=1-$, we have $V_{1} =  - F^{(c)}_{1-} > 0$ and 
$u_{1-} =\U\cdot \N_{1-} < 0$.
The final expression of the substep is 
\be
\cijk^{n,l+1} =  \cijk^{n,l} (1  +  u_{m+} +  u_{m-} )  - F^{(c)}_{m-} - F^{(c)}_{m+} \,,
\label{sumflag}
\nd
which shows that the approximation of the derivative is 
$\partial_m^h u_m = u_{m+}  + u_{m-}$ and the compression coefficient is
\be
c_m = \cijk^{n,l} 
\label{cmle}
\nd
In order to remove spurious asymmetries in the flow it is important
to change the order of split advections at each timestep. Then
the compression coefficient in the horizontal advection, $m=1$, can
be associated to the first substep, $l=0$, in timestep $n$, and to
the last substep, $l=2$, in the next timestep.
The sequence of three Lagrangian substeps (\ref{sumf2}) 
does not result in volume conservation
\be
{\cijk^{n+1} - \cijk^{n}} = - \sum_{\rm{faces}\, f} F^{(c)}_f 
+ \sum_{m=1}^{3} c_{m}\,(u_{m+}  + u_{m-}) \,.
\label{sumfall}
\nd
While the flux terms cancel upon integration over the domain, the sum of the compressive 
terms does not vanish since $c_m$ changes at each substep $l$.

\begin{figure}
\begin{center}
    \includegraphics[width=\textwidth]{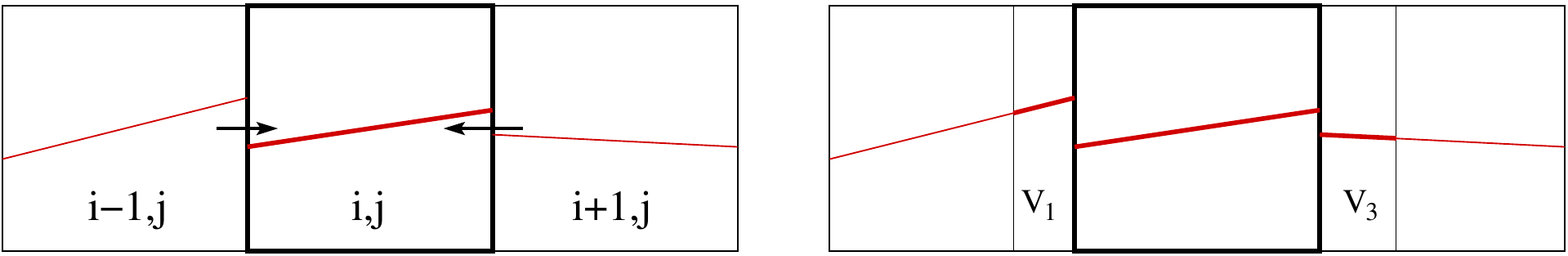}
\end{center}
\caption{Eulerian flux representation for advection in the horizontal direction: 
same initial reconstruction and horizontal velocities of Fig. \ref{lagfig} (left); 
fluxes, or areas $V_1$ and $V_3$, are calculated directly from the interface
reconstruction in each cell (right)}
\label{eulflux}
\end{figure}

\subsubsection{Weymouth and Yue's advection}

The (WY) split advection is exactly mass-conserving  \cite{Weymouth:2010hy}. 
In this method the compression coefficient is independent of direction $m$, 
so that $c_m = c$, and is defined as
\be
c = H \big( \cijk^n - 1/2 \big) 
\label{cmwy}
\nd
where $H$ is a one-dimensional Heaviside function, 
that is $c=0$ if $\cijk^n < 1/2$ and $c=1$ if  $\cijk^n \ge 1/2$. 
The fluxes $F^{(c)}_f$ are also defined differently. 
The reference phase fluxed through the left face in direction $m=1$ 
is equal to the volume fraction in a cuboid of width $u_{i-1/2,j,k}$ 
adjacent to the left face $f=1-$. This fluxed volume 
corresponds to ``Eulerian Implicit'' (EI) advection in the terminology 
of \cite{Scardovelli02} and is represented in 2D by the area $V_1$ of 
Fig. \ref{eulflux}. Using these definitions,  
Weymouth and Yue were able to show that the final result obeys 
$C$-bracketing \cite{Weymouth:2010hy}. 

This split advection scheme conserves volume at machine accuracy. 
Indeed the summation of three substeps (\ref{sumf2}) results in 
\be
{\cijk^{n+1} - \cijk^{n}} = - \sum_{\rm{faces}\, f} F^{(c)}_f 
+ c \sum_{m=1}^{3} \partial_{m}^h u_m 
\label{sumfall2}
\nd
Since $\sum_{m=1}^{3} \partial_{m}^h u_m = \sum_{m=1}^{3} (u_{m+}  + u_{m-}) $ is 
the finite-volume expression for $\nabla \cdot \U$, it disappears and mass 
is conserved at the 
accuracy with which condition (\ref{divu}) is satisfied. 
}
\subsubsection{Clipping}
\label{clipping}

The algorithm that has been coded involves a number of additional
steps designed to avoid unwanted effects of arithmetic floating point
round-off error. The most important one is clipping: at the end of
each directional advection, the values of $\cijk$ are reset so that
$C_{ijk}$ is set to $0$ if $C_{ijk} < \eps_c$ or to
$1$ if $C_{ijk} > 1 - \eps_c$. When there is no surface tension 
the choice $\eps_c = 10^{-12}$ works well.
Otherwise $\eps_c = 10^{-8}$ gives more stable results with smoother
interface shapes. This stronger clipping is a necessity for some
simulations with WY, as we observe that WY produces many more ``wisps'', 
i.e. cells with tiny values of $1-\cijk$
inside fluid 1 or $\cijk$ inside fluid 2.
We have not yet been able to determine the origin of this need for a more
forceful clipping with WY, but it could be related to the fact that the
CIAM method has a geometrical interpretation, while WY is intrinsically algebraic
in nature.

\subsection{Momentum-advection methods}
\label{mom}
\newcommand{\dpt}[1]{\frac{\partial #1}{\partial t}}
\newcommand\pijk{\phi_{i,j,k}}

\subsubsection{Advection of a generic conserved quantity}

Consider the advection of a generic {\em conserved} quantity $\phi$ by a continuous velocity field
\be
\dert \phi + \nabla \cdot ( \phi \,\U)  = 0. 
\label{phiconv}
\nd
We assume that $\phi$ is smoothly varying except on the interface where it may be discontinuous.
Indeed finding a correct scheme for the advection of this discontinuity, at the same speed as the 
advection of the volume fraction, is the goal of the present study.
The smoothness of the advected quantity away from the interface
is verified for the density $\rho$, the momentum $\rho \U$ or the internal energy 
$\rho e$. We first integrate \eqref{phiconv} in time
\be
{\pijk^{n+1} - \pijk^{n}} = - \sum_{\rm{faces}\, f} F^{(\phi)}_f \,. 
\label{sumfp}
\nd
The sum on the right-hand side is the sum over faces $f$ of cell $i,j,k$ 
of the fluxes $F^{(\phi)}_f$ of $\phi$, which are defined in the same way as the color 
function fluxes $F^{(c)}_f$ in \eqref{faceint}
\be
F_f^{(\phi)} = \int_{t_n}^{t_{n+1}} {\rm d}t \int_{f} u_f(\X,t) \,\phi(\X,t) \,
{\rm d}\X.
\label{pfaceint}
\nd
In order to ``extract'' the discontinuity we introduce the 
 characteristic function $H(\X,t)$
\be
F_f^{(\phi)} = 
\int_{t_n}^{t_{n+1}} {\rm d}t \int_{f} [ u_f  \,H \,\phi  +  u_f \,(1-H) \,\phi ] 
\, {\rm d}\X \,, 
\label{fluxphi}
\nd
and rewrite it as
\be
F_f^{(\phi)} = 
\bar \phi_1 \int_{t_n}^{t_{n+1}} {\rm d}t \int_{f} u_f \,H \,{\rm d}\X + 
\bar \phi_2 \int_{t_n}^{t_{n+1}} {\rm d}t \int_{f} u_f \,(1-H) \,{\rm d}\X \,,
\label{barphi}
\nd
where the face averages $\bar \phi_s$, $s=1,2$, are
\be 
\bar \phi_s = \frac{\int_{t_n}^{t_{n+1}} {\rm d}t \int_{f} \phi \,u_f \,H_s  
{\rm d}\X}{\int_{t_n}^{t_{n+1}} {\rm d}t \int_{f}  
u_f \,H_s\,{\rm d}\X} \,,  
\label{barphi2}
\nd
and $H_1=H$, $H_2= 1-H$. Expression \eqref{barphi} can be written in terms 
of the fluxes $F^{(c)}_f$ and $F^{(1-c)}_f$, this second one being obtained by 
replacing $H$ with $1-H$ in \eqref{faceint}
\be
F_f^{(\phi)} = \bar \phi_1 \,F_f^{(c)} +  \bar \phi_2 \,F_f^{(1-c)} \,.
\label{fluxphi12}
\nd

\subsubsection{Cloning the tracers}

When a cell is cut by the interface, and the field $\phi$ is not smooth, 
it becomes difficult to estimate the integrals in (\ref{barphi}). A possibility is to 
define two new fields $\phi_{1}$ and $\phi_{2}$, with $\phi_s = \phi$ 
inside phase $s$, then
\be
\phi = H \,\phi_1 + (1-H) \,\phi_2 \,.
\nd
This is more costly in memory usage but simplifies considerably the computation of the 
averages in (\ref{barphi}). The two equations \eqref{interfadv} and \eqref{phiconv} 
are now replaced by three equations,
the same volume fraction equation (\ref{interfadv}) and 
\be
\dert \phi_1 + \nabla \cdot ( \phi_1 \U) = 0 \,\,,\qquad
\dert \phi_2 + \nabla \cdot ( \phi_2 \U) = 0 \,.
\label{pcv1}
\nd
The three equations \eqref{interfadv} and \eqref{pcv1} now imply (\ref{phiconv}). 
The addition of a pair of ``cloned'' variables to deal with large 
density contrasts is similar to the methods used for the resolution of the 
momentum and energy equations for compressible flow. For example Saurel and Abgrall 
used two density, momentum and energy variables in their seven-equations model 
\cite{Saurel99}, while Allaire, Clerc and Kokh use two density variables 
in their five-equations model \cite{allaire02}. The addition of a cloned 
tracer variable in incompressible isothermal flow was also implemented by Popinet in 
the ``Basilisk'' code \cite{basilisk}. 

\subsubsection{Advection of the density field}

The density $\rho(\X,t)$ obeys (\ref{phiconv}) with $\phi = \rho$. 
Moreover we consider a divergence-free velocity field, with constant density 
in each phase. We can extract the density trivially from the integrals 
\eqref{barphi2} to obtain {\em exactly} $\bar \rho_s = \rho_s$. 
The flux of $\rho$ is then
\be
F_f^{(\rho)} = \rho_1 F_f^{(c)} +  \rho_2 F_f^{(1-c)}.\label{fluxrho}
\nd
Using this flux definition for $\rho$, and any VOF method for the fluxes of the color 
function, one obtains a conservative method for $\rho$, since eq. (\ref{sumfp}) 
evolves $\rho$ as a difference of fluxes. Thus the total mass is conserved. 
However this result is not consistent with the advection of the color function in the 
CIAM case, as CIAM does not conserve volumes exactly (see  \eqref{sumfall}).
As a result the advection of $\rho$ 
is not consistent with the advection of $C$. 

This paradox may be resolved if one notices that the compression term is missing 
in \eqref{phiconv}. For consistency the compression term should be kept,
and the advection equation for a conserved quantity becomes
\be
\dert \phi + \nabla \cdot ( \phi \,\U)  = \phi \, \nabla \cdot \U \label{phiconv2}
\nd
It is then possible to define the evolution of $\phi$ through a sequence of 
directionally-split operations which are equivalent to the operations performed on 
the color function 
\be
{\pijk^{n,l+1} - \pijk^{n,l}} = - F^{(\phi)}_{m-} - F^{(\phi)}_{m+} 
+ \Big( \tilde \phi_1^m c^{(1)}_m + \tilde \phi_2^m c^{(2)}_m \Big) \,
\partial_{m}^h u_m \label{sumfpconsistent}
\nd
where $F^{(\phi)}_{m\pm}$ are defined in \eqref{fluxphi12}, 
the cell averages $\tilde \phi_s^m$ are
\be
\tilde \phi_s^m = \frac{\int_{t_n}^{t_{n+1}} {\rm d}t \int_{\Omega}  \phi  H_s  
\partial_{m}^h u_m  \,  {\rm d}\X}
{\int_{t_n}^{t_{n+1}} {\rm d}t \int_{\Omega} H_s  \partial_{m}^h u_m \,{\rm d}\X} \,,
\label{cellphi2}
\nd
and $c^{(1)}_m=c_m$ is the compression coefficient of the VOF advection,
while $c^{(2)}_m= 1 -c_m$ is that of the symmetric color fraction  $1 - C$.
Specifically for $\rho$ this gives 
\newcommand\rijk{\rho_{i,j,k}}\be
{\rijk^{n,l+1} - \rijk^{n,l}} = - F^{(\rho)}_{m-} - F^{(\rho)}_{m+} + C_m^{(\rho)}\,,
\label{sumfrho}
\nd
where the fluxes are given by (\ref{fluxrho}) and the compression term is
\be
C_m^{(\rho)} =  \Big( \rho_1 c^{(1)}_m + \rho_2 c^{(2)}_m \Big) \,\partial_{m}^h u_m  
\label{central}
\nd
with no implicit summation on $m$ and $c_m^{(s)}$ given by (\ref{cmle}) or
(\ref{cmwy}). 
For the WY method, the compression terms eventually cancel out and mass is conserved 
at the same accuracy as the discrete incompressibility condition
$\sum_{m=1}^3 \partial_{m}^h u_m=0$ is verified. 

\subsubsection{Momentum advection: basic expressions}

For momentum advection we consider the transport of the scalar quantities  
$\phi=\rho u_q$, where $q=1,2,3$ is the component index. 
With definition (\ref{barphi2}), we obtain for the face weighted averages 
$\bar \phi_s = \overline { \rho u_q}_s$ the expression
\be 
\overline{\rho u_q}_s = \rho_s \bar u_{q,s} 
\nd
where 
\be \bar u_{q,s} =  \frac{\int_{t_n}^{t_{n+1}} {\rm d}t \int_{f}  u_q u_f  H_s  \, {\rm d}\X}{\int_{t_n}^{t_{n+1}} {\rm d}t \int_{f}  u_f  H_s\,{\rm d}\X} \label{barudef}
\nd
We term $\bar u_{q,s}$ the ``advected interpolated velocity'' and 
explain below how it is computed. 
Thus the evolution of the momentum is given by
\newcommand\mijk{(\rho u_{q})_{i,j,k}}
\begin{eqnarray}
{\mijk^{n,l+1} - \mijk^{n,l}}  = - F^{(\rho u)}_{m-} - F^{(\rho u)}_{m+}
 + \Big( \rho_1 \tilde u_{q,1}^m  c^{(1)}_m +  \rho_2 \tilde u_{q,2}^m c^{(2)}_m 
 \Big) \, \partial_{m}^h u_m 
\label{sumfrou}
\end{eqnarray}
where
\be
 F^{(\rho u)}_{f} =  \rho_1 \,\bar u_{q,1}  \,F^{(c)}_{f}  +  
 \rho_2 \,\bar u_{q,2}  \,F^{(1-c)}_{f} \,,
\nd
and the ``central interpolated velocity'', corresponding to the 
averages $\tilde \phi_s^m$ of \eqref{cellphi2}, are
\be
\tilde u_{q,s}^{m} = \frac{\int_{t_n}^{t_{n+1}} {\rm d}t \int_{\Omega} u_q \, H_s \,  
\partial_{m}^h u_m   \,  {\rm d}\X}
{\int_{t_n}^{t_{n+1}} {\rm d}t \int_{\Omega} H_s \, \partial_{m}^h u_m \,{\rm d}\X} 
\label{tildeudef}
\nd
From now on we omit the superscript $m$ 
for $\tilde u_q^m$ to avoid too complex notations. 
Notice that ``cloning'' the advected velocities 
$\bar u_{q,1}$ and $\bar u_{q,2}$
would make it easier to advect a velocity field with a jump on the interface. 
However in viscous flow without phase change the velocity is continuous on the 
interface, and to avoid an excessively complicated method we 
approximate the velocity field as continuous and we choose 
$\bar u_q =  \bar u_{q,1} = \bar u_{q,2}$ for the 
``advected interpolated velocity'' and $\tilde u_q =  \tilde u_{q,1} = \tilde u_{q,2}$
for the ``central interpolated  velocity''. An important simplification is then 
\be
 F^{(\rho u)}_{f} = \bar u_q F^{(\rho)}_{f} \label{frou}
\nd
(which is the central equation in this development) and thus
\be
{\mijk^{n,l+1} - \mijk^{n,l}} =  -\bar u_q  F^{(\rho)}_{m-} - \bar u_q  F^{(\rho)}_{m+} 
+ \tilde u_q C_m^{(\rho)},
\label{sumfmom2}
\nd
where the density fluxes are defined in (\ref{fluxrho}) and the compression term 
$C^{(\rho)}$ in (\ref{central}). 
In the above expression the face-weighted average velocities $\bar u_q$ are defined
using (\ref{barudef}) on the corresponding
left face $m-$ or right face $m+$. \red{It is important to note that up to this point
the weighted averages  $\bar u_q$ and $\tilde u_q$ have been defined but the method
in which they are estimated in the numerical method will be given only in what follows.
}

If we combine the scheme above with the CIAM scheme, the compression coefficient
in the volume fraction advection from \eqref{cmle} is $C^{n,l}$, and for the 
central interpolated velocity we take $\tilde u_q = u_q^{n,l}$. The compression
term in (\ref{sumfmom2}) does not cancel out when the final momentum is computed 
after three directionally-split advections and the result is not exactly conservative. 
On the other hand with the WY scheme, the compression coefficient
in the volume fraction advection from \eqref{cmwy} is $c$,
that is independent of direction $m$ and \eqref{central} becomes
\be
C_m^{(\rho)} =  \Big( \rho_1 c + \rho_2 (1-c ) \Big) \,\partial_{m}^h u_m  \,.
\label{central2}
\nd
Since there is no bracketing on the velocity components, we take 
$\tilde u_q = u_q^{n}$ which is independent on the substep $l$.
Provided the velocity field is incompressible, that is 
$\sum_{m=1}^3 \partial_{m}^h u_m =0$, after the three split advections (\ref{sumfmom2}) 
one obtains a cancellation of the compression terms and 
\be
{\mijk^{n,3} - \mijk^{n}} =  - \sum_{\rm faces \, f}  \bar u_{q}  F^{(\rho)}_{f} . 
\label{sumfmomtotfracstep}
\nd
The momentum transport coupled with WY advection is thus exactly 
conservative. \red{ The advected interpolated velocity $\bar u_q$ and the velocity
$u_f$ normal to face $f$ are discussed in the next section.}

\subsubsection{Momentum advection: interpolations and flux limiters}

The momentum equation (\ref{sumfmom2}) can be approximated either
 1) in the bulk of the phases or 2) in the neighborhood of the interface.
In the first case the expression simplifies considerably since both the density
and the color fraction are constant and the spurious compression terms cancel out
\be
{u^{n,3}_q - u_q^{n}} =  - \sum_{\rm faces \, f}  \bar u_{q} u_f . \label{sumfmomtot}
\nd
We distinguish an ``advecting'' velocity $u_{f}= \U \cdot \N_f $ and an ``advected velocity'' 
component $\bar u_{q,f}$, involving an average over face $f$. Both velocity components
require an interpolation from their position in the staggered grid to where they are needed.
Thus the scheme in the bulk is 
\be
{u^{n,3}_q - u_q^{n}} =  - \sum_{\rm faces \, f}  \bar u_{q}^{\rm (advected)} 
u^{\rm (advecting)}_f \,,  
\label{sumfmomtotbulk}
\nd
while near the interface is 
\be
{(\rho u_q)^{n,3} - (\rho u_q)^{n}} =  - \sum_{\rm faces \, f}  \bar u_{q}^{\rm (advected)}  
F^{(\rho)}_{f} + \sum_{m=1}^3 \tilde u_q C_m^{(\rho)} \,. 
\label{advect-ed-ing-2}
\nd
Momentum advection in our model is described by these two equations which are
solved on a cubic grid with a finite-volume method. In the previous sections
we have derived a new expression for the momentum fluxes and the
compression term, i.e. the RHS of \eqref{advect-ed-ing-2}, that is consistent
with the volume fraction advection.  

To estimate the advecting velocities $u_f^{\rm (advecting)}$ we use a centered scheme.
The staggered 2D grid of Fig. \ref{stag-grid} has the same variables
arrangement that is found in 3D on a plane perpendicular to the $z$-axis and through the 
pressure point $p_{i,j,k}$. To illustrate the procedure we consider a face perpendicular
to the horizontal direction $1$, in particular $f=1-$. 
There are two cases. In the first case the advected component is not aligned with the 
face normal, this corresponds to $q=2$. The $u_2$ control volume in 
Fig. \ref{stag-grid} is centered on $i,j+1/2,k$, and face $f=1-$ is then 
centered on $i-1/2,j+1/2,k$. The advecting velocity  $u_{1-}^{\rm (advecting)}$ 
is not given on this point and has to be interpolated 
\be
 u_{1;i-1/2,j+1/2,k}^{\rm (advecting)} = \frac12 \big( u_{1;i-1/2,j,k} + u_{1;i-1/2,j+1,k} 
\big) \,.
\label{uf2}
\nd
In the second case the advected component is aligned with the 
face normal, this corresponds to $q=1$. The $u_1$ control volume 
in Fig. \ref{stag-grid} is centered on $i+1/2,j,k$ and face $f=1-$ is then 
centered on $i,j,k$. The interpolation is now
\be
 u_{1;i,j,k}^{\rm (advecting)} = \frac12 \big( u_{1;i-1/2,j,k} + u_{1;i+1/2,j,k} \big) \,.
\label{uf1} 
\nd
 
\begin{figure}
\begin{center}
    \includegraphics[width=\textwidth]{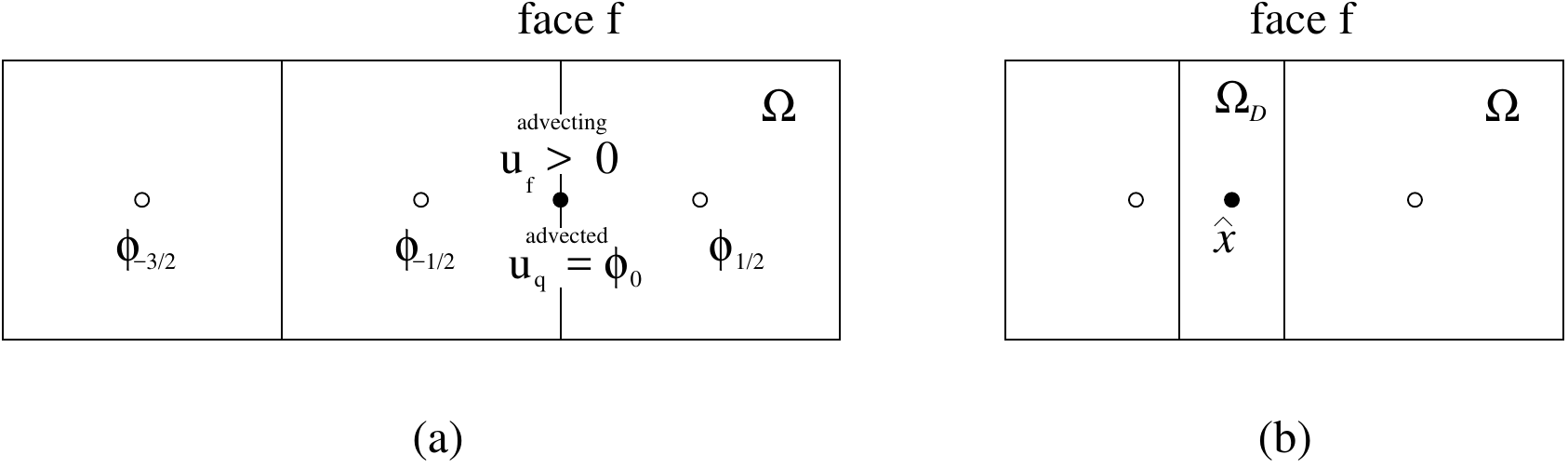}
\end{center}
\caption{The reference control volume $\Omega$ for the advected velocity component 
$\phi=u_q$ is shown. A horizontal advection is here considered and both the advecting velocity 
$u_f$ and the advected velocity require an interpolation for their value on the left face 
$f=1-$: (a) the value $\bar u_q = \phi_0$ (full circle) is interpolated 
(see \ref{appinterp}) from the values $\phi=u_q$ on the nodes (open circles);
(b) a more sophisticated interpolation predicts the value $\phi(\hat x)$ where $\hat x$ is
at the center of the ``donating'' region $\Omega_D$ (see \ref{appinterp}).}
\label{advect-ed-ing-fig}
\end{figure}

Now we turn to the interpolation of the {\em advected} velocity $\bar u_{q}$ in
(\ref{advect-ed-ing-2}). The interpolants we use in this case are
one-dimensional and operate on the velocities $u_q$, on the center of their 
control volume, that are regularly spaced on a segment aligned 
with the direction of the advection, that is perpendicular to face $f$. 
We still consider an advection along the horizontal direction $1$. 
In Fig. \ref{advect-ed-ing-fig}, with the lighter notation $\phi = u_q$,
we need to interpolate the advected velocity on the left face
of the reference control volume $\Omega$.
For the advected velocity $u_1$ and the advecting velocity \eqref{uf1} on face
$f=1-$ the correspondence with the $\phi$ values in Fig. \ref{advect-ed-ing-fig} is
\be
\phi_{-3/2} = u_{1;i-3/2,j,k}, \quad \phi_{-1/2} = u_{1;i-1/2,j,k}, \quad   
\phi_{1/2} = u_{1;i+1/2,j,k}, \;\cdots
\nd
while for the advected velocity  $u_2$ and the advecting velocity \eqref{uf2} is
\be
\phi_{-3/2} = u_{2;i-2,j+1/2,k}, \quad  \phi_{-1/2} = u_{2;i-1,j+1/2,k}, 
\quad  \phi_{1/2} = u_{2;i,j+1/2,k}, \;\cdots
\nd
The extension of these results to an advection along the other two directions $q=2,3$ 
follows easily. We need to predict $\phi_0$ on face $f=1-$ in Fig. \ref{advect-ed-ing-fig}
to serve as an approximation of $\bar u_q$ given in (\ref{barudef}). We consider 
an interpolation function $f$ that computes 
this value as a function of the four nearest points,
and in an upwind manner based on the sign of the {\em advecting} velocity $u_f$
\be
\phi_0 = f \big( \phi_{-3/2}, \phi_{-1/2}, \phi_{1/2},\phi_{3/2},{\rm sign}(u_f)
\big) \,.
\label{simpleinterp}
\nd

In this study we have extensively tested two kinds of interpolations:
\begin{enumerate}
\item a scheme that uses a QUICK third-order interpolant in the bulk, away from the interface 
and a simple first-order upwind flux near the interface. We call this scheme QUICK-UW;
\item a scheme that uses a Superbee slope limiter \cite{roe1985some} for the flux in the 
bulk and a more complex Superbee limiter tuned to a shifted interpolation point near the 
interface.  We naturally call this scheme ``Superbee''. 
\end{enumerate}
The details of the two schemes are given in \ref{appinterp}.
\label{tunedinter}

\red{
\subsubsection{VOF-consistent momentum advection on staggered grids}

In order to apply the above method on the staggered grid, we need
the color fraction data in the velocity control volumes. 
At the start of the velocity advection operations, summarized by the operator 
$\LLL^h_{\rm conv}$, each velocity control volume overlaps two pressure/VOF 
control volumes, for example $\Omega_{i+1/2,j}$ overlaps $\Omega_{i,j}$ and $\Omega_{i+1,j}$ 
in the 2D case of Fig. \ref{halffractions}. An estimate of the shifted volume fraction
$C_{i+1/2,j}$ in $\Omega_{i+1/2,j}$ is then obtained by performing the usual reconstructions
in $\Omega_{i,j}$ and $\Omega_{i+1,j}$  and adding the two half-fractions.
The following operations are then performed at each time step and are summarized
in \textbf{Algorithm 1}:

\begin{algorithm}
\caption{Summary of the algorithm for the momentum and VOF time step}
\label{resumealgo}
\begin{algorithmic}[]
   \State Reconstruct interface from volume fractions $C^{n}$
   \For {each component $q$}
      \State Compute ``shifted'' volume fraction $C^{n,0}_q$ in the staggered control volumes 
      \State Compute density $\rho^{n,0}_q$ (Eq. \ref{muH})
      \State Compute momentum component $(\rho_q u_q)^{n,0}$
   \EndFor
   \For {each substep $l$}
      \For {each component $q$}
         \State Momentum advection in the $x_l$ direction to compute $(\rho_q u_q)^{n,l+1}$ 
                (Eq. \ref{sumfmom2})
         \State (the $x_l$ coordinate direction changes with the time step)
         \State VOF advection of ``shifted'' $C_q$ in the $x_l$ direction to compute 
                $C^{n,l+1}_q$
         \State Compute density $\rho^{n,l+1}_q$ (Eq. \ref{muH})
         \State Update velocity component $u_q^{n,l+1} = (\rho_q u_q)^{n,l+1}/\rho^{n,l+1}_q$
      \EndFor
      \State VOF advection of $C$ in pressure control volumes in the $x_l$ direction
             to compute 
      \State $C^{n,l+1}$
   \EndFor
\end{algorithmic}
\end{algorithm}

\begin{enumerate}
\item Reconstruction of the interface at time $t_n$ from the data $C^{n}$,
and computation of the shifted fraction of Fig. \ref{halffractions} to obtain
the ``shifted'' data $C^{n}_q$ and $\rho^{n}_q$ in the staggered control volumes
($q=1,2,3$ is the component index), for example $\rho_1$ in $\Omega_{i+1/2,j,k}$ for the 
horizontal momentum component $\rho_1 u_1$.
\item Computation of the three momentum components $(\rho_q u_q)^{n}$
at time $t_n$.
\item Advection of the three momentum components along one coordinate direction,
say $x$ direction, using \eqref{sumfmom2} to obtain the updated momentum components 
$(\rho_q u_q)^{n,1}$ after the first substep.
\item Advection with the VOF method of the ``shifted'' volume fraction data $C^{n}_q$ 
of the staggered control volumes along the $x$ direction to obtain the updated volume 
fractions $C^{n,1}_q$ and from \eqref{muH} the densities $\rho_q^{n,1}$ after the first 
substep. 
\item Extraction of the provisional velocity components $u_q^{n,1}$
after the first substep, $u_q^{n,1} = (\rho_q u_q)^{n,1}/\rho_q^{n,1}$.
\item Repeat the previous operations for momentum components, shifted
volume fractions and densities, and velocity components for the next two substeps
with split advections along the $y$ and $z$ directions.
Eventually obtain $(\rho_q u_q)^{n+1} = (\rho_q u_q)^{n,3}$ and 
$\tilde \rho_q^{n+1} = \rho_q^{n,3}$. At each time step, the sequence $x, y, z$
is permuted. 
\item In parallel, computation of $C^{n+1} = C^{n,3}$ on the pressure control volumes using
the VOF method. 
\end{enumerate}

\begin{figure}
\begin{center}
    \includegraphics[width=0.5\textwidth]{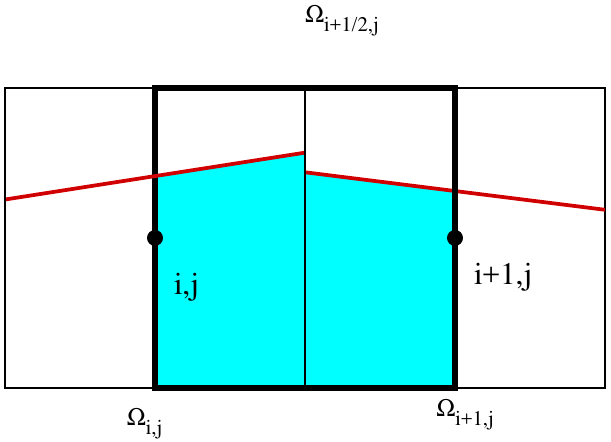}
\end{center}
\caption{Computation of the shifted volume fractions from the half-fractions.}
\label{halffractions}
\end{figure}

The interface reconstruction, the computation of volume fraction fluxes and the 
interpolation of ``advected'' and ``advecting'' velocity components have been 
detailed in the previous sections. 
We remark that the advected velocity components $u_q$ are updated at each substep,
while the advecting velocities $u_f$ are interpolated from the initial
velocity field $\U^n$ at time $t_n$. The shifted fractions of Fig. \ref{halffractions}
are computed by the same routine that is computing the Eulerian fluxes $V_1$ and
$V_3$ of Fig. \ref{eulflux}.

The three momentum components $(\rho_q u_q)^{n+1}$ at the beginning of next time step
$t_{n+1}$ require the computation of the three ``shifted''  volume fractions $C^{n+1}_q$
and densities $\rho^{n+1}_q$ starting from $C^{n+1}$. However, these densities
$\rho^{n+1}_q$ are different from the densities $\tilde \rho_q^{n+1} = \rho_q^{n,3}$
computed in the previous  time step by directly advecting the 
``shifted'' volume fractions $C^{n}_q$. The reason for this difference is that the
linear reconstruction is approximate and it is not even continuous on the 
boundary of its control volume. As a matter of facts, at each substep we have
four slightly different interface reconstructions. This implies that momentum is
not conserved between two time steps. We note that  attempting to always use only the three 
sets $C^{n}_{q}$  and evolve them by the VOF method on the staggered cells 
would maintain conservation but result in the three staggered grids evolving independently 
of each other and eventually diverging.
A diagram of the whole scheme is presented in Fig. \ref{figscheme}.
\begin{figure}
\begin{center}
    \includegraphics[width=\textwidth]{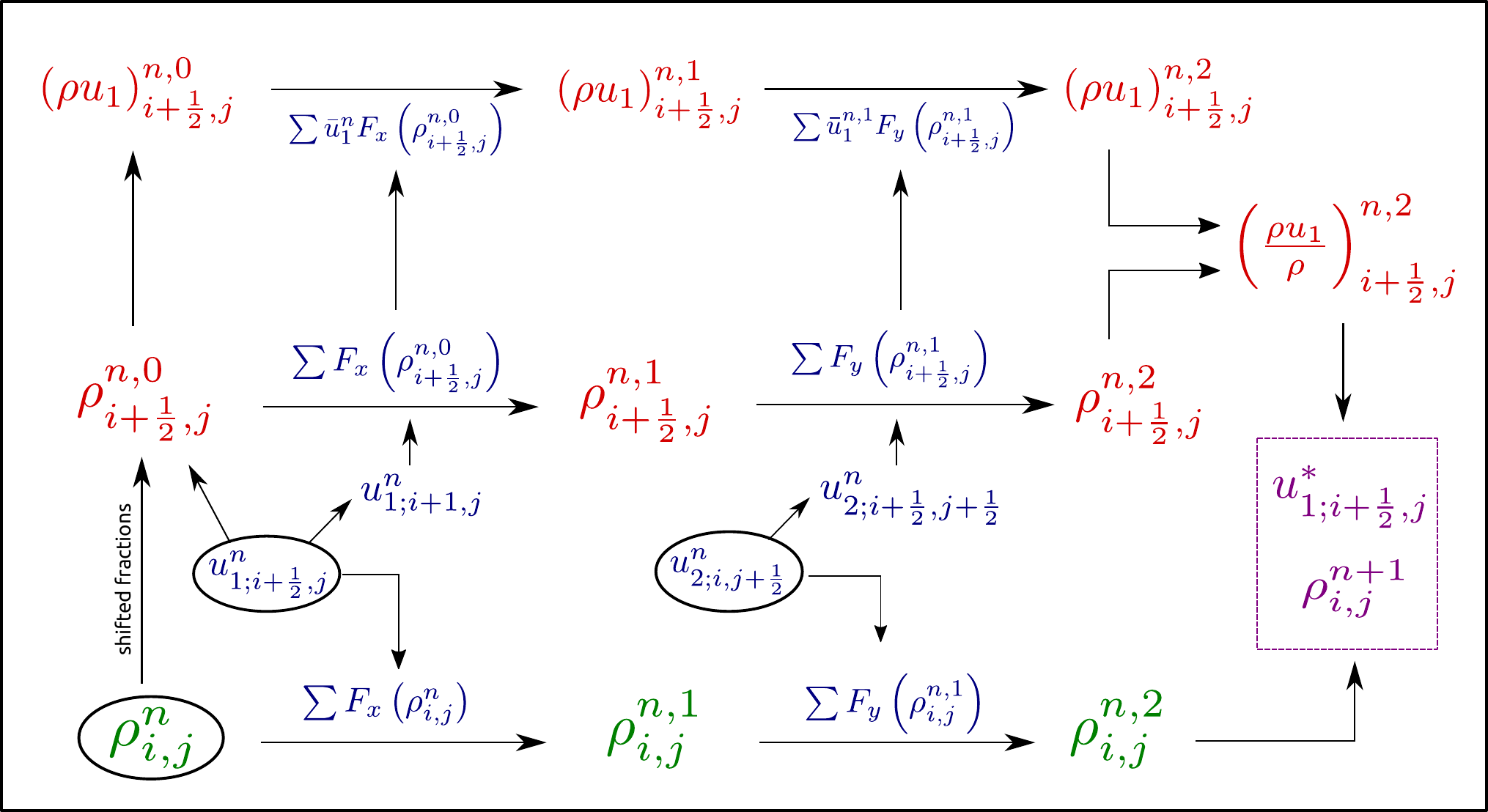}
\end{center}
\caption{\red{Diagram of the time step. For simplicity is represented the 2D case for the
density on the grid $i,j$ and horizontal velocity $u_1$ on the staggered grid $i+1/2,j$.
The evolution of the velocity component $u_2$ on the staggered grid $i,j+1/2$ is similar.
The initial variables $\rho^n$, $u_1^n$, $u_2^n$ are inside the ellipses. The interpolated
``advecting'' components $u_1$ and $u_2$ have superscript $n$. The shifted density $\rho^{n,0}$
is constructed with the shifted fractions of $C$ to initialize the momentum component
$(\rho u_1)^{n,0}$. The first split advection is along the $x$ direction  to 
variables with superscript $n,1$, the second one is along the $y$ direction to
variables with superscript $n,2$. The updated density is $\rho^{n,2}=\rho^{n+1}$
while the horizontal velocity $u_1^{n,2}=u_1^{*}$ enters the RHS of the Poisson-like
equation \eqref{Pois}.}}
\label{figscheme}
\end{figure}
}

\subsection{Description of the other time-split terms}

The other time-split terms in equation (\ref{conspredictedvel}) and in the projection
step (\ref{fotm}) are solved in a standard centered way. The density on the faces
of the central cells $\Omega_{i,j,k}$ is estimated using a simple average
$\rho_{i+1/2,j,k} = (\rho_{i,j,k} + \rho_{i+1,j,k})/2$. Although this is less accurate and 
consistent than the usage of the densities $\rho_q$, computed from
the shifted fractions as described above, the simple average 
is used both for simplicity and because tests have shown that the usage
of $\rho_q$ leads to less stable simulations. 

The velocities in the diffusion term are introduced in an explicit way. Although this requires small
time steps of the order $\rho h^2/\mu$, the capillary restriction on time steps
is usually even smaller, being of order $\tau = (\rho h^3/\sigma)^{1/2}$. The two restrictions
become of the same order when $h \sim l_{\mu \sigma}$, where $l_{\mu \sigma} = \mu^2 / (\sigma \rho)$ 
is the length at which the viscous and capillary terms balance. For water, this length is 
of the order of 10 nanometers, and grids of that size are not used in the flows we consider. 
However, should the velocities be treated in an implicit manner, we do not believe this would
change the conclusions of this paper. 

Surface tension is computed using the Continuous Surface Force method proposed by 
\cite{brackbill92}, together with an estimate of the curvature through the computation
of height functions, in a manner that closely follows the method of \cite{popinet09}. 
The external forces in equation (\ref{conspredictedvel}) 
are only gravity and are computed in a trivial manner with 
$\frac 1 {\rho^{n+1}} \LLL_{ext} =  \G$, where gravity $\G$  is a constant.

\section{Testing and Validation}
\label{test}
\subsection{Consistent cylinder advection}

\begin{figure}
\centering
\begin{minipage}{0.48\textwidth}
\centering
\includegraphics[width=0.98\textwidth]{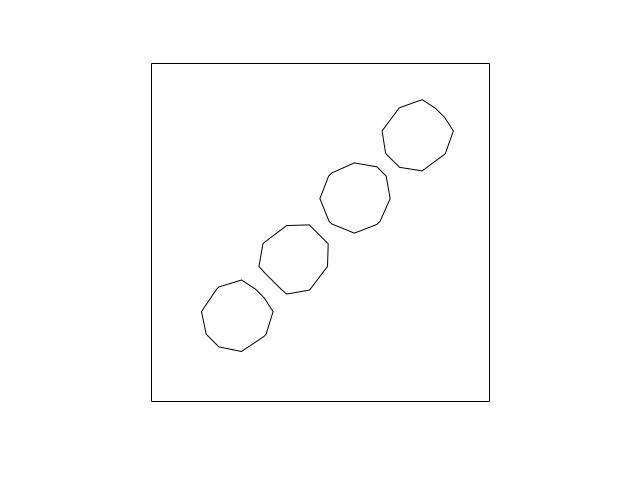}
\end{minipage}
\begin{minipage}{0.48\textwidth}
\centering
\includegraphics[width=0.98\textwidth]{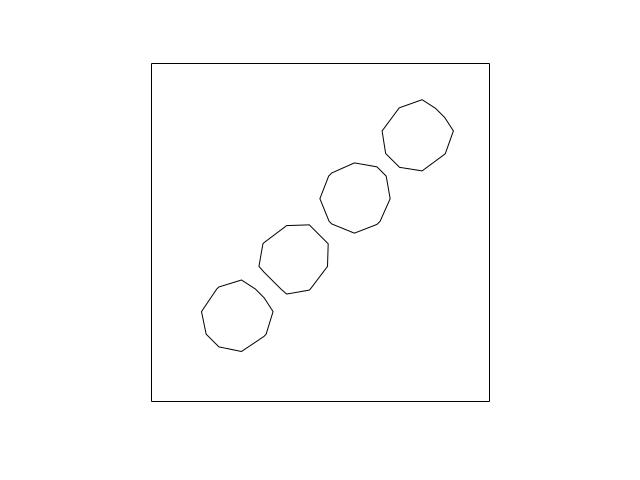}
\end{minipage}
\caption{Large-density-ratio droplet in a uniform velocity field: a droplet 
with $D/h=3.2$ grid points per diameter is advected in 3D in the plane $z=0$ 
(see text); left:  density ratio $10^9$ without viscosity,
right: density ratio $10^3$ with viscosity}
\label{CylAdv}
\end{figure}

An elementary test of our method, that mostly verifies that the coding 
has been performed correctly, considers a uniform planar velocity field 
$u_1 = u_2 = 1.6 \times 10^{-2}$ and a droplet of density $\rho_l = 10^9$ 
in gas at density $\rho_g=1$ with a CFL number of $0.0256 \sqrt 2$.  
Viscosity and surface tension are set to zero in this first test.  
The number of grid point in the diameter of the droplet is $D/h=3.2$. The unit domain is 
discretized on a $16 \times 16$ grid. The droplet shapes that result are shown 
on the left of Fig. \ref{CylAdv}. 
The irregularities seen in the advected droplet are due to the roughness of 
the VOF approximation at such low resolutions. We repeat the test with 
conditions close to air/water: now viscosities are $\mu_l = 0.1$, $\mu_g=0.002$ 
and  densities are $\rho_g=1$, $\rho_l = 10^3$, while there is still no
surface tension. We get identical results: a viscosity contrast 
will not generate numerical instabilities on a uniform velocity field, as shown 
on the right of Fig. \ref{CylAdv}. 

\subsection{Kelvin Helmholtz Instability}
\label{sec_khi}

The Kelvin-Helmholtz instability arises between unequal velocity fluid
streams. It is closely related to the issues addressed in the current
paper since it arises in many of the flows for which the current
method is designed, such as atomisation. Moreover, the
Kelvin-Helmholtz instability is particularly strong on a vortex sheet,
since (as we show in the next section) it has for an infinitely thin sheet a
divergent growth rate as the wavenumber goes to infinity. Compounding
the issue, the baroclinic term of the vorticity equation leads for unequal densities to the
creation or strengthening of a vortex sheet on the interface.
In previous papers some of us have studied the
Kelvin-Helmholtz instability in viscous flows with surface tension
\cite{yecko02,boeck05,bague10}.

We focus here, in contrast to these earlier papers,
on the inviscid, no surface tension case.
Indeed we want to focus on the new discretization of the advection terms.
Moreover the inviscid, no surface-tension case is
a kind of ``worst-case scenario'' without the stabilizing effects of viscosity
and capillarity.

\begin{figure}
\begin{center}
\includegraphics[width=0.6\textwidth]{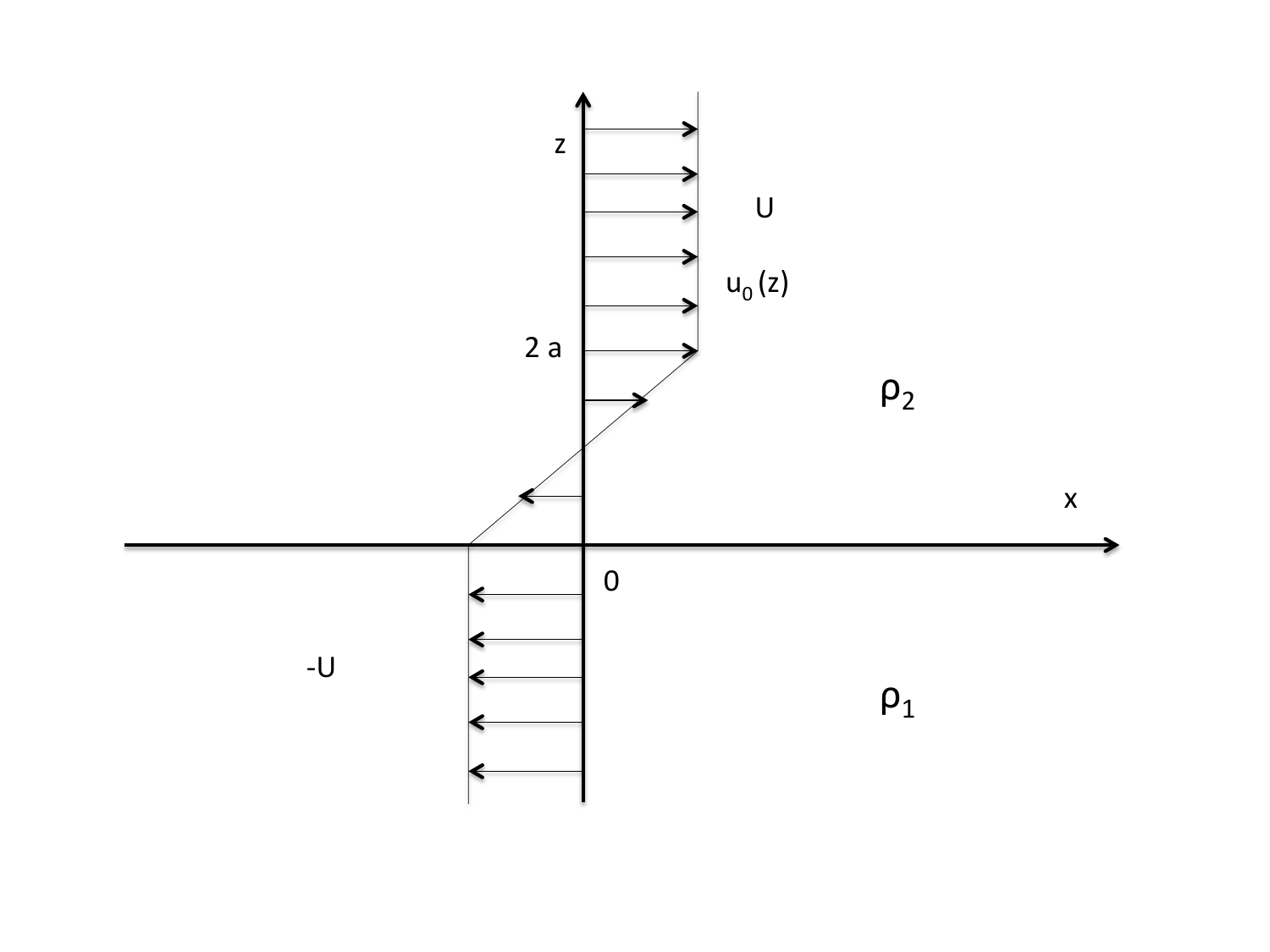}
\caption{Base velocity profile. \label{khappfig} }
\end{center}
\end{figure}
\begin{figure}
\centering
\includegraphics[width=0.3\textwidth]{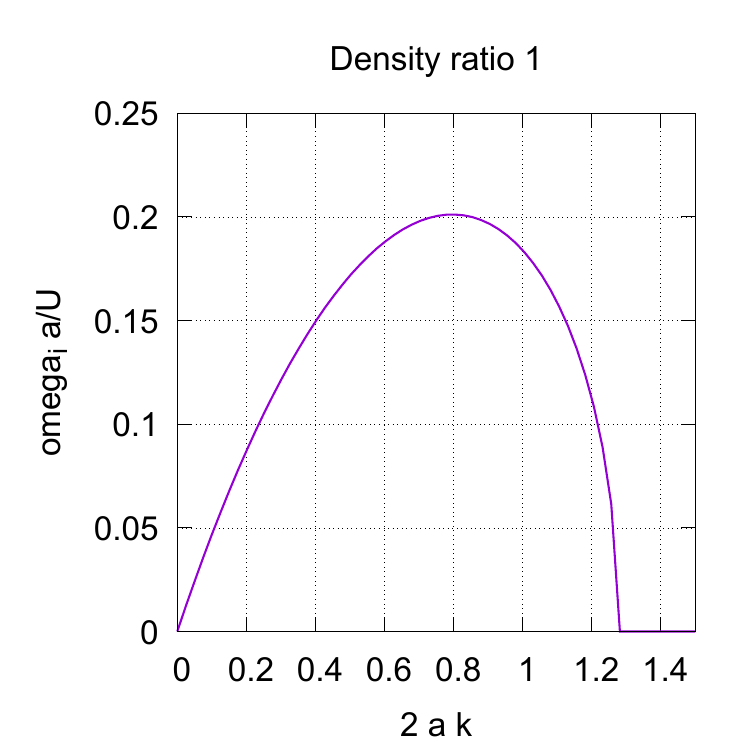}\quad
\includegraphics[width=0.3\textwidth]{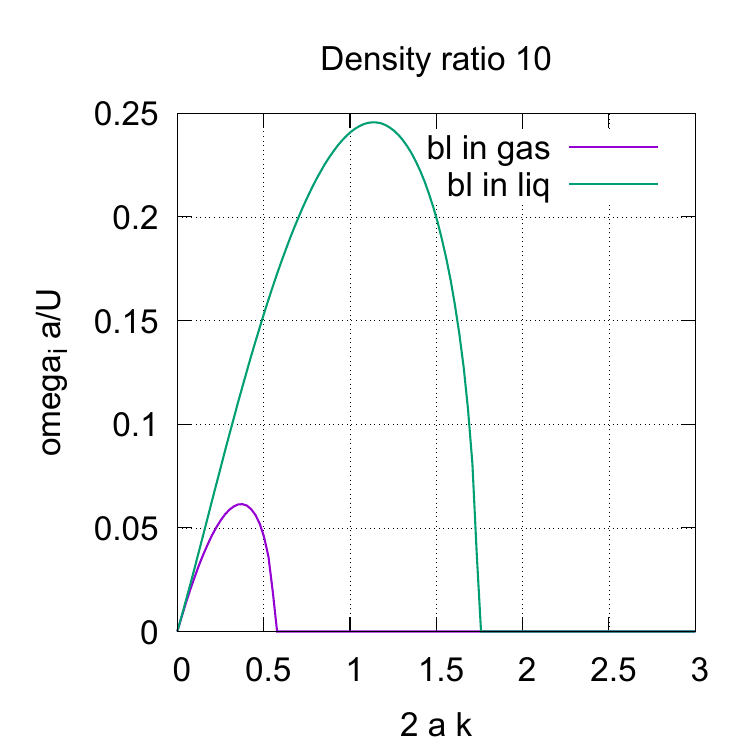}\quad
\includegraphics[width=0.3\textwidth]{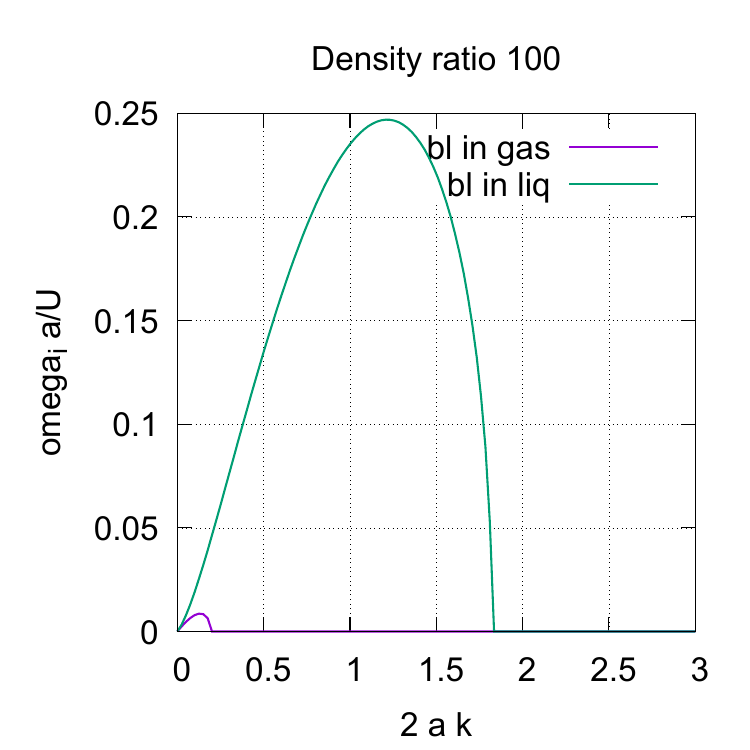}
\caption{The reduced growth rate as a function of the dimensionless wavenumber.}
\label{grr}
\end{figure}

\subsubsection{Problem setup}

The simplest setup is that of a vortex sheet, for which the growth rate is
given by
\be
\om_i = \frac{2 \sqrt r}{r+1}  k U  \label{omkazi}
\nd
where $r=\rho_g/\rho_l$.
This vortex-sheet setup is however leading to an ill-posed problem since for increasing $k$,
exponentially large amplitudes are predicted at finite time. It is thus desirable
to regularize the growth rate by having  a boundary layer.
The theory of the inviscid Kelvin-Helmholtz instability with a boundary layer is
summarized in  \ref{appKH}. We consider two kinds of boundary layers: (i) a boundary layer
in the liquid phase, (ii) a boundary layer in the gas phase. Both cases may be represented
by the sketch in Fig. \ref{khappfig}.  The resulting growth rate
is a function of the dimensionless number $\kappa = 2 a k$. The dimensionless
growth rate $\Omega_i = \omega_i a / U$ is plotted on Fig. \ref{grr} for three different
density ratios, and for both cases (i) and (ii) above.
It is seen that the growth-rate curve has a maximum at $\Omega_{i,max}$
for $\kappa=\kappa_{max}$ and vanishes above the threshold $\kappa_c$.
As the density contrast increases, the behavior is different depending on the case. (i) For a boundary layer in the liquid, the interfacial problem becomes in the limit $r \rightarrow 0$ a free-surface problem and the growth rate goes to a limit. (ii) On the other hand for a boundary layer
in the gas, the maximum growth rate and  the corresponding maximum reduced wavenumber $\kappa_{max}$
both tend to zero and
\be
\Omega_{i,max} \sim \sqrt r,\quad \kappa_{max} \sim \sqrt r
\nd
We initialize all simulations with one perturbed wavelength in the computational domain, so that
$k=2\pi/L_x$ and we choose parameters so that $k=k_{max}$. Then 
for $\kappa_{max} = 2 a k_{max} = 4 \pi a / L_x$ the number of points in the boundary layer is $2a/h= L_x \kappa_{max}/(2\pi h)=n_x \kappa_{max}/(2\pi)$ where $n_x=L_x/h$ is the number of grid points in the horizontal. 
In the $r=0.01$ case with the boundary layer in the gas,
we have $\kappa_{max} \simeq 1.2$ and thus $a \simeq 0.2 L_x$ for $k$ near $k_{max}$
and $2 a/h \sim n_x/5$.
We initialize the simulation with the following method. We use the solution obtained in
 \ref{appKH} for the mode corresponding to $\kappa_{max} = 2 ak_{max}$. The modes computed in
\ref{appKH} are extending to $\pm \infty$ in the vertical, so a sufficiently large box
in the vertical has to be selected. We thus use an $L_x \times L_z$ domain with $L_z = 2 L_x$.
The error on the boundary at $z=\pm L_z/2$ is of order $e^{-kL_z/2} = e^{-2\pi} \simeq 2 \, 10^{-3}$.

We choose the boundary layer thickness and the initial amplitudes of the perturbations
such that the excited wavelength corresponds to the box size. Moreover it is important to notice
that wavenumbers are
quantized in the box so that $k_n =2\pi n /L_x$.
Since $2 \kappa_{max} > \kappa_c$ as can easily be seen on
Fig. \ref{grr} all the quantized modes above the first one are stable. This would not be the case if
a narrower boundary layer were used. In the other limit, for a vortex sheet setup ($a=0$), all
modes are unstable in stark contrast with the above situation.

The flow field $u_0(x,z,t=0),w_0(x,z,t=0)$ in the initial condition is initialized using a discrete
stream function formulation. The motivation for this is the desire to avoid a projection of the
velocity field at the first time step. Indeed, less careful initialization methods may start with a velocity field that has a large localized divergence. One may wonder how the pressure gradient in the projection step rearranges this velocity field, possibly inadvertently transferring energy to modes other than the theoretically selected one.
As a result, the theory in \ref{appKH} computes the stream function $\psi$ at initial time.
The code computes the finite differences
\be
u_0 = \derz^h \psi, \qquad w_0 =- \derx^h \psi
\nd
allowing the resulting field to be discretely divergence free. 
This means that it obeys \refeq{cont-eq1} at machine accuracy.
The unperturbed interface, according to the theory, is located in the mid-horizontal-plane $z=0$ of the $(1\times2)$ box
thus exactly at the boundary of two ``centered'' $C_{ijk}$-cells.
However it is possible and interesting to shift vertically by $\Delta z = h/2$ the origin
of the coordinate system to locate the interface in the {\em middle} of a row of
``centered'' $C_{ijk}$-cells.
In addition in what follows we uniformly set $L_x=1$.

\subsubsection{Results for boundary layer in the liquid phase}
\begin{figure}
\centering
 \includegraphics[width=0.49\textwidth]{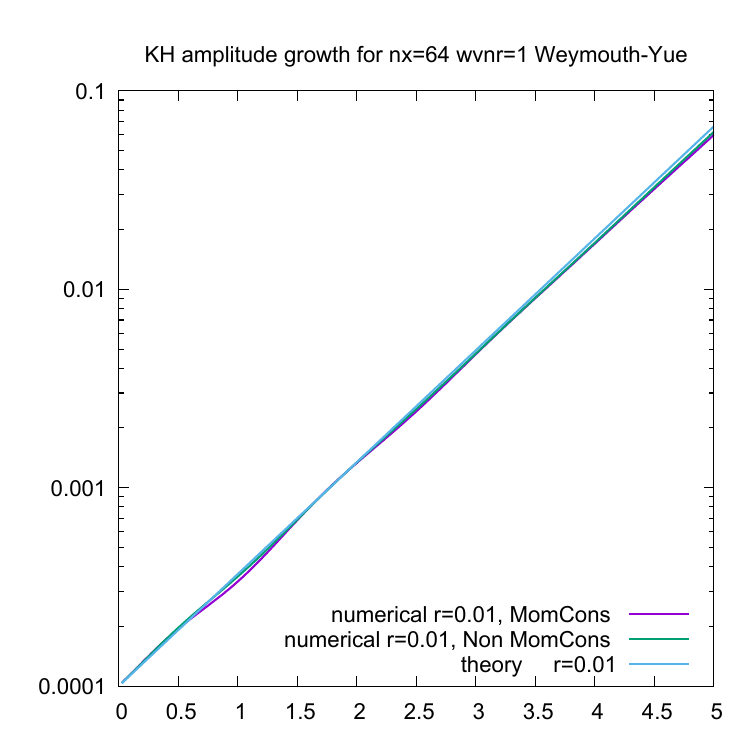} \includegraphics[width=0.49\textwidth]{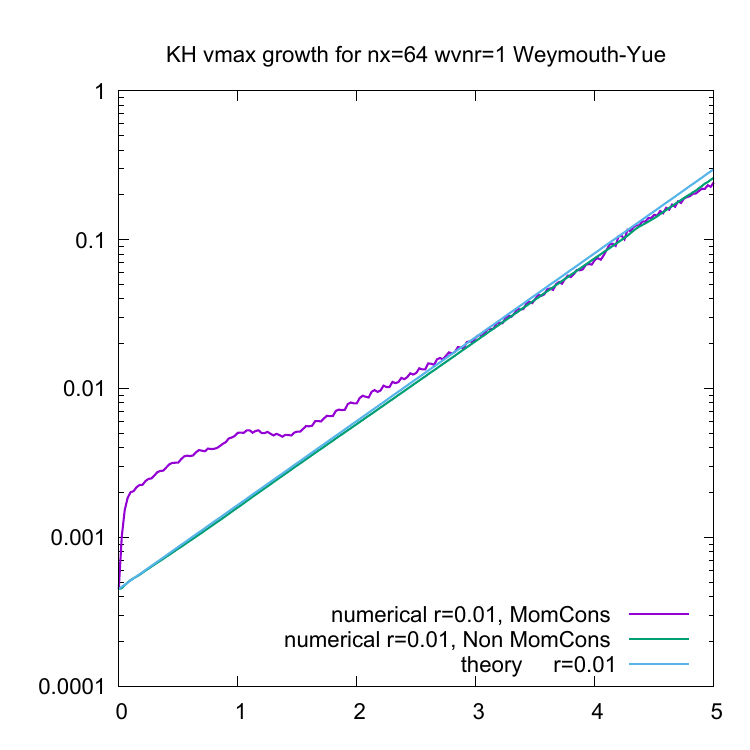} \\
\centering (a) \hskip 5.5cm  (b)
\caption{Comparison of the theoretical and numerical growth for a boundary layer in the liquid
as a function of time with $n_x=64$ grid points, 13 points in the boundary layer and shift $\Delta z=0$ (see text): 
(a) amplitude plot; (b) maximum velocity plot. The signal ``jumps'' by one order of magnitude for 
the consistent method (nicknamed ``MomCons'') while it remains close to the theory for the 
non-consistent method.
\label{gr1}}
\end{figure}

In order to verify that the numerical scheme is consistent with the theory, we first compute the growth rate
for case (i). 
In what follows we first show results for the WY/QUICK-UW combination of VOF and momentum advection methods, and both for the classical method and with the VOF-momentum-consistent method. 
The case for $a=\kappa_{max} L_x/(4\pi) = 0.095$ is shown in Fig. \ref{gr1}. The numerical results are plotted in
two manners. The ``amplitude growth'' plots follow the logarithm of the amplitude of the Fourier mode of
wavenumber $k=2\pi$. This has the effect of filtering any contribution of the other modes (the other modes
should not grow according to the theory but could still appear because of numerical approximations).
The second plot is the ``maximum velocity'' plot which follows the logarithm of the max norm
of the vertical velocity $w_{max}(t) = ||w||_{\infty}$. The maximum velocity
corresponds to a superposition of the amplitude of all the modes. 

We first show the results for $n_x=64$ which corresponds to $2a/h \simeq 13$ points in the boundary layer
with $\Delta z=0$. The initial interface perturbation amplitude is
$A_h=10^{-4}$ and that of the velocity is  $w_{max}(0) = | \omega A_h |$
(see Appendix for the definition of $A_h$).
The results for amplitude growth are good, showing exponential growth over nearly three orders of magnitude. However the results for the maximum velocity growth show a glitch, with an nonphysical jump in
the velocity in the case of the VOF-consistent method. The glitch is created by small scale vortical structures
with large wavenumber near the interface. These structures have a small amplitude and are overtaken by the physical growth of the $k=2\pi/L_x$ mode after some time, with a recovery of the predicted growth rate.
In the case of the non-consistent method the glitch is not seen. If now one places the interface in
the middle of the cell the glitch disappears. 
\begin{figure}
\centering
 \includegraphics[width=0.49\textwidth]{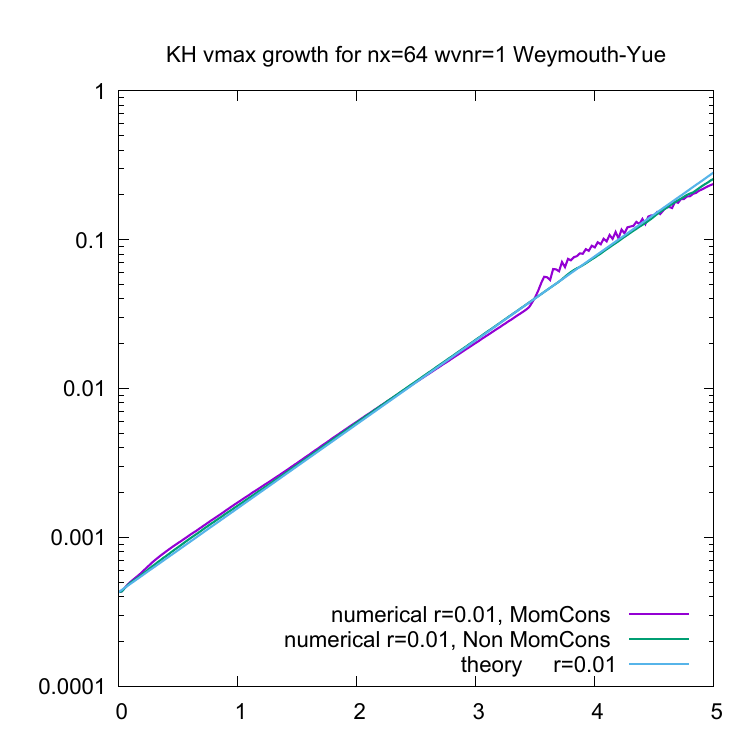} \includegraphics[width=0.49\textwidth]{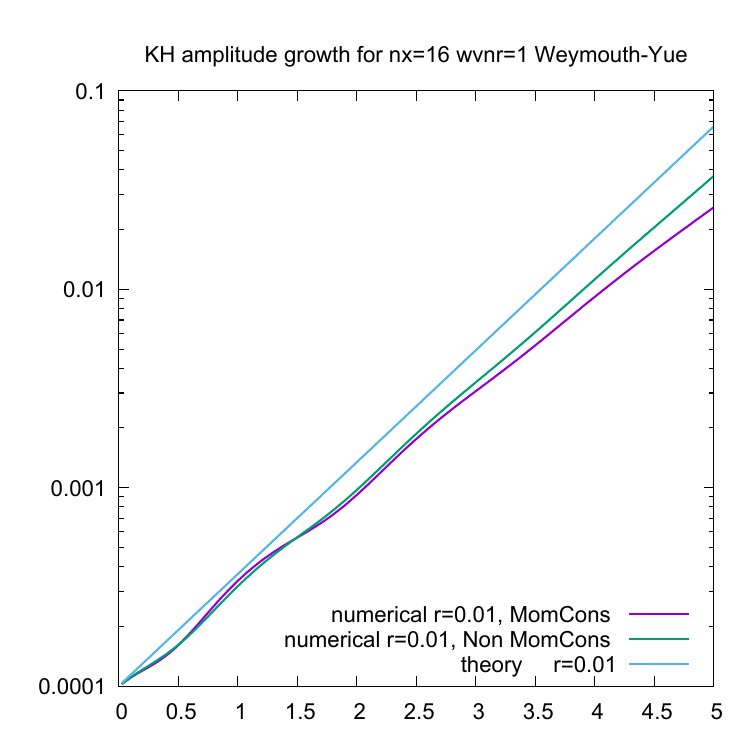} \\
\centering (a) \hskip 5.5cm  (b)
\caption{Comparison of the theoretical and numerical growth for a boundary layer in the liquid
as a function of time:
(a) maximum velocity plot for shift $\Delta z=h/2$ and other parameters as in Fig. \ref{gr1}.
The nonphysical ``jump'' for the consistent method disappears; (b) amplitude plot for a 
smaller number of grid points $n_x=16$ (with $2a/h=3$ points in the boundary layer)  and 
shift $\Delta z=h/2$. 
\label{gr1c}}
\end{figure}
\begin{figure}
\centering
 \includegraphics[width=0.49\textwidth]{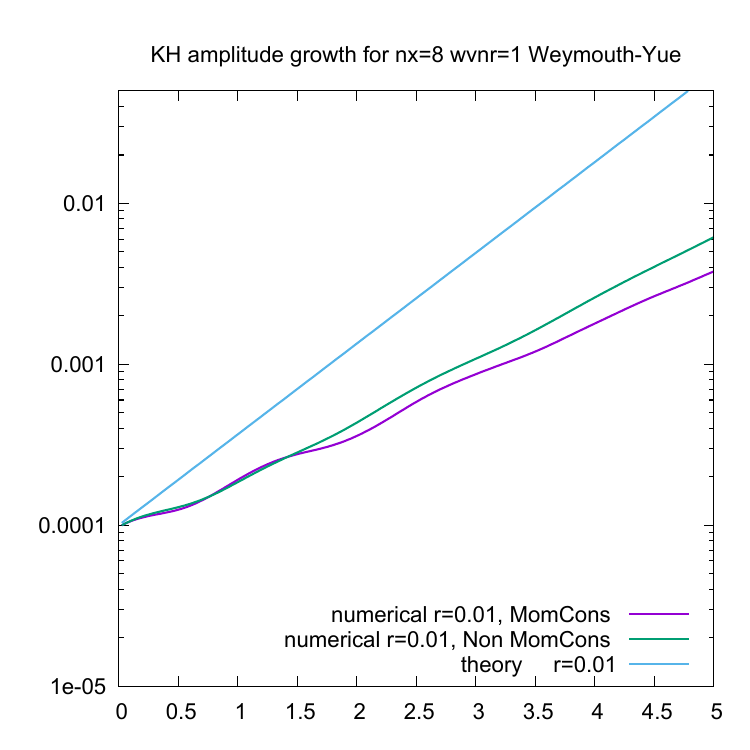} \includegraphics[width=0.49\textwidth]{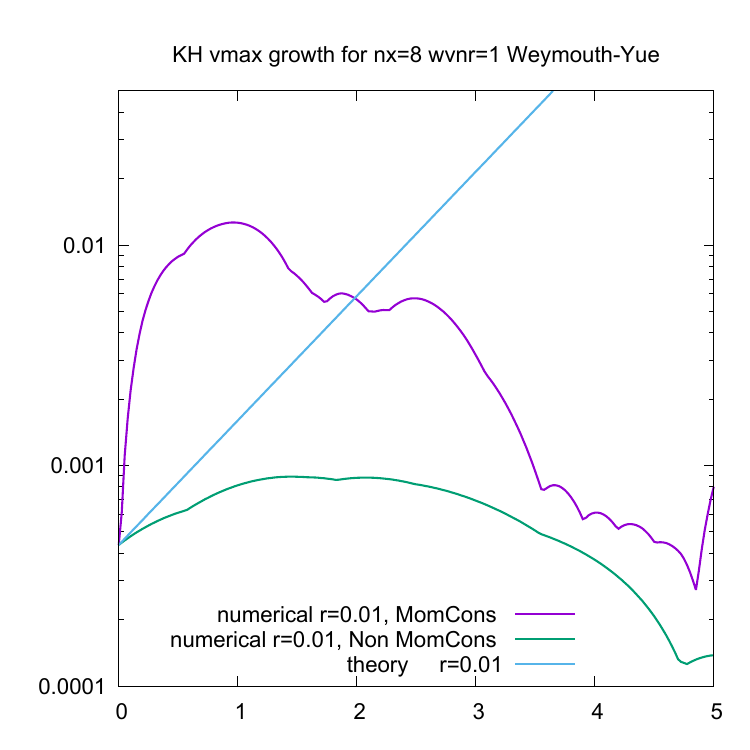}\\
\centering (a) \hskip 5.5cm  (b)
\caption{Comparison of the theoretical and numerical growth for a boundary layer in the liquid
as a function of time with $n_x=8$ grid points (with $2a/h=1.5$ points in the boundary layer):
(a) amplitude plot with shift $\Delta z=h/2$; (b) maximum velocity plot in the
``worst case'' scenario with shift $\Delta z=0$. The nonphysical ``glitch'' for the consistent 
method is marked, but both methods eventually result in the damping of the perturbation. 
\label{gr3}}
\end{figure}

As expected, the results become worse when the boundary layer contains a smaller number of grid points. For $n_x=16$  exponential growth is still observed (Fig. \ref{gr1c}b)
over nearly three orders of magnitude, but with a growth rate approximately 13\% smaller than the theoretical one. For $n_x=8$ there is not much more than a single grid point in the boundary layer,
and the growth rate is approximately 50\% smaller (Fig. \ref{gr3}a). Finally the $n_x=8$ case for $\Delta z= 0$ and with
the maximum velocity plot is a kind of  worst case scenario. However even in that case the
code does not diverge (Fig. \ref{gr3}b), and the growth is eventually damped,
which despite the disagreement with theory
has the advantage of stabilizing the computation.

\subsubsection{Results for boundary layer in the gas phase}
\begin{figure}
\centering
 \includegraphics[width=0.49\textwidth]{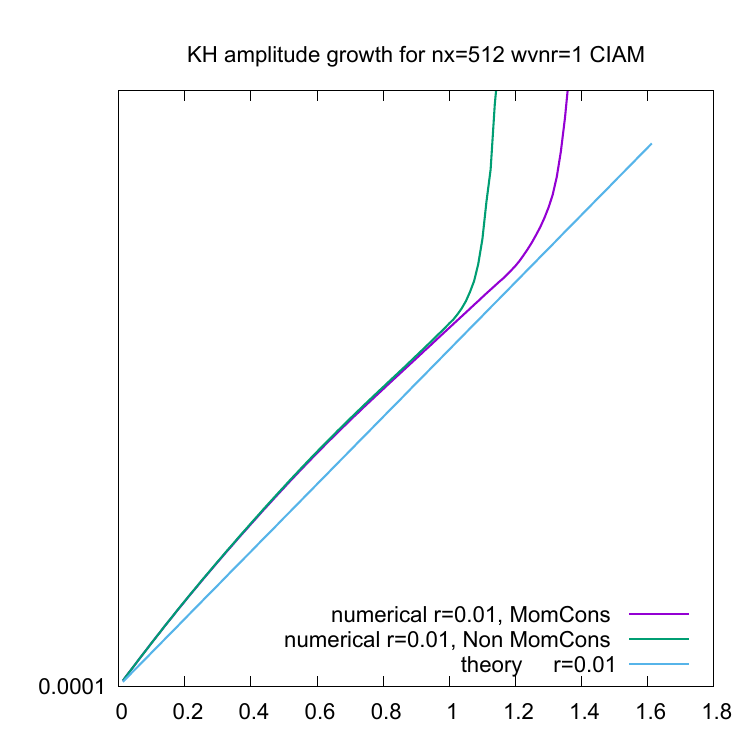} \includegraphics[width=0.49\textwidth]{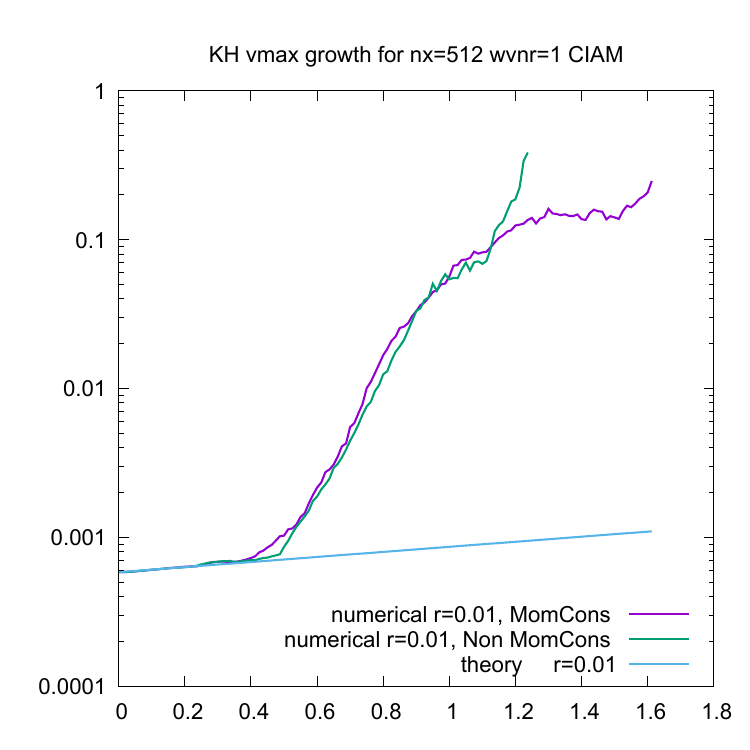}\\
\centering (a) \hskip 5.5cm  (b)
\caption{Comparison of the theoretical and numerical growth for a boundary layer in the gas
as a function of time with $n_x=512$ grid points, $a/h\simeq 12$ points in the boundary layer,
shift $\Delta z=h/2$ and $U\tau/h=0.32$: (a) amplitude plot. The simulation follows 
approximately the theory but blows up after time $t\simeq 1$, with a faster blow up for the 
non-consistent method; (b) maximum velocity plot. The rapid growth starting at
$t\simeq 0.7$ is due to the appearance of small structures that are stable in the linear theory 
but not in the numerics.
\label{gr4}}
\end{figure}

We now turn to the case where the boundary layer is in the gas phase. For $r=100$ the maximum
growth rate is obtained for $\kappa_{max} \simeq 0.143$. We now have
a reduction by approximately one order of magnitude of $\kappa_{max}$
and thus also of $a$ and $a/h$. Now if $L_x$ is the wavelength with
maximum growth, the boundary layer thickness must be $a \simeq 0.011$ and
the number of grid points in the boundary layer is much smaller
than with the boundary layer in the liquid phase, with $a/h \simeq 0.02 n_x$.
Indeed for $n_x=64$ there is only about one point in the boundary layer.
It is interesting to notice that if a larger $a$ were used, the gas boundary layer would become
stable, and the only growth would result from the spurious numerical growth of modes which
in theory should be stable. 
With $n_x=512$ there are
about 12 points in the boundary layer. The computation is beset by numerical instabilities
that lead to the blowup of the simulation, so we use $\Delta z=h/2$ and the CIAM/Superbee method,
as the later is more stable. The results are shown in Fig. \ref{gr4}a. Both the consistent and non-consistent method give identical results at short times, close to the theory, but with less accuracy than when
the boundary layer is in the liquid phase. At times around $t=0.7$ a small scale instability starts appearing in
the maximum velocity graph (Fig. \ref{gr4}b) and eventually causes the demise of the simulation.

\begin{figure}
\centering
  \includegraphics[width=0.49\textwidth]{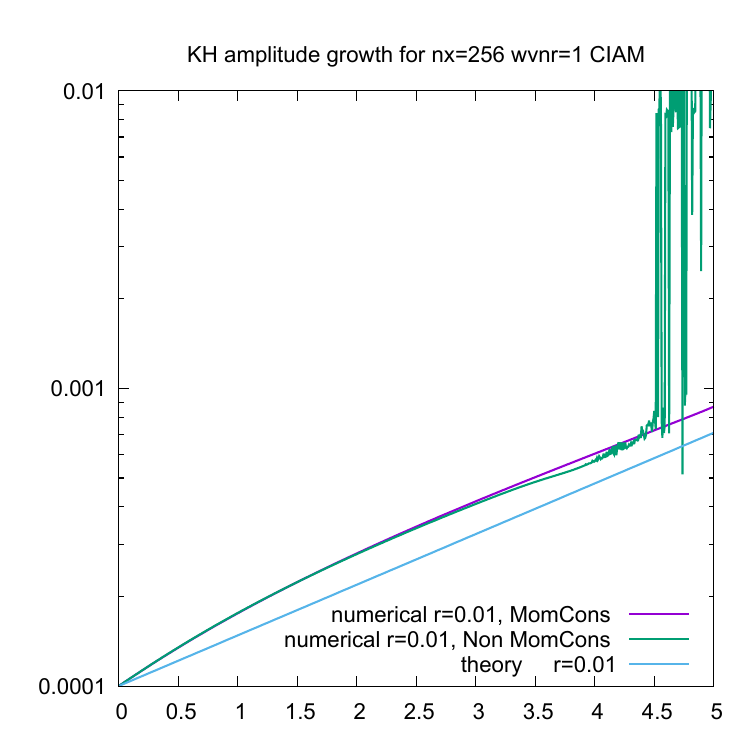} \includegraphics[width=0.49\textwidth]{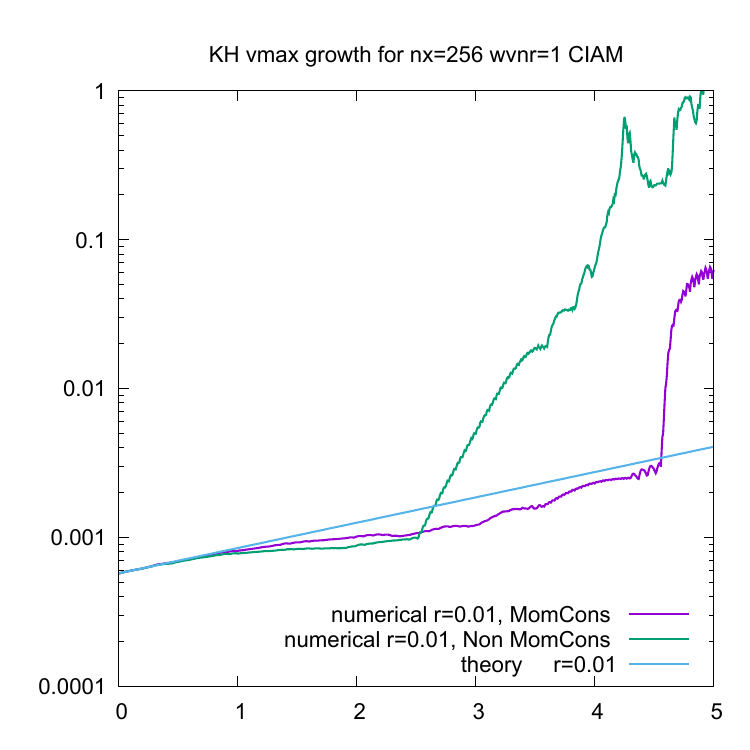}\\
\centering (a) \hskip 5.5cm  (b)
\caption{Comparison of the theoretical and numerical growth for a boundary layer in the gas
as a function of time with $n_x=256$ grid points, $a/h\simeq 6$ points in the boundary layer,
shift $\Delta z=h/2$ and $U\tau/h=0.08$: (a) amplitude plot; (b) maximum velocity plot.
\label{gr5}}
\end{figure}

With a smaller number of points, $n_x=256$ and $a/h=6$ points in the boundary layer, the results shown
in  Fig. \ref{gr5} display again a spurious growth due to a small scale numerical instability
for the non-consistent method around $t=2.5$.

\subsubsection{Results for the vortex sheet case}
\label{subsec_khi}
\begin{figure}
\centering
  \includegraphics[width=0.49\textwidth]{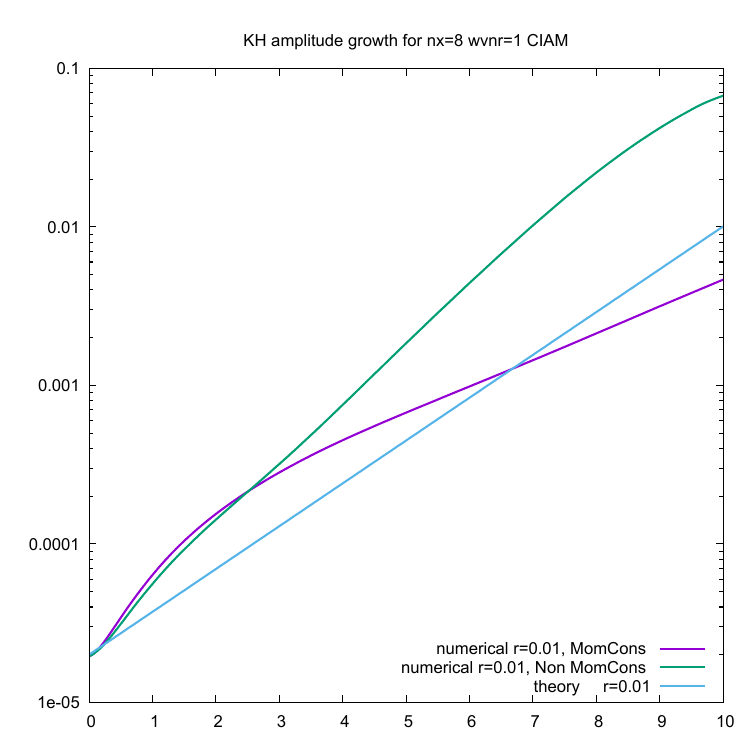} \includegraphics[width=0.49\textwidth]{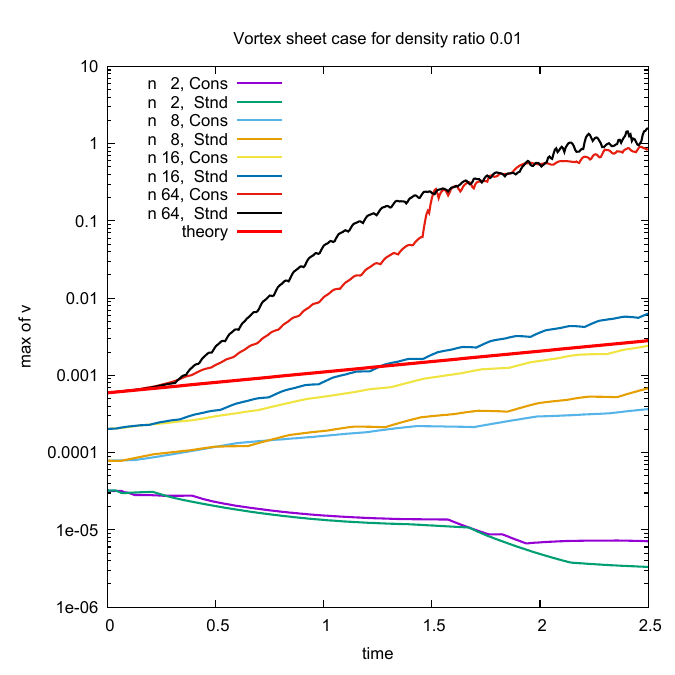}\\
\centering (a) \hskip 6cm  (b)
\caption{
Kelvin-Helmholtz instability in the vortex sheet case with density ratio $r=0.01$:
(a) amplitude plot for $n_x=8$; (b) maximum velocity plot. The number $n$ indicates the 
ratio $\lambda/h$ of wavelength to grid size, ``Cons'' stands for consistent method and 
``Stnd'' for non-consistent method. 
\label{gr6}}
\end{figure}

In the vortex sheet case, there is no boundary layer thickness and there are no remaining length scales
from the theoretical (continuum) point of view.
Various ratios of wavelength to grid size are plotted on  Fig. \ref{gr6}.
In this case the results are best for $\lambda/h=8$ and adding more grid points does 
not improve the results, except for a short interval of time for  $\lambda/h=64$.
The theoretical growth rate is indeed obtained for $n_x=64$ in a small interval of time for
$t < 0.25$.
For the least resolved case, the highest grid (Nyquist) frequency corresponding to
$\lambda/h=2$, there is no amplification but rather damping,
indicating that numerical dissipation
defeats the instability at this wavenumber.
This indicates that the mechanism that makes the computation diverge in other cases is not
represented by the Kelvin-Helmholtz instability at near-Nyquist
wavenumbers $kh ={\cal O}(1)$. The latter is
true at least in the
current setup of this instability.  Other setups, with for example different $\Delta z$, or
with interfaces not aligned on the grid, may give different results.

\begin{figure}
\centering
 \includegraphics[width=0.49\textwidth]{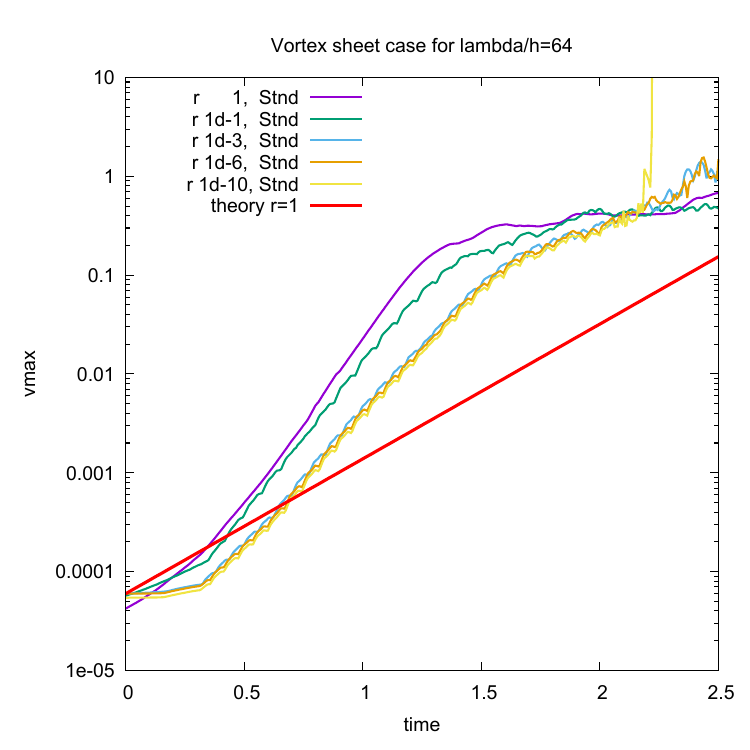} \includegraphics[width=0.49\textwidth]{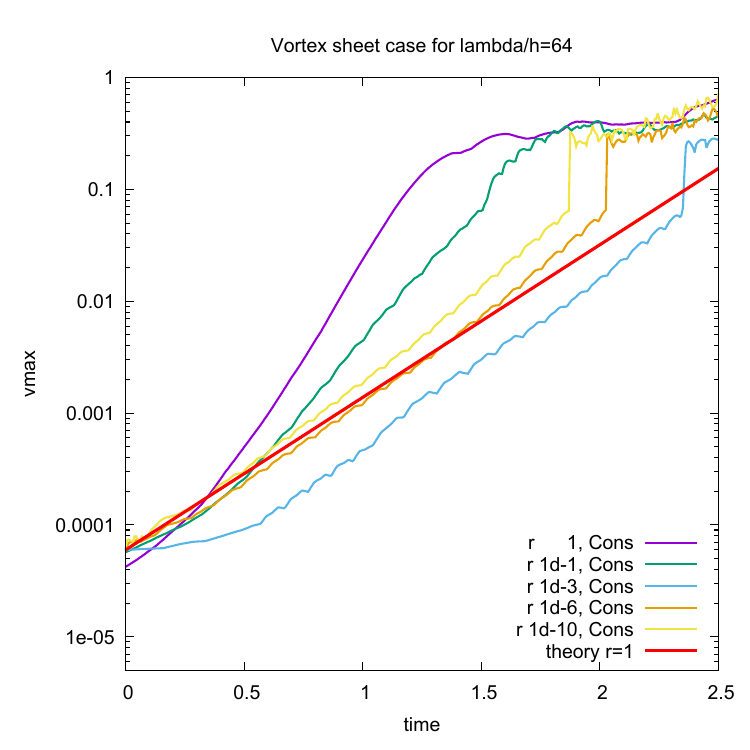} \\
\centering (a) \hskip 6cm  (b)
\caption{
  Kelvin-Helmholtz instability in the vortex sheet case, with shift $\Delta z=h/2$,
  $\lambda/h=64$ and different values of $r$. The theory for $r=1$ is also plotted.  
  (a) Non-consistent method. The growth rate is always much larger than in the theory, 
  and it decreases with $r$, reaching a minimum around $r=10^{-3}$. Around $t=2.2$ the 
  maximum of the velocity diverges for $r=10^{-10}$. This divergence may be suppressed 
  by using a smaller tolerance for the iterative Poisson solver for the pressure, 
  but we keep it here for illustrative purposes. 
  (b) Consistent method. The agreement between the $r=10^{-6}$ case and
  the theory for $r=1$ is coincidental. As the value of $r$ is decreased,
  the growth rate first decreases, reaches a minimum around $r=10^{-3}$ and then 
  increases. 
  \label{gr7}}
\end{figure}

\begin{figure}
\centering
 \includegraphics[width=0.49\textwidth]{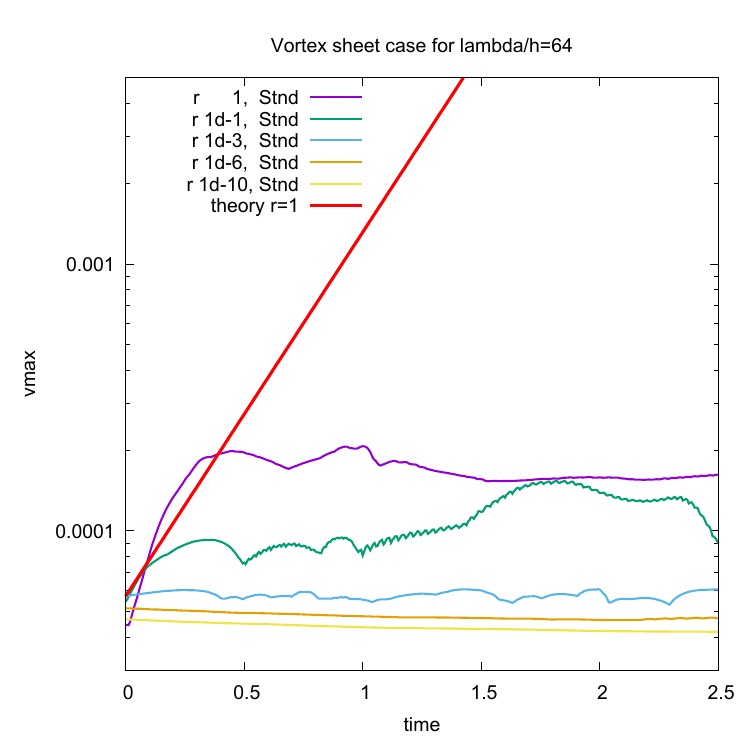} \includegraphics[width=0.49\textwidth]{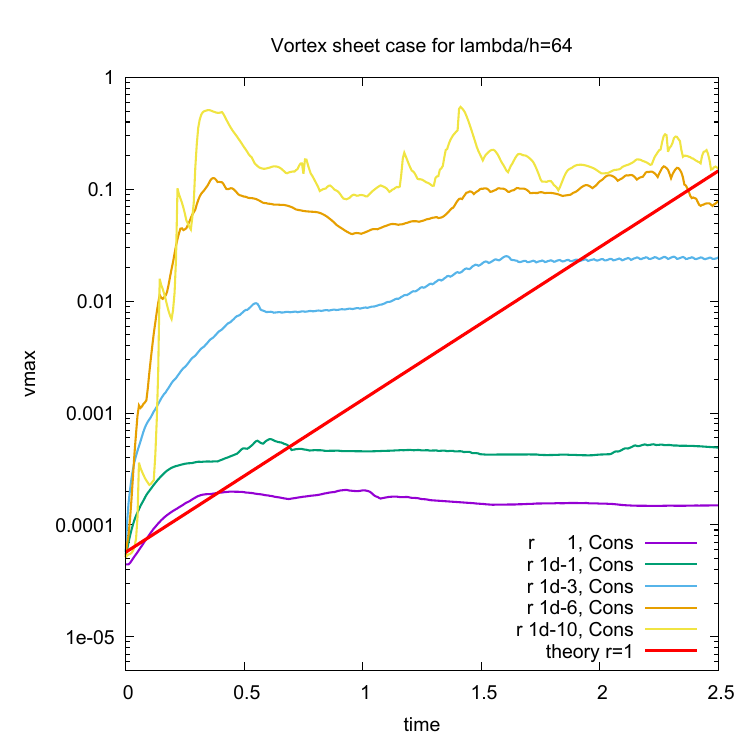} \\
\centering (a) \hskip 6cm  (b)
\caption{ Same parameters of Fig. \ref{gr7} but with shift $\Delta z=0$. 
  (a) Standard method. At large $r$, the expected vanishing growth rate is achieved. 
  (b) Consistent method. The amplitude of the maximum vertical velocity increases rapidly 
  with increasing density contrast. For $r=10^{-10}$ a super-fast increase is seen. 
  This fast increase is made smaller when the clipping value $\eps_c$ is increased and 
  is made even faster when $\eps_c$ is decreased.
\label{gr8}}
\end{figure}

However, it is still possible that the excitation of relatively large wavenumbers by the
initial perturbation could explain the uncontrolled growth of numerical instabilities at large
density contrasts. To test this hypothesis, we plot on  Fig. \ref{gr7} the 
numerical growth of $|| v ||_\infty$ compared to the theoretical growth for $r=1$,
for a fixed value of $\lambda/h$. We first describe the case $\Delta z= h/2$.
It is seen  on Fig. \ref{gr7} 
that the growth rate is larger than the theoretical one even for $r=1$. This
is explained by the leakage of energy from the initial $k=k_1$ mode
into modes with larger $k=k_n$ that have a larger growth rate
as expressed by (\ref{omkazi}). 
Also on Fig. \ref{gr7}  it is seen that for smaller values of $r$, the growth rates are reduced.
However, for the consistent
method, a stronger reduction is achieved. It is obvious that this reduction is still very far 
from being as strong as that predicted by the theory, which would yield a growth rate 
$\om_i \sim 10^{-5}$ for the smallest $r$.
We now turn to the case $\Delta z =0$. In this case, as shown on Fig. \ref{gr8}, 
the situation is reversed: the non-consistent method grows much more slowly than the 
consistent method. For $r=1$ the growth of the non-consistent method, after a fast
initial transient, is altogether stopped. For $r=10^{-10}$ a super-fast increase is seen.
This fast increase is made smaller when the clipping value (see Section \ref{clipping}) $\eps_c$ is increased and is made even faster when $\eps_c$ is decreased.
In the set of tests described in this section $\eps_c$ was set to $10^{-8}$.

\subsection{Sudden acceleration of a cylinder at large density contrast}
\label{sda}
A test that is often included in studies of momentum-conserving or mass-momentum-consistent methods
\cite{bussmann2002modeling,desjardins10,raessi12,le13,Vaudor:2017ip,zuzio2020new}
and other methods designed to improve the stability of two-phase flow
computations \cite{Fuster2013energy} is to initialize a droplet of very
high density at velocity $\U_l(\X)=U_0 \E$ with the other, lighter fluid, at
rest, so that $\U_g(\X)=0$. Surface tension and viscosity are not present
as in the previous test, the only difference being the
discontinuity of the initial velocity on the interface, which
amounts to a vortex sheet on the surface
of the cylinder. After the first time step, the projection method (\ref{fotm})
adds a dipole potential flow so that $\U_g = \tau \nabla p / \rho_g$
in the gas around the droplet, identical to the dipole flow around a solid object
and with a slightly different velocity $U$ of the droplet.
Indeed, the dipole flow absorbs some of the initial momentum of the liquid in the gas, and results
in a reduction from $U_0$ to $U$ of the droplet velocity during the first time step, similar to the momentum transfer
after the traversal of the droplet by a shock wave. This velocity shift, which is estimated
in \cite{marcotte2019density}, is small, of order $r=\rho_g/\rho_l$.
There are at least three ways to explain the perturbations created on the interface.

\subsection{Sudden acceleration of a cylinder at large density contrast}
\label{sda}
\begin{figure}
\begin{center}
\includegraphics[width=0.4\textwidth]{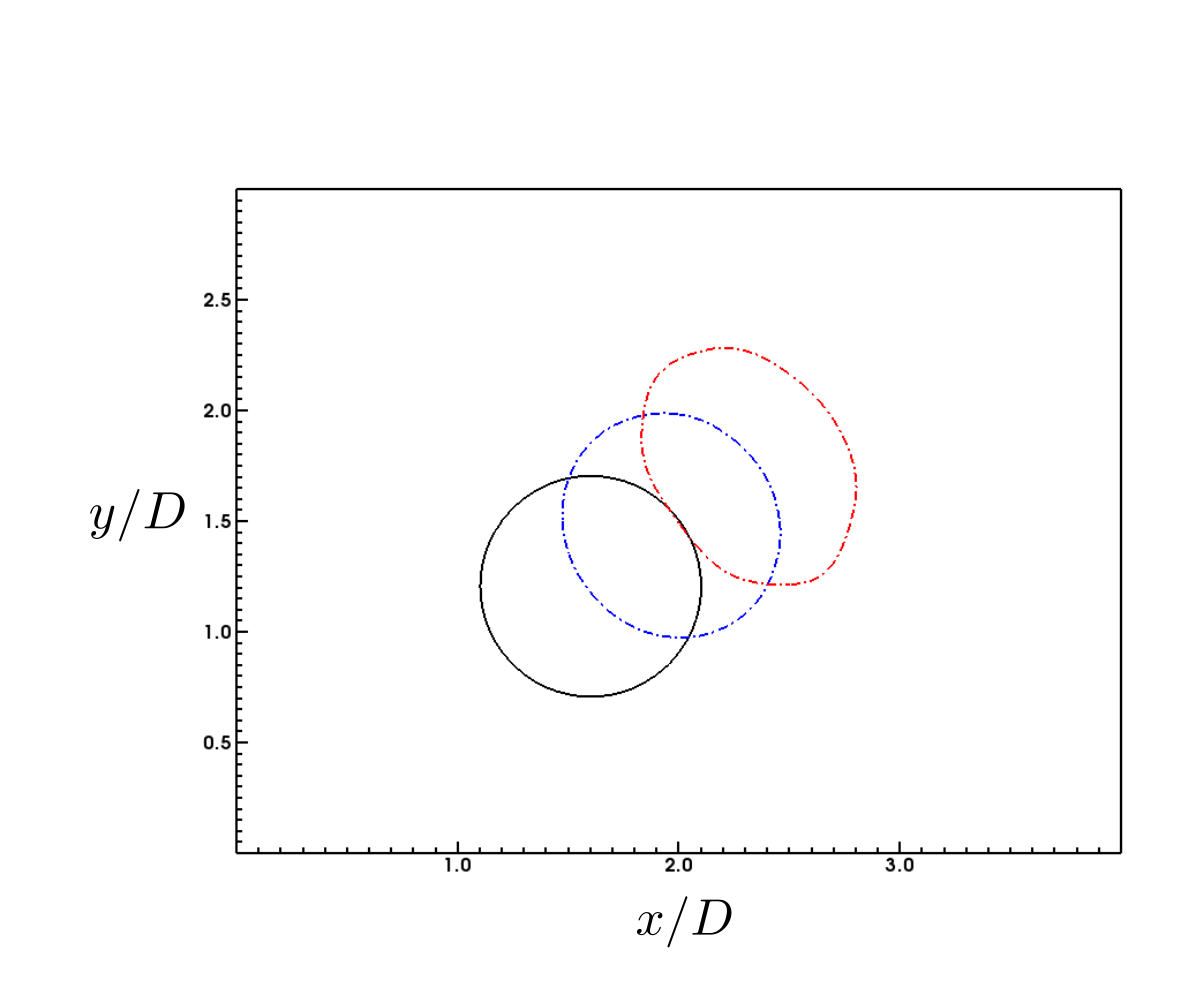}
\quad\includegraphics[width=0.4\textwidth]{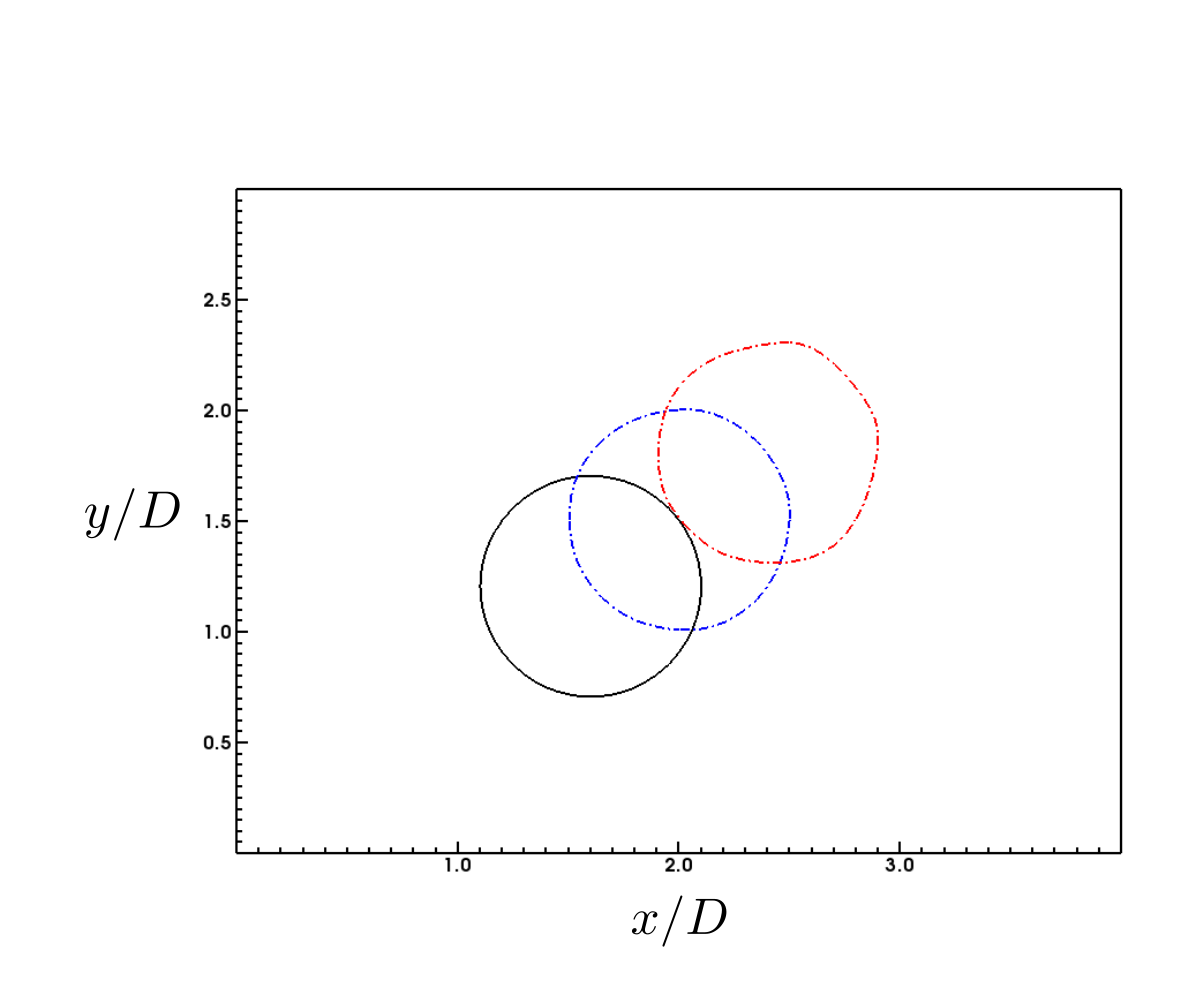} \\
\includegraphics[width=0.4\textwidth]{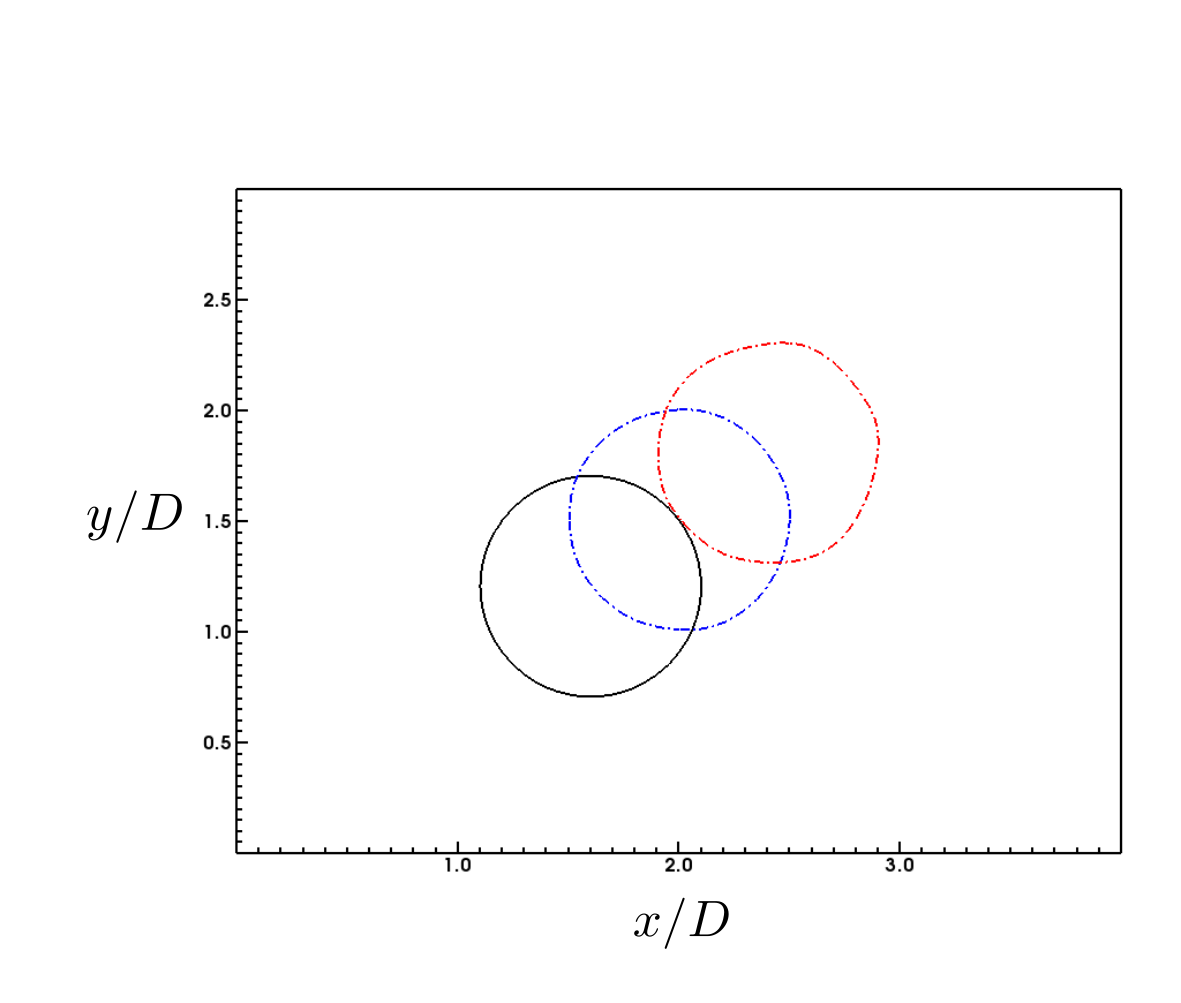}
\quad\includegraphics[width=0.4\textwidth]{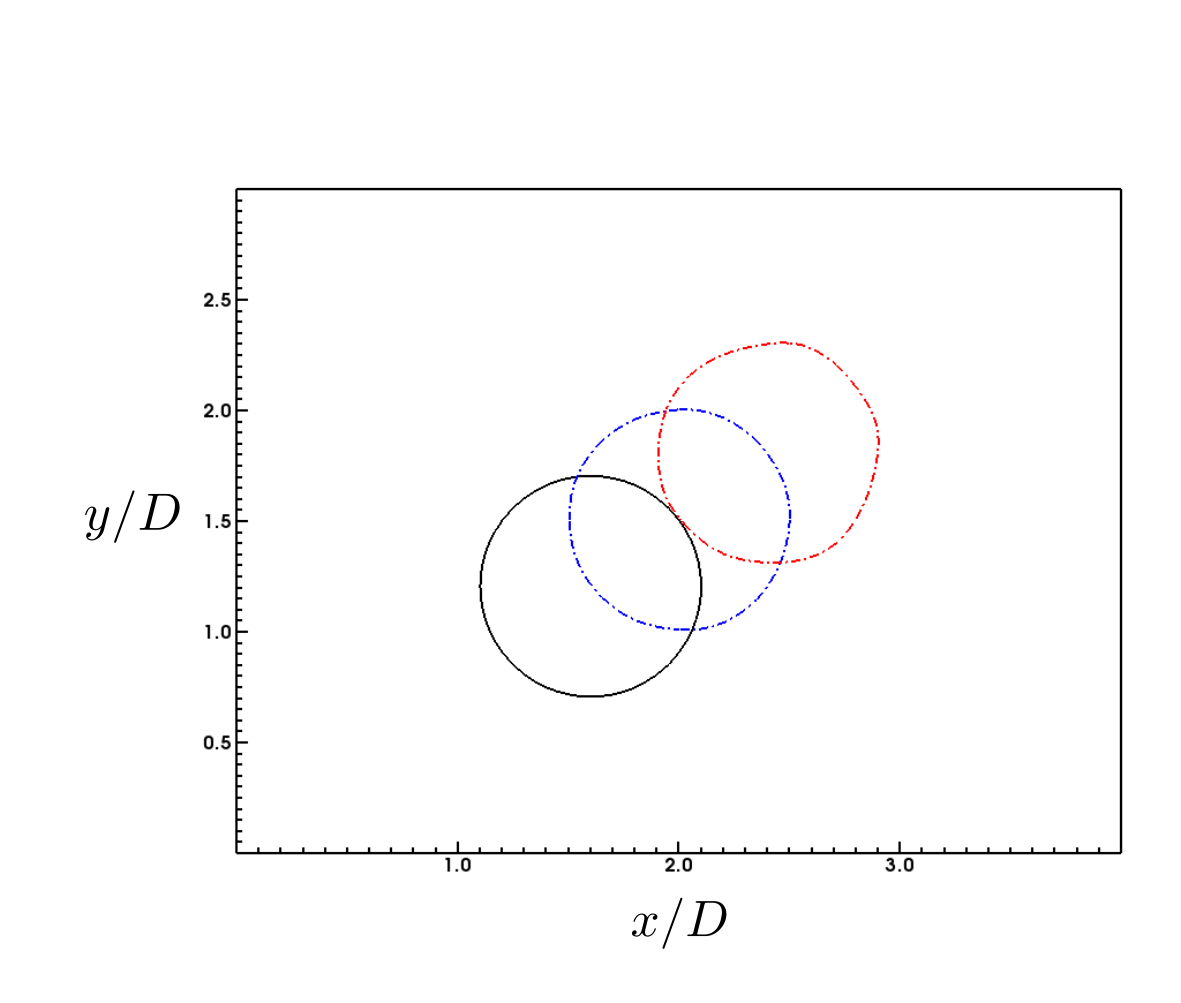} \\
\end{center}
\caption{Deformation of a droplet moving through a light fluid. 
The initial droplet resolution is $D/h=20$.
Black shape at $t=0$, blue at $t=D/(2U)$ and red at $t=D/U$. 
Density ratios: $r=10^{-1},10^{-3},10^{-6},10^{-9}$ 
(left to right and top to bottom).}
\label{rich}
\end{figure}

\begin{figure}
\begin{center}
\includegraphics[width=0.4\textwidth]{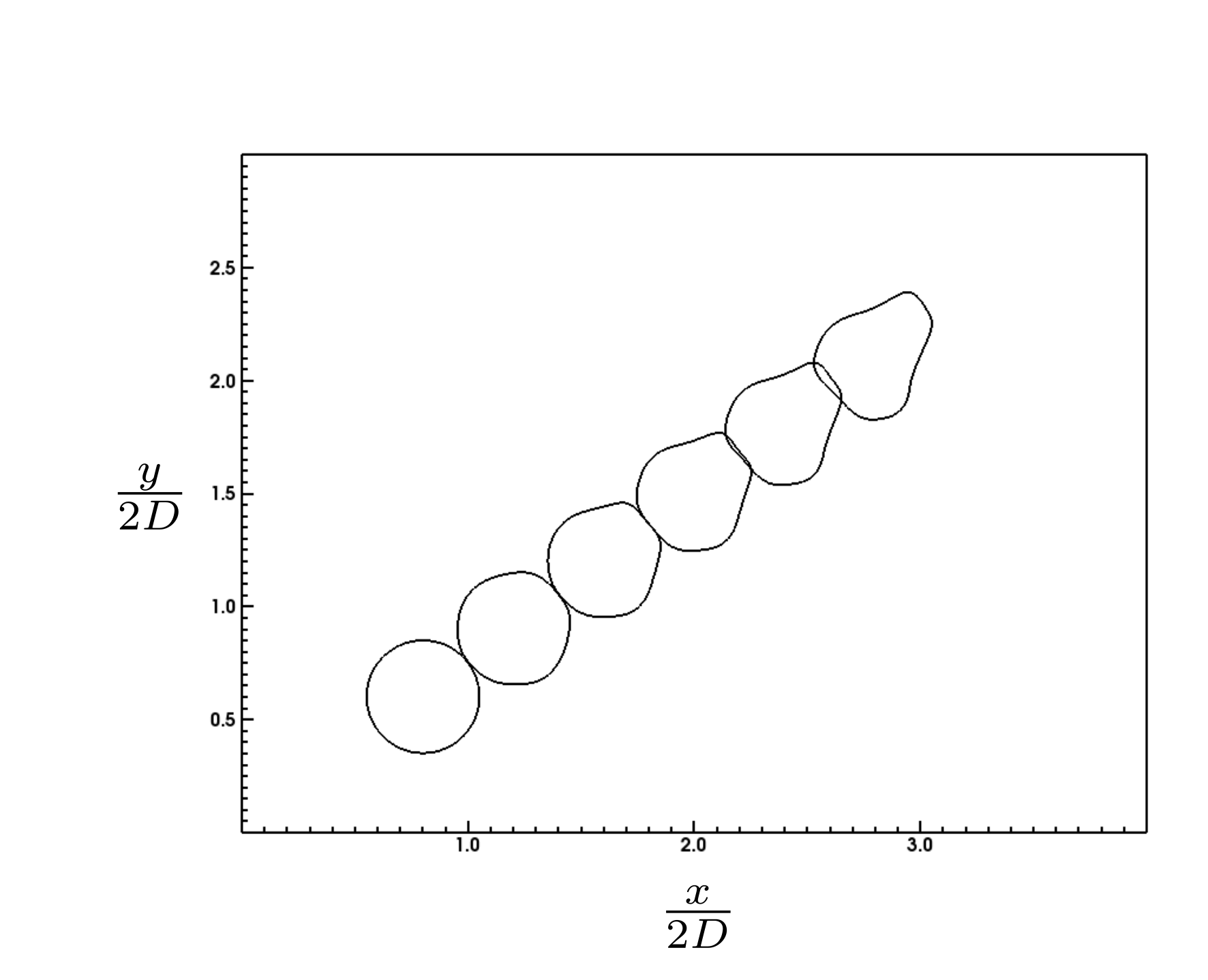}
\end{center}
\caption{Deformation of a droplet moving through a light fluid at $r=10^{-9}$.
The initial droplet resolution is $D/h=20$.}
\label{long}
\end{figure}
\red{First, as seen in the previous section and in the Appendix,
a vortex sheet is unstable with respect to the Kelvin-Helmholtz instability. 
The perturbation amplitude $A(t)$  grows exponentially
as $A(t) \sim A(0) \exp ( \omega_i t )$ with $\omega_i$ given by
equation (\ref{omkazi}). In the limit of small $r$ the growth rate becomes small, on
the order of $\sqrt r$.
Since the maximum wavenumber on the grid is $\pi/h$, an estimate of the growth rate
of the small wavelength instabilities is $\pi U N \sqrt r /D$,
where $N$ is the number of grid points per diameter.
After advection by a droplet diameter, the 
elapsed time is $\Delta t = D/U$. For typical orders of magnitude
in the literature of $r=10^{-6}$
and }
$N=32$ the amplitude growth would be 
\be
\exp (\om_{i,\rm max} \Delta t) =  \exp\left( {\pi N}{ \sqrt r} \right) 
= \exp ( 0.032\pi) = 1.1058\ldots  \label{rmmax}
\nd
which means the amplitude should grow by 10\% after advection by one diameter 
and by $\sqrt {\rm e}$ after advection by 5 diameters. 

Second, beyond the linear growth stage of the Kelvin-Helmholtz instability, there is a 
self-similar, non-linear growth stage for which dimensional analysis implies
that $A(t) \sim  \sqrt r U  t $ \cite{hoepffner11}. By this argument 
also the perturbation of the cylinder should remain small, of order  
$A(\Delta t) \sim D \sqrt r$, after advection by a droplet diameter. 

Third, physical deformation is expected from the spatial pressure variation 
induced by the dipole gas flow. This variation involves a larger pressure at 
the aft and fore stagnation points and a lower pressure along the equator
of the droplet \cite{Clift78}. The resulting integrated stress is of order 
$\rho_g U^2$ resulting in the growth of the droplet deformation as 
$A(t) \sim r U^2 t^2 / D$ and after advection by a droplet diameter as
$A(\Delta t) \sim r D$. 
This growth is observed experimentally \cite{opfer14} and results in an 
elliptically shaped (mode 2) drop, albeit of much smaller amplitude than the two former 
Kelvin-Helmholtz-related growth mechanisms, so we can exclude this mechanism
in the present case except perhaps in the $r=1/10$ case where a mode 2 is apparent
in Fig. \ref{rich}a.

The results are shown on Figs. \ref{rich} and \ref{long} for several times and
density ratios in a manner comparable to \cite{bussmann2002modeling}. For these
numerical experiments, the initialization of the velocity fields is rather 
important in order to not allow in the first time step some gas 
velocity in the liquid, and hence to avoid the situation of a boundary layer in the liquid
in which the growth rate is independent of $r$ as explained in
Section \ref{sec_khi}. 
Thus the density is initialized to $\rho_l$ to machine accuracy using the 
{\sc Vofi} library \cite{bna2015numerical,bna2016vofi}
within a disk implicitly defined by $x^2 + y^2 < R^2$ 
and the {\em velocity} is initialized to 1 for all the velocity nodes inside
a disk implicitly defined by $x^2 + y^2 < (R+nh)^2$, where $n$ is the 
size of the ``halo'' in number of grid points. The velocity in the other
nodes is initialized to 0. The tests shown were performed with $n=1$. 
Increasing the size of the halo from $n=1$ to $n=2$ improves the results at
early times ($\Delta t \le D/U$) of the droplet motion, but not at late times
 ($\Delta t > D/U$). 

The droplet momentum has been oriented along the diagonal as in 
\cite{bussmann2002modeling}. The WY scheme is used with a QUICK-UW
interpolant. The droplet deforms little after advection by one droplet 
diameter (Fig. \ref{rich}). For longer advection the deformation is more pronounced
but, as explained above, this is to be expected
\red{ at high resolution, except in the $r=10^{-9}$ case for which we show advection 
by 5 diameters in Fig. \ref{long}. 
If the non-consistent method is used, the
droplet deforms rapidly and the simulation breaks down.
Our results with the consistent method are better or on a par with previous
results for early times ($\Delta t \le D/U$) but for late times, $r=10^{-9}$ and
low order modes the results are somewhat worse
than those of \cite{bussmann2002modeling,desjardins10,raessi12,le13,zuzio2020new}  
(see Fig. \ref{long}).
Finally for late times and more moderate density ratios it is difficult to
asses the results because of the expected {\em physical} growth of perturbations
as expressed for example in (\ref{rmmax}).
As a comparison, in the results of the recent reference \cite{zuzio2020new}
there was no significant deformation after one ``transit time'' $\Delta t = L_1/U_1$
which with the parameters of  \cite{zuzio2020new} corresponds to three diameters
($\Delta t = 3D/U$).

Further comments can be made on the causes of the droplet deformation, depending
on whether the theoretical dynamics of the Kelvin-Helmholtz instability are
correctly reproduced by the numerics or not.
First, if the theory is approximately reproduced by the numerics, and
 the gas momentum numerically diffuses inside the liquid, then some vorticity}
may penetrate into the liquid despite the fact that in inviscid flow
vorticity should remain confined on the interface. If this happens,
then by the analysis of section \ref{sec_khi} 
the growth rate becomes independent
of the density ratio $r$ (Figure \ref{grr}), and has a much larger value of order
$\om_i \sim  U k$,  without the $\sqrt r$ factor.
Moreover, the instability seen for the $r=10^{-9}$ case for the long advection case in  Figure \ref{long} is 
a mode 3.
If a low $k$ mode such as mode 3 grows faster than the high frequency modes related to the
grid, this indicates that a boundary layer has grown by numerical diffusion.
The boundary layer is located at least in part
in the liquid side otherwise it would have a much smaller growth
proportional to $\sqrt r$.

Second, there could be an effect of the deviation from the Kelvin-Helmholtz theory.
Indeed we observe  spurious growth, especially  at the end of subsection  \ref{subsec_khi}.
This could also be the cause of the 
deformation of the drop.

This points to two separate routes to the improvement of the results of this test:
1) by improving the method's performance for the vortex sheet case of the Kelvin-Helmholtz
instability and minimising numerical diffusion;
2) by maximising numerical diffusion of vorticity into the gas phase and minimising
diffusion of vorticity into the liquid phase. The first option would lead to
a more accurate method, while the second one, even if successful for the advected droplet test,
would lead to an exceedingly diffusive code.

We note also that the tests of the Kelvin-Helmholtz instability in  Section  \ref{sec_khi}
are performed for interfaces aligned with the grid, while in this section interfaces may
have any orientation with respect to the grid, so that even if 
vortex sheet dynamics were perfectly reproduced for interfaces aligned with the grid there
would be no guarantee to have inviscid droplet dynamics represented correctly in all configurations
and thus to have good results for the large density ratio droplet. 

\subsection{Sheared layer}
In order to better analyze the behavior of the methods in flow under shear, 
such as vortex sheets and Kelvin Helmholtz instabilities, we set up 
a planar, parallel shear flow in a $(-1/2,1/2)^2$ domain with the following 
initial conditions
\bea
u_1& = 15 &\qquad {\rm if \;\;} |x_2| > 1/10 \nonumber \\
u_1& = 1  &\qquad {\rm if \;\;} |x_2| < 1/10\,, \nonumber 
\nda
$$
u_2  =  0.01\, \sin(2 \pi x_1) \exp(-20 x_2^2) \,,
$$
$\rho=1$  if $|x_2| > 1/10$, and $\rho=10^3$ otherwise. 
This flow is similar to a liquid sheet in high velocity gas. The flow is 
simulated until time $t_f=2$ unless the simulation blows up at an earlier time
$t_b <t_f$.
\red{The crashes seem connected to the spurious growth of the Kelvin-Helmholtz instability
described in Section \ref{sec_khi}. We note that  the physical instability could lead to interface deformation,
but not to the catastrophic increase of the total kinetic energy that is witnessed when the code crashes.}

\begin{figure}
\begin{center}
\includegraphics[width=\textwidth]{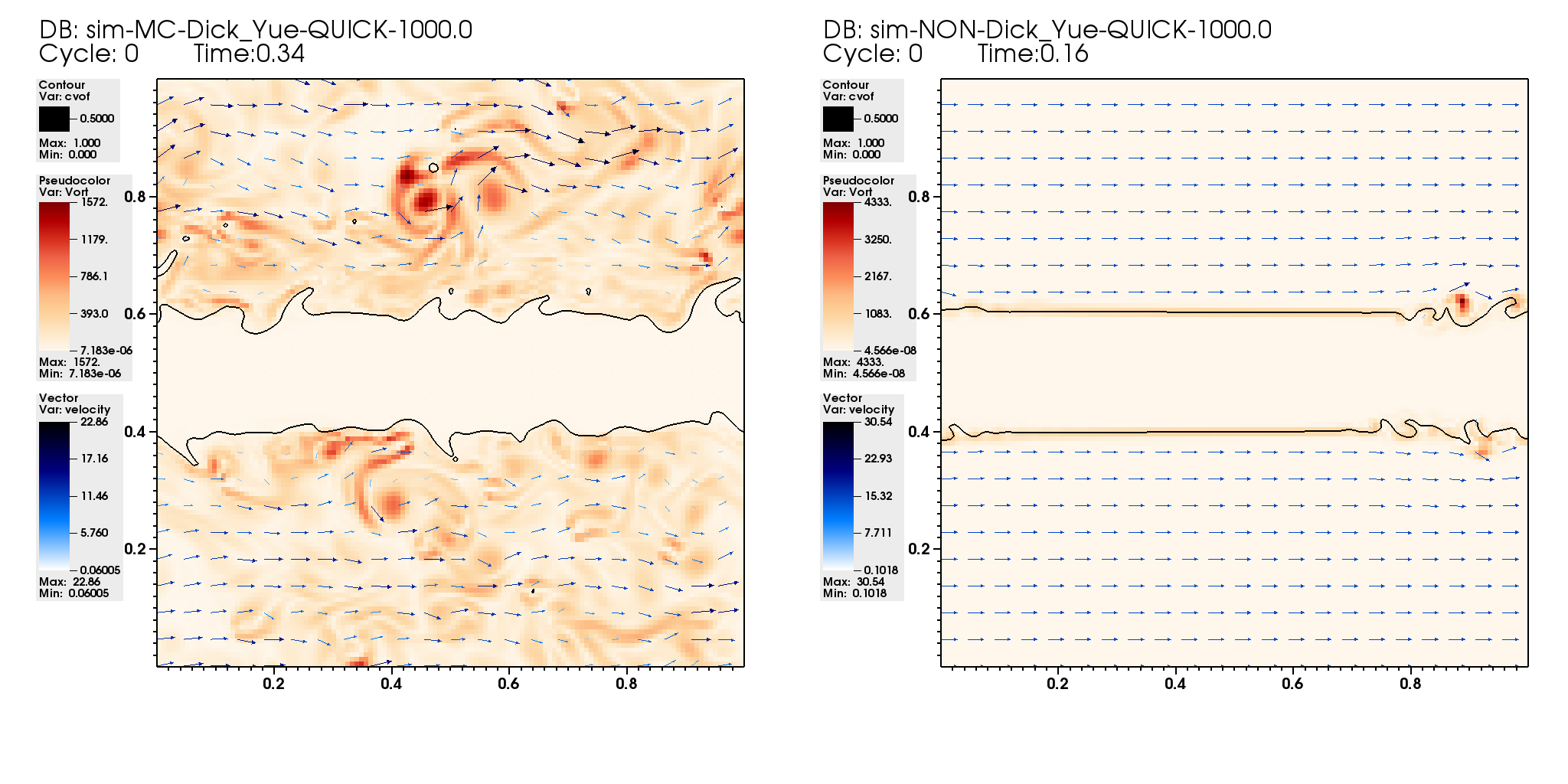}
\end{center}
\caption{The state of the simulation of the sheared layer just before its
breakdown with the combination of WY VOF advection scheme
and QUICK-UW velocity interpolation: consistent method (left), non-consistent 
method (right).}
\label{shelay}
\end{figure}

Table \ref{shearmtable} shows the numerical schemes that have been used in a few
simulations together with the fraction $t_{b}/t_f$ of the final time that has 
been reached. All simulations have been performed on a $128\times128$ grid with 
a CFL of 0.03 and a density ratio of 1000. The consistent method is
systematically more stable than the non-consistent one for the two combinations 
shown in Tab. \ref{shearmtable}: a) the WY VOF scheme and the QUICK-UW velocity 
interpolation schemes and b) the CIAM VOF with the Superbee slope limiters. 
The state of the two simulations with the first combination of schemes is shown
in Figure~\ref{shelay}, just before breakdown at time $t_b = 0.34$ 
(or $t_b/t_f = 0.17$) for the consistent method, and at time  $t_b = 0.16$ 
(or $t_b/t_f = 0.08$) for the non-consistent method. We have also tested a number of other
combinations, for example  WY \& Superbee that turns out to be very unstable.  
Furthermore, we have performed simulations with smaller  $64\times64$ and $32\times32$ 
grids that yield similar results. 
Finally, we note that this case is also unstable in a single-phase configuration 
when using the QUICK third-order velocity interpolation.
\begin{table}
\begin{center}
\begin{tabular}{cccc}
\hline
Method	       &VOF Scheme & velocity scheme &	fraction completed\\
\hline
non-consistent &    WY	   &	QUICK-UW     &	0.08\\ 
consistent     &    WY	   &	QUICK-UW     &	0.17\\ 
non-consistent &    CIAM   &	Superbee     &	0.50\\ 
consistent     &    CIAM   &	Superbee     &	1. \\ 
\hline
\end{tabular}
\end{center}

\caption{Percentage of completion of the shear layer test with various methods. 
All simulations are performed on a $128\times128$ grid with a CFL of 0.03 
and a density ratio of 1000.}
\label{shearmtable}
\end{table}

\subsection{Falling raindrop}
\begin{figure}
\begin{center}
\includegraphics[width=0.99\textwidth]{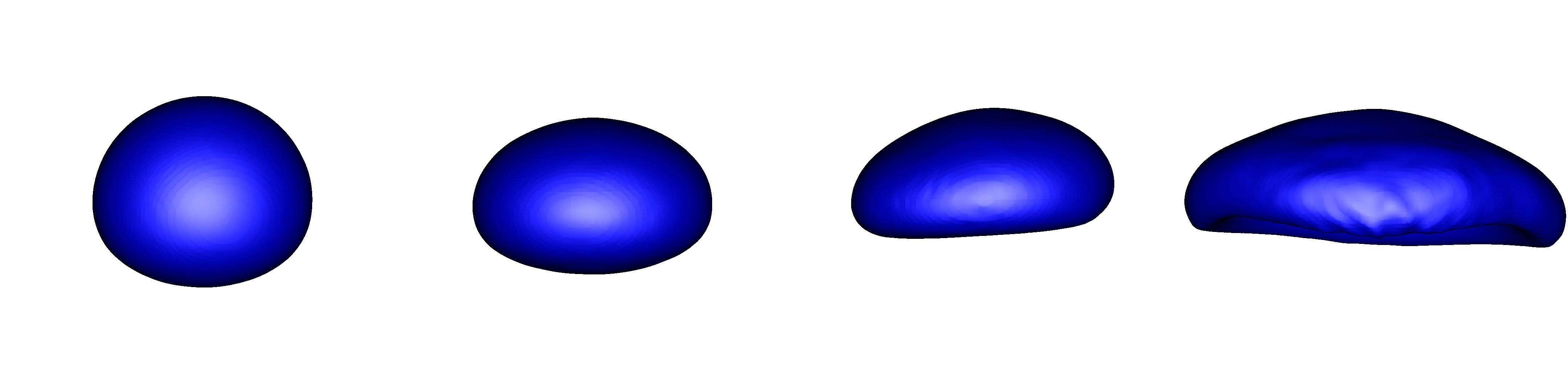}
\end{center}
\caption{Flattening of the droplet with increasing equivalent diameter.
  From left to right $D_e=3, \,4.6,\, 6.4$ and $8\, mm$.}
\label{flatten}
\end{figure}

\begin{figure}[h!]
\begin{center}
\includegraphics[width=0.5\textwidth]{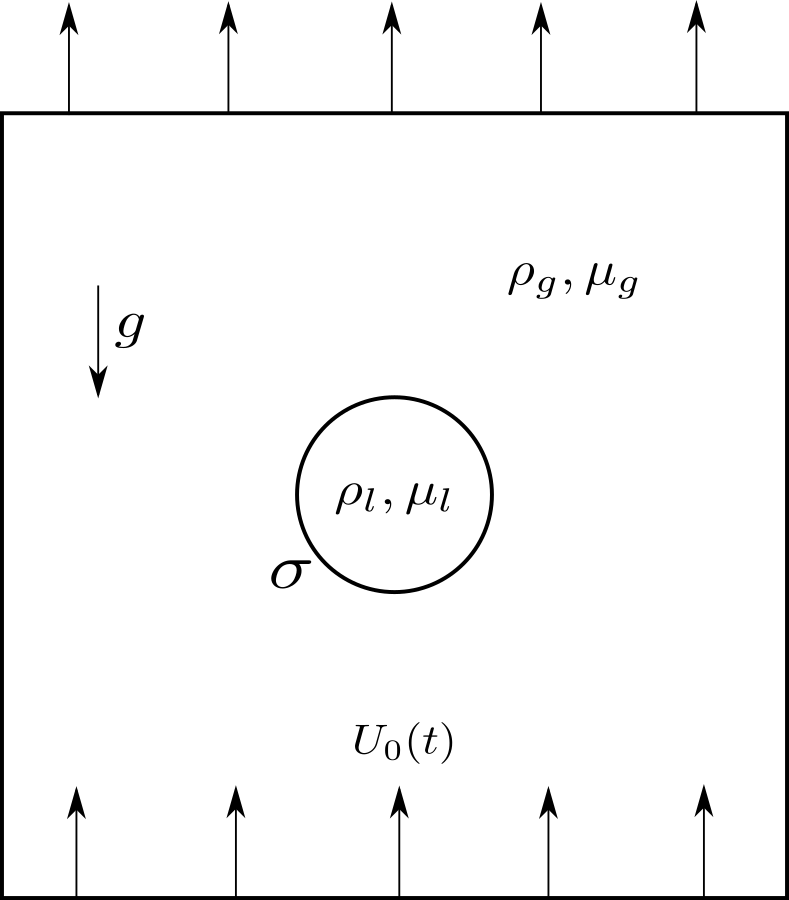}
\end{center}
\caption{The numerical setup for the falling raindrop. 
	A droplet of diameter $D$ is placed at the center of a cubic domain 
	of side $L$, with $L/D = 4$. The liquid properties $\rho_l$ and $\mu_l$ 
	correspond to water, and the gas properties $\rho_g$ and $\mu_g$ 
	correspond to air. We apply a uniform inflow velocity condition 
	$u_1 = U_0(t)$ on one face and an outflow velocity condition on the 
	opposite face, corresponding to zero normal gradient.
	Boundary conditions on the side walls are free-slip (no shear stress).}
\label{setup}
\end{figure}

\newcommand\DDD{{\cal D}}
\red{
A flow configuration that combines the complexities of large 
density contrasts with the interaction between capillary, viscous and 
inertial stresses is that of a water droplet falling in the 
air under the influence of gravity. From our experience it is
this combination that leads to the blowups observed
in many air-water simulations when a non-consistent method is used.
We chose a falling water droplet in air of diameter $D=3$mm,
essentially a raindrop near the largest diameters (Fig. \ref{flatten}) for which
sphericity is approximately maintained.
This choice is motivated by the paradigmatic value of
a near-spherical raindrop simulation, and by the fact that
the corresponding Weber number (given below) is the same as in
similar air-water suddenly-accelerated-droplet (or ``secondary atomisation'') simulations of
\cite{xiao2012large,Xiao:2014vs}.
We first define the problem in general terms and describe several variants of the
simulation setup, then discuss the physics of droplet fall and
the numerical results. 


\subsubsection{Problem Setup}

The droplet is placed at the center of a cubic domain 
	of side $L$, with $L/D = 4$. The liquid properties $\rho_l$ and $\mu_l$ 
	correspond to water, and the gas properties $\rho_g$ and $\mu_g$ 
	correspond to air. We apply a uniform inflow velocity condition 
	$u_1 = U_0(t)$ on the bottom face and an outflow velocity condition on the 
	top face, corresponding to zero normal gradient.
        The droplet is subject to a gravity field $g$. 
	Boundary conditions on the side walls are free-slip (no shear stress).
        


The subscripts $l$ and $g$ represent water and air phases respectively. 
The parameters in the problem setup are given in Tab. \ref{raindropprop}, 
and a schematic representation is given in Fig. \ref{setup}. 
For $D=3$mm experiments  \cite{gunn1949terminal} indicate a terminal velocity close to $U_t=8.06 m/s$ which
 corresponds to a Weber number
$\We =\rho_g U_t^2 D / \sigma \simeq 3.2$ similar to that tested by Xiao \cite{xiao2012large}.

We perform two main types of simulations
\begin{itemize}
\item Simulations in which the inflow velocity $U_0(t)$ is adjusted at each time step by a controller that
aims at keeping the droplet at the center of the box.
\item Simulations in which the inflow velocity has a constant value $U_0$.
\end{itemize}
The first type of simulations allows to keep the droplet near the
center of the box for a long time without particular difficulties, but
the controller
adds some unwelcome jitters to the inflow velocity
and as a result
the frame of reference is not Galilean anymore,
while the missing acceleration terms had not been
added to the Navier--Stokes equations. Thus the simulations have only
heuristic value. 
The second type has
the droplet leaving the domain after a certain time, but is convenient
for relatively short-time investigations. A third possible type of
simulation, arguably the most natural one consists in using a much larger domain
filled with air at rest, with zero inflow velocity and to let the
droplet fall from the top of the domain.  Note that Dodd and Ferrante
\cite{dodd2014} used such a setup. It is much more expensive and in
fact smaller droplet diameters were investigated in \cite{dodd2014}.

In order to better grasp the physics of droplet fall, some understanding may be obtained from the fall of
a solid object. Indeed, because of the large inertia
of the liquid compared to the gas, during the first instants the droplet internal flow is
comparatively negligible.
Experiments and numerical simulations on falling solid spheres show that the
drag coefficient is {\em larger} than the drag coefficient on a sphere held fixed,
for example in a wind tunnel. 
At $\Re \simeq 10^3$
on finds $C_D \simeq 0.6$ to $0.7$ for a falling sphere compared to $C_D \simeq 0.45$ for a fixed sphere
(see for example \cite{krishnan2016transient} and references therein).
(The difference between fixed and falling solids is probably due to the fluctuating center-of-mass velocity of the
falling solids.) Drag coefficients $C_D \sim $ 0.6--0.7 would result in a fall velocity of 6.8 to 7.3 m/s.
The resulting difference between the falling velocity of water droplets and that of falling spheres may be explained as follows.
The internal vortical motion probably results in a diminished friction between the droplet
surface and the air flow and a resulting smaller drag coefficient, yielding the larger fall velocity.
We detail the time scales related to the acceleration of
the droplet and the internal motion below, and we perform numerical experiments in the two above setups.
For the second setup we use two different inflow velocities.
}

\subsubsection{Results with controlled inflow velocity}
\label{secdr1}
\begin{figure}
\begin{center}
\begin{tabular}{cc}
\includegraphics[width=0.45\textwidth]{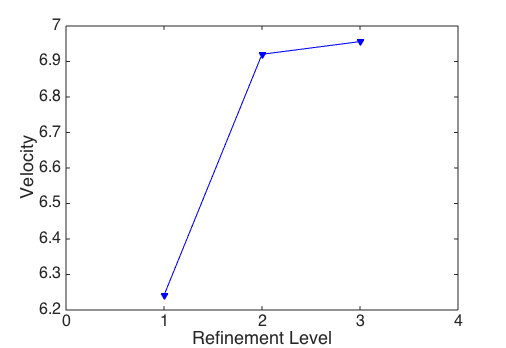}
& \includegraphics[width=0.45\textwidth]{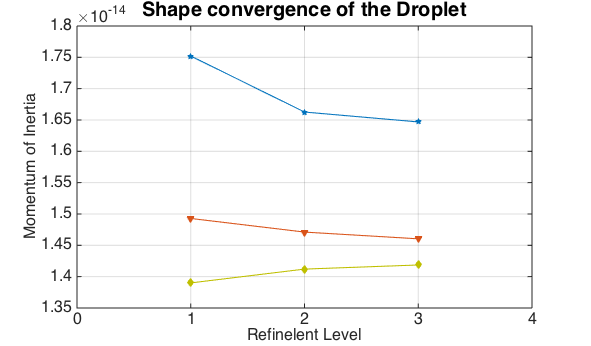} \\
(a) & (b)
\end{tabular}
\end{center}
\caption{Convergence of simulations. (a) Evolution of the terminal velocity 
with grid refinement. (b) Evolution of the three moments of inertia with 
grid refinement.}
\label{converge}
\end{figure}

We use the controller setup with an initial inflow velocity $U_0(0) = 8$m/s.
Three
grids are used, with $D/h = 15, 30$ and 60.
Numerical simulations of this test case at all resolutions 
carried out \textit{without} the 
consistent scheme described in this paper result in the catastrophic 
deformations of the droplet that will be further investigated below. 
Using the consistent scheme, we avoid the instability and may
observe the droplet for a long time with enough CPU resources (up to 200 ms
for the $D/h=32$ case) and may
study the convergence of the terminal velocity and that of the shape. 
For the latter, we use as a descriptor of the shape the three moments of 
inertia $I_{mm}$ defined by
\be
I_{mm} = \int_\DDD H x_m^2 {\rm d}\X \;, \quad  1 \le m \le 3, \label{immde}
\nd
where $\DDD$ is the domain used for the computation and $x_m$ is relative to
the center of mass. The convergence of the moments of inertia and terminal 
velocity is shown on Figure \ref{converge}. The velocity seems to converge 
to a value around $7\, m/s$, which is consistent with the solid-sphere fall velocity.

%

\begin{table}[h!]
\begin{center}
\begin{tabular}{ccccccc}
\hline\hline
$\rho_{g}$ & $\rho_{l}$ & $\mu_{g}$ 
& $\mu_{l}$ & $\sigma$ & $D$ & $g$\\
$\left(kg/m^3\right)$ & $\left(kg/m^3\right)$ & $\left(Pa \, s\right)$ 
& $\left(Pa \,s \right)$ & $\left(N/m\right)$ & $(m)$ & $(m /s^{2})$ \\
\hline
1.2 & $0.9982 \times 10^3$ & $1.98 \times 10^{-5}$ & 
$8.9 \times 10^{-4}$ & $0.0728$ & $3 \times 10^{-3}$ & $9.81$\\
\hline\hline
\end{tabular}
\caption{Parameter values used in the simulation 
	of a falling water droplet in air. \label{raindropprop}}
\end{center}
\end{table}

\begin{figure}
\begin{center}
\includegraphics[width=0.75\textwidth]{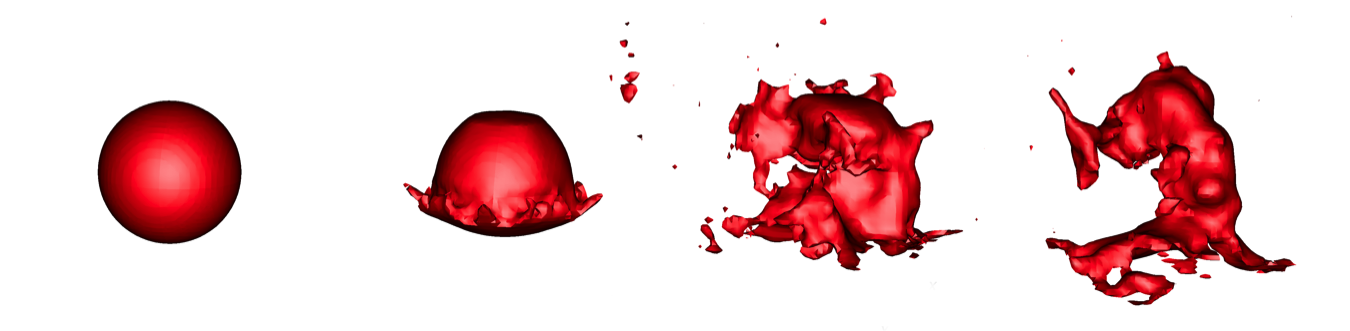}
\end{center}
\caption{Catastrophic deformation of a falling raindrop while using a 
	version of our method that does not ensure consistency 
	between mass and momentum transport ($D/h=30$).}
\label{cata}
\end{figure}
\red{
\subsubsection{Fixed 8 m/s inflow velocity}

We now use the fixed velocity setup with $U_0(t)=8$m/s and perform
simulations for very short times (of the order of 1 ms). 
Numerical simulations of this test case at moderate resolution 
(from $D/h=16$ to $D/h=64$) carried out \textit{without} the 
consistent scheme described in this paper result in catastrophic 
deformations of the droplet illustrated in Fig. \ref{cata}, 
a kind of `fictitious' or `artificial' atomization.
These simulations display marked peaks
or spikes in kinetic energy as a function of time,
associated with massively-deformed interface shapes (see Fig. \ref{explode_compare}).
Additionally, our studies suggest that certain combinations
of the VOF advection method and the velocity interpolation scheme are
numerically more robust than others,
in particular the most stable combinations
are the CIAM advection with the
Superbee slope limiter, and the WY advection with the QUICK-UW interpolation.
This is in agreement with the results reported in Tab. \ref{shearmtable}. 

We propose the following explanation in order to account for such numerical artifacts. 
To start with, we neglect gravity and viscous effects at this relatively large Reynolds number. 
Also, we are interested in steady-state flow. 
On the axis and near the hyperbolic stagnation point 
at the front of the droplet the tangential velocity is $u_2=0$ 
and the axial momentum balance is
\be
u_1 \partial_1 u_1 = - \frac 1 \rho \partial_1 p.
\nd
Due to the large viscosity and density contrasts, it is not possible for 
the air flow to immediately entrain the water, so the fluid velocity 
is significantly smaller inside the droplet. 
In the air the acceleration near the stagnation point is 
of the order $U^2/D$, whereas the pressure gradient is
\be
\partial_1 p \sim \rho_{g} U^2/D.
\nd
The pressure gradient in the water is much smaller, 
however, in the case of a mixed cell the water density 
multiplies the air acceleration $U^2/D$, so that
\be
\partial_1 p \sim \rho_{l} U^2/D,
\nd
then a large pressure gradient results in the mixed cells. 
This large pressure gradient results in a large pressure inside the 
droplet near the front stagnation point, as shown in Fig. \ref{pressure_1}. 
This large pressure is balanced by surface tension only for a sufficiently 
large curvature near the droplet front. 
This explains the presence of a ``dimple'' often observed in low 
resolution simulations of the falling drop. 
This artifact has been observed by Xiao \cite{xiao2012large} in a similar case 
involving the sudden interaction of a droplet at rest with a uniform gas flow. 
The resulting large nonphysical pressure gradients across the interface 
eventually lead to its rapid destabilization and concomitant breakup.  
\begin{figure}
\begin{center}
\begin{tabular}{cc}
\hspace*{-1.0cm}
\includegraphics[width=0.5\textwidth]{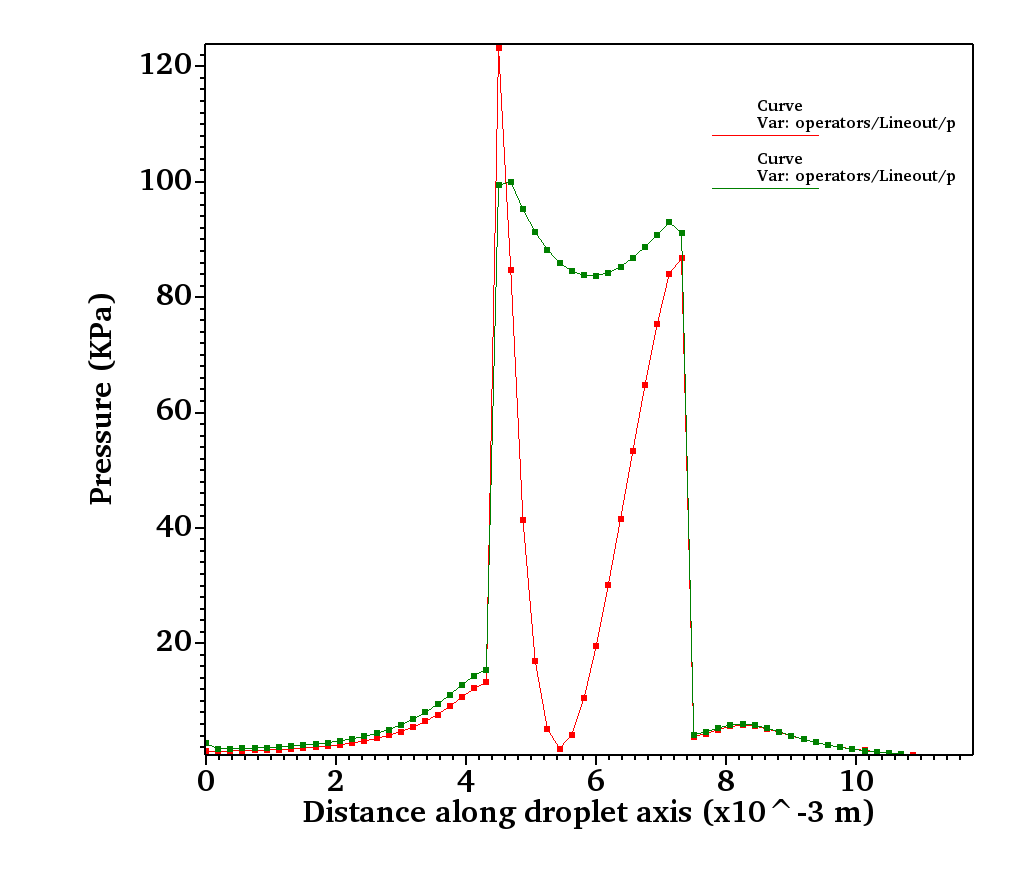} &
\hspace{-0.4cm}%
\includegraphics[width=0.5\textwidth]{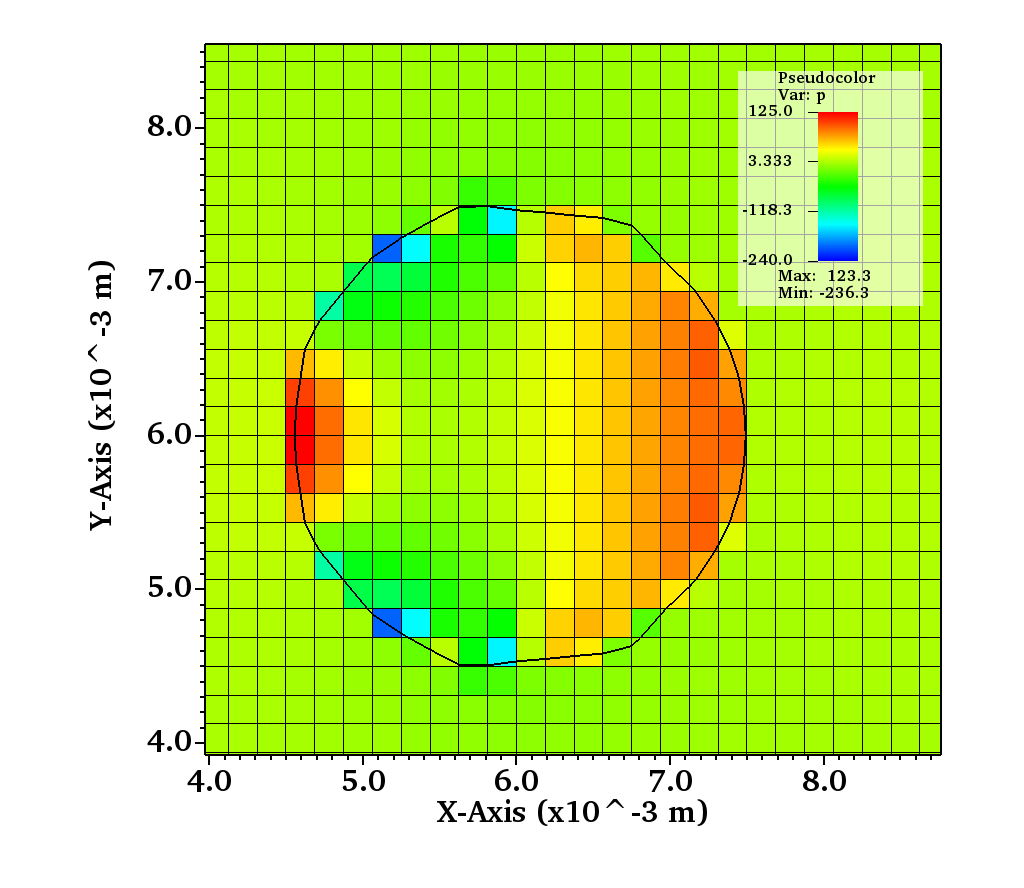}\\
\hspace{-0.8cm}%
(a) & (b)
\end{tabular}
\end{center}
\caption{The origin of the pressure peak in the front of the droplet. 
(a) Pressure profiles on the axis a few timesteps after initialization 
with the non-consistent method (red curve) and the consistent one (green curve). 
Much larger pressure gradients are present across the interface with the first method. 
(b) Pressure distribution immediately after the start of the simulation 
with the non-consistent method. The pressure peak has not yet resulted
in the formation of a dimple. The initial droplet resolution is $D/h = 16$.
The simulations are carried out with the CIAM advection method and the Superbee 
slope limiter.}
\label{pressure_1}
\end{figure}

\begin{figure}
\begin{center}
\begin{tabular}{cc}
\hspace*{-1.0cm}
\includegraphics[width=0.5\textwidth]{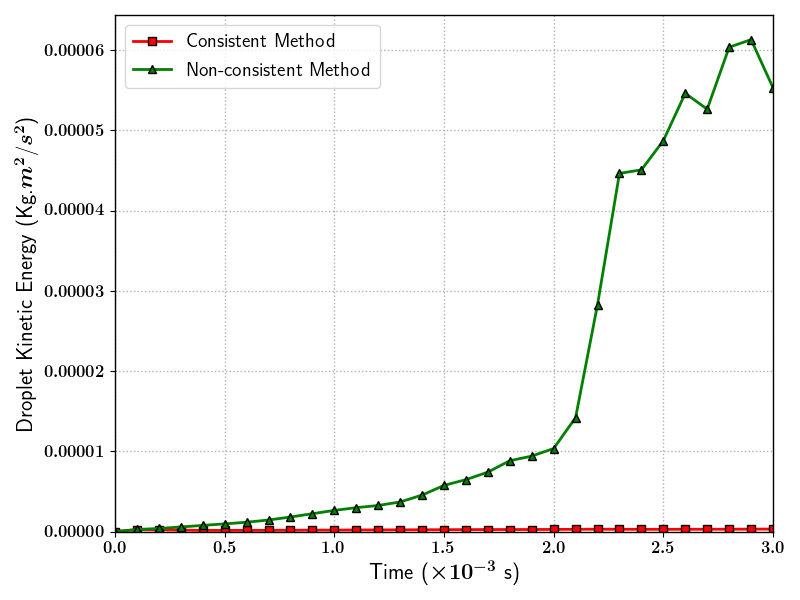} &
\hspace{-0.4cm}%
\includegraphics[width=0.5\textwidth]{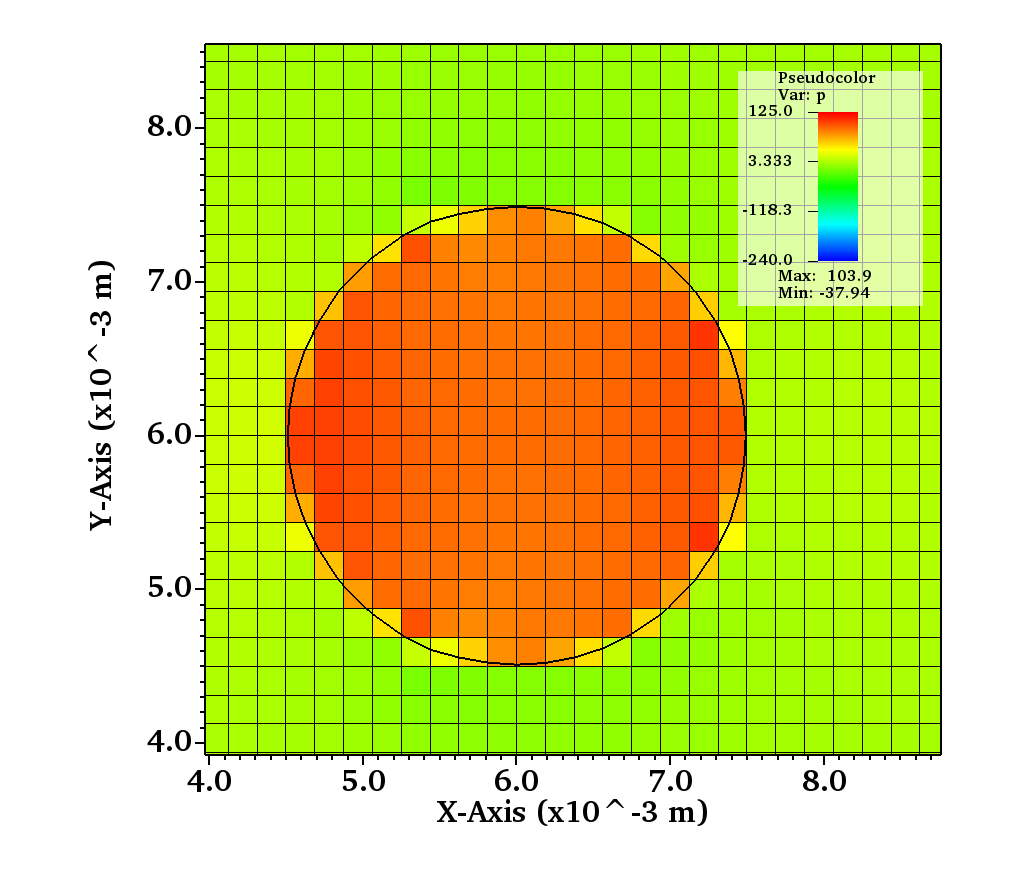}\\
\hspace{-0.8cm}%
(a) & (b)
\end{tabular}
\end{center}
\caption{ (a) Comparison of the temporal evolution of droplet kinetic energy. 
The non-consistent method displays spikes in the kinetic energy that are 3 
orders of magnitude larger than with the consistent method, 
leading to rapid destabilization. 
(b) Pressure distribution immediately after the start of the simulation 
with the consistent method. The initial droplet resolution is $D/h = 16$.
The simulations are carried out with the CIAM advection method and the Superbee 
slope limiter.}
\label{pressure_2}
\end{figure}

\begin{figure}[!h]
\begin{center}
\begin{tabular}{cc}
\hspace*{-1.0cm}
\includegraphics[width=0.5\textwidth]{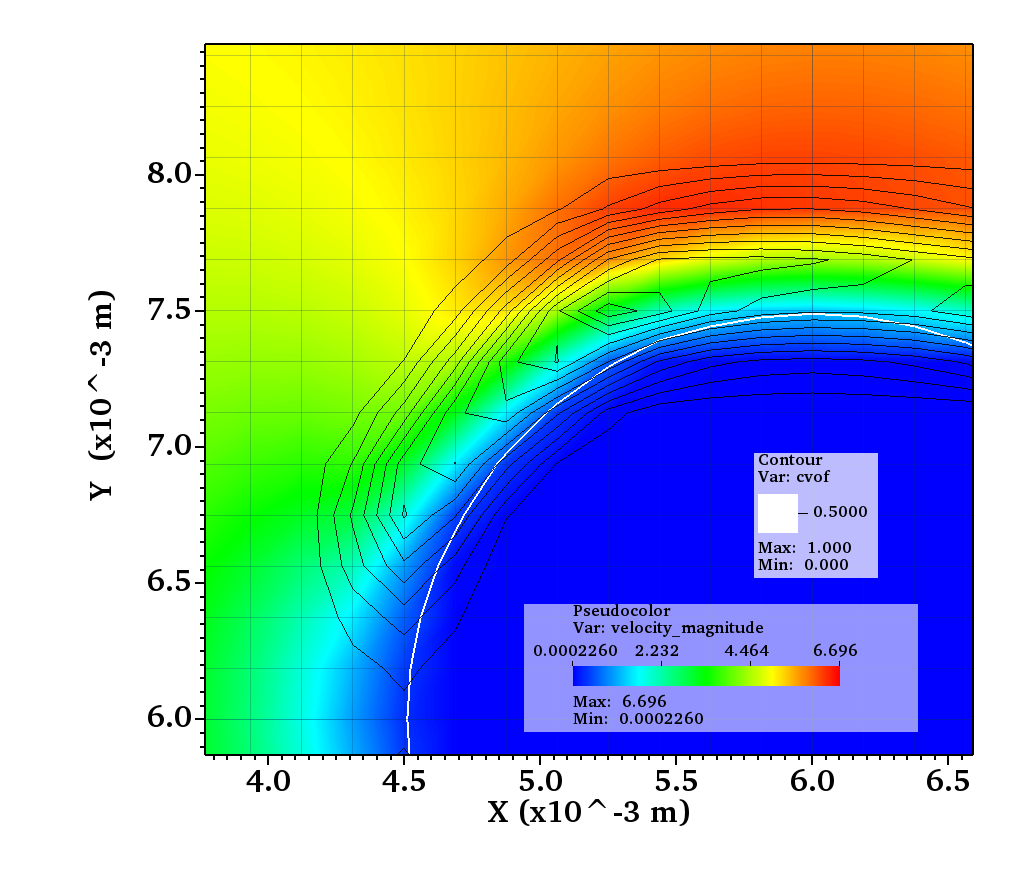} & 
\hspace{-0.4cm}%
\includegraphics[width=0.5\textwidth]{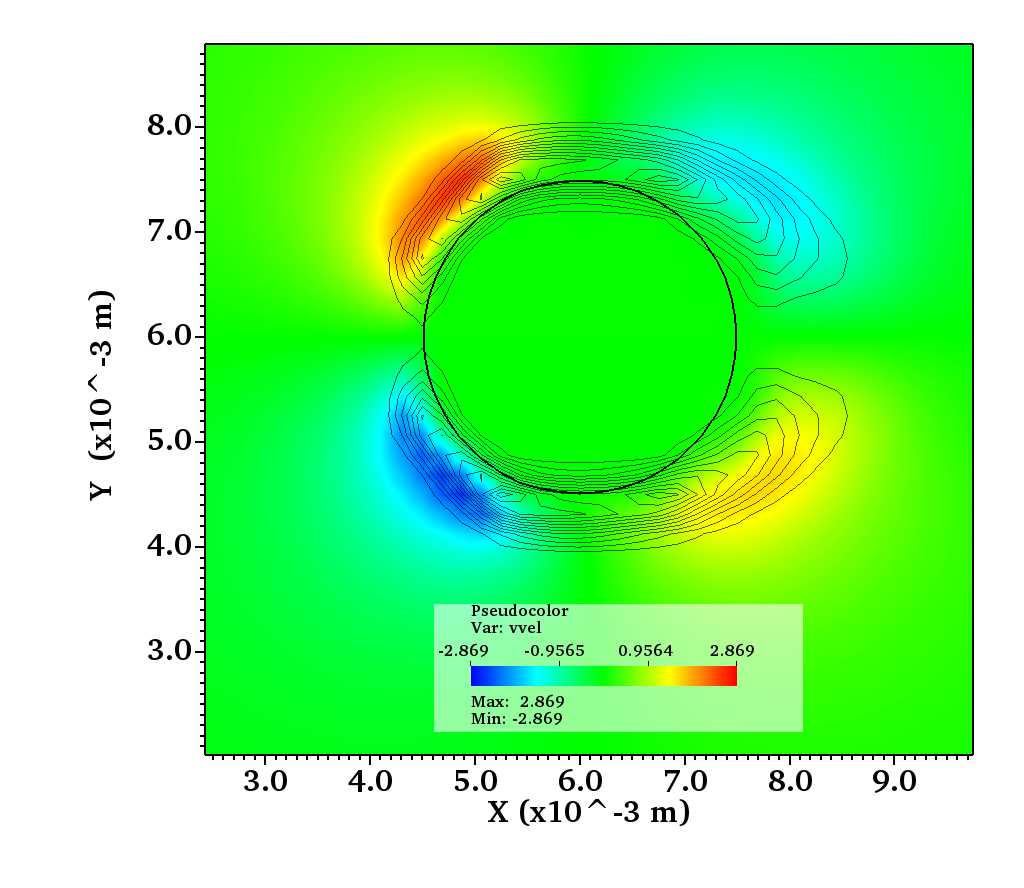} \\ 
\hspace{-0.8cm}%
(a) & (b)
\end{tabular}
\end{center}
\caption{Flow field around the $3$ mm droplet immediately 
after the start of the simulation with the consistent method, 
demonstrating the contours of the magnitude of the vorticity field (black lines). 
The 2D cross-section in these two figures corresponds to the 
mid-plane slice along the $z$ axis, with the inflow along 
the positive $x$ axis and gravity opposite to it. (a) Velocity magnitude. 
The boundary layer is resolved by only 2-3 cells. 
(b) The velocity component in the $y$ direction, perpendicular to the flow. 
As the flow develops further, a marked separation of the boundary layers is observed with 
a more complex vortical region in the wake.
The initial droplet resolution is $D/h = 16$.
The simulations are carried out with the CIAM advection method and the Superbee 
slope limiter.}
\label{flow_field}
\end{figure}
Visualization of the flow around the droplet in Fig. \ref{flow_field} 
illustrates the challenging nature of the flow configuration, 
even for such a seemingly simple physical problem. 
As one can observe, the boundary layers are extremely thin. 
We observe that applying the numerical method described in this 
paper brings a considerable and systematic improvement over a 
spectrum of different velocity interpolation schemes
and CFL numbers, as evidenced by comparing Figs. \ref{pressure_1} and \ref{pressure_2}.

\begin{figure}[h!]
\begin{center}
\includegraphics[width=1.0\textwidth]{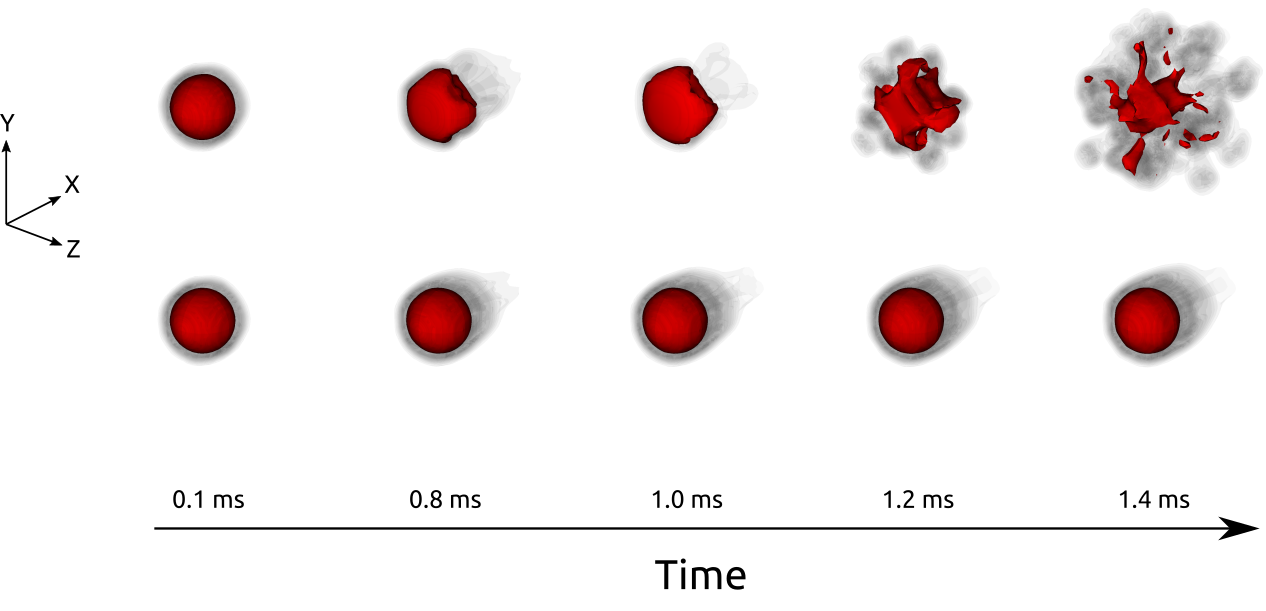}
\end{center}
\vspace*{-0.5cm}
\caption{A comparison of the temporal evolution between the 
non-consistent method (top row) and the consistent one (bottom row). 
The flow is along the positive $x$ direction, with gravity opposite to it. 
The red contour indicates the isosurface of the volume fraction 
field corresponding to $C=0.5$, whereas the black contours 
surrounding the drop represent isosurfaces of the magnitude of vorticity. 
The raindrop with the non-consistent method displays massive deformations 
leading to artificial breakup as a result of rapidly growing numerical instabilities. 
The initial droplet resolution is $D/h = 16$.
The simulations are carried out with the CIAM advection method and the Superbee 
slope limiter.}
\label{explode_compare}
\end{figure}

\subsubsection{Convergence study at 5 m/s inflow velocity}

In this section we use the fixed inflow velocity setup but with smaller initial velocity.
We systematically vary the resolution from  $D/h = 8, 16, 32 $ and $64$. Despite using
the consistent method, simulations at  $D/h = 8$ are sometimes unstable, so we use a workaround
and use a lower fixed inflow velocity of $U_0=5$ m/s, which differs from the expected
long term terminal velocity
$U_{t} \simeq $8 m/s used in the previous section and also from the solid sphere terminal
velocity. It however offers a milder initial condition and allows to observe the first phase of the (physical)
acceleration towards the final statistical steady state. We perform simulations for $t_S=5 $ms (for reasons of CPU cost and because we do not want the droplet to get too close to the domain boundaries) and
examine the convergence properties
of the numerical system in this time frame. 

In order to better understand the setup, it is useful to quickly review the relevant time scales.
\begin{itemize}
\item The time scale $t_a=D/U_0$ of the air flow around the droplet, around 0.6 milliseconds, much shorter
than the simulation time. 
\item The time $t_w = L/[2(U_t - U_0)] = 3$ ms that the droplet would take to travel by half the domain
  the domain once it had reached the terminal velocity. This time is not relevant here since one needs to wait first for the next two times
  before terminal velocity is reached. 
\item The time scale $t_c \simeq 15.1$ ms \cite{rayleigh79} of capillary oscillations of the droplet shape.
\item The time scale $t_i$ of relaxation to terminal velocity. Using the dynamics \refeq{borda} below, 
this time is $\rho_l/\rho_g D/U_t = 215$ ms much longer than the simulation time. 
\item The time scale $t_\mu$ needed to entrain the internal vortical motion of the liquid under the action of the gas. An estimate this time is  $D^2/\nu_l =  \Re D/U_t = 400$ ms.
\end{itemize}
The time of relaxation may be estimated using a simple square-velocity drag law for the droplet. We model the droplet motion as a one-dimensional dynamics under the effect of gravity and drag as
\be
\rho_l \frac{\pi D^3}6 \romandt{U} =  \rho_l \frac{\pi D^3}6 g - C_D \rho_g \frac{\pi D^2} 8  U^2, \label{borda}
\nd
hence
\be
 \romandt{U} = -  \frac 34 \frac {r}{D} (U^2 - U_t^2) = -\frac{U-U_t}{t_i}, \label{dtu}
\nd
where for $U \simeq U_t$
\be
t_i =  \frac 23 \frac {D}{r U_t},
\nd
which gives 205  ms. 

It is interesting that the simulation time is inserted in the set of relevant time scales as
$t_a \ll t_S \lesssim t_c \ll t_i < t_\mu$. This means we can see the effect of $t_c$ in the data
(especially the variation in time of the moments $I_{mm}$) but not the effect of the other time scales.



\begin{figure}[h!]
\begin{center}
\includegraphics[width = 1.0\textwidth]{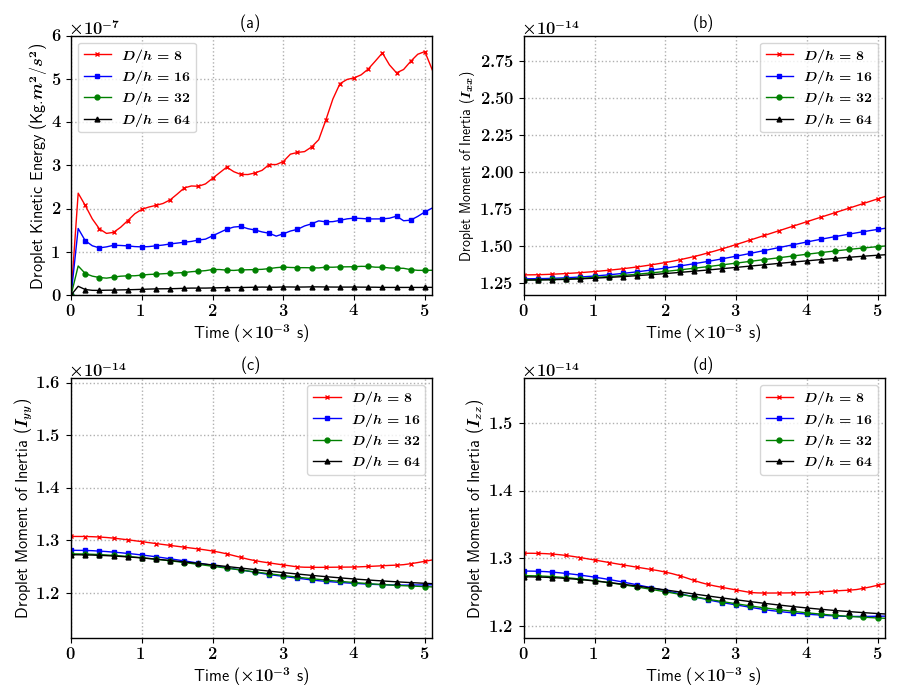}
\end{center}
\vspace*{-0.5cm}
\caption{Temporal evolution of a few physical quantities to evaluate the 
performance of our present method for different droplet resolutions $D/h$: 
(a) kinetic energy of the droplet;  
(b) moment of inertia $I_{xx}$ of the droplet along the flow direction ($x$ axis); 
(c) and (d) moments of inertia $I_{yy}$ and $I_{zz}$.}
\label{multi}
\end{figure}

\begin{figure}[h!]
\begin{center}
\includegraphics[width = 1.0\textwidth]{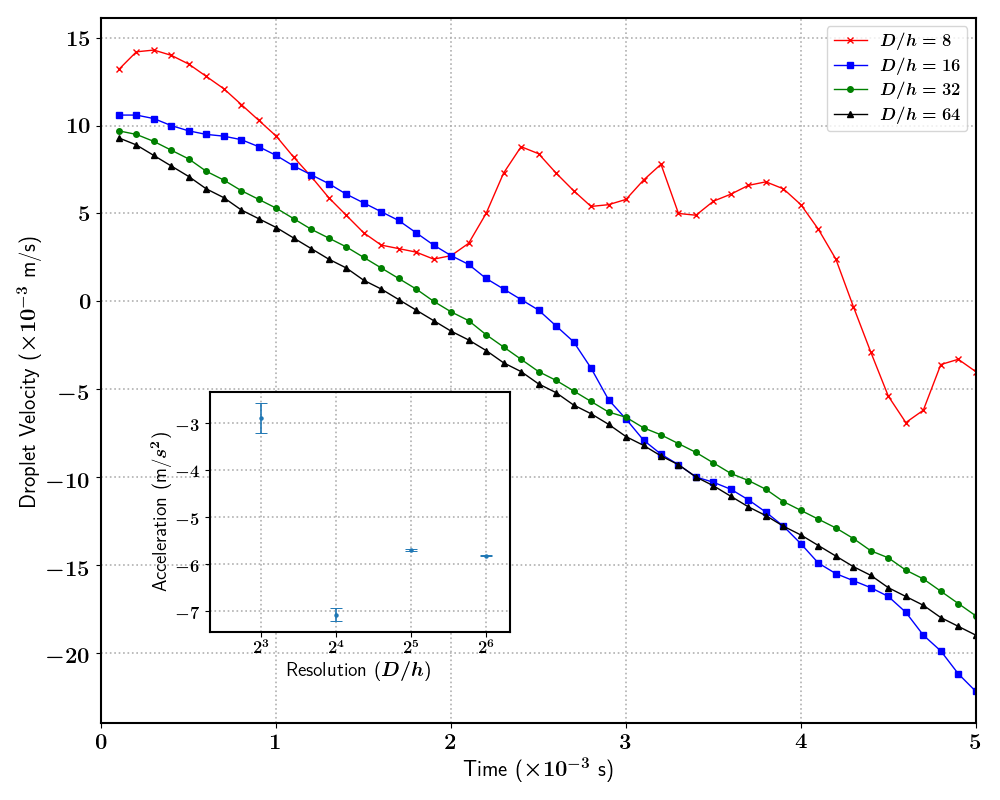}
\end{center}
\vspace*{-0.5cm}
\caption{Droplet velocity as a function of time, 
for different droplet resolutions $D/h$. 
The droplet velocity corresponds to that of the center of mass. 
Inset: convergence of the droplet acceleration as a function of its resolution, 
computed with the best linear fit over the temporal variation 
of the velocity. 
The error bar signifies the asymptotic standard 
error (least-squares) corresponding to the linear fit.} 
\label{drop_vel}
\end{figure}

\begin{figure}[!h]
\begin{center}
\begin{tabular}{cc}
\hspace*{-1.0cm}
\includegraphics[width=0.5\textwidth]{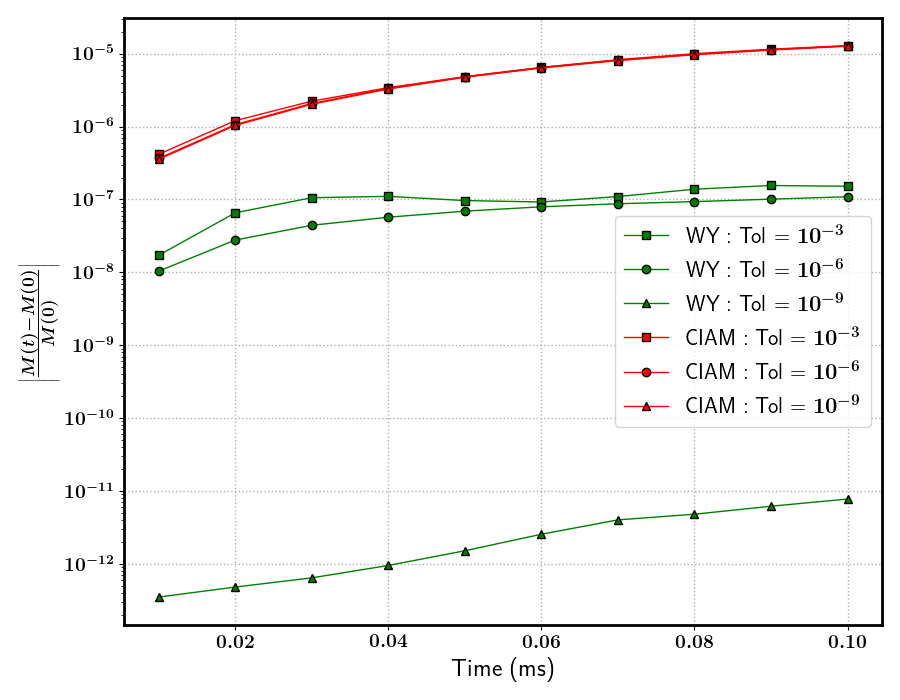} & 
\hspace{-0.2cm}%
\includegraphics[width=0.5\textwidth]{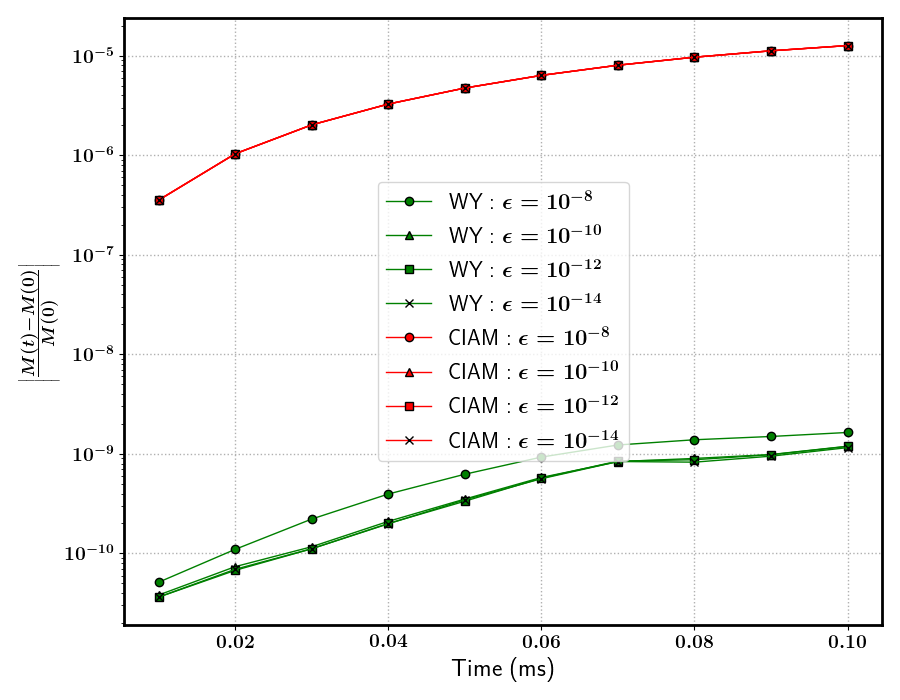} \\ 
\hspace{-0.2cm}%
(a) & (b) \\
\hspace*{-1.0cm}
\includegraphics[width=0.5\textwidth]{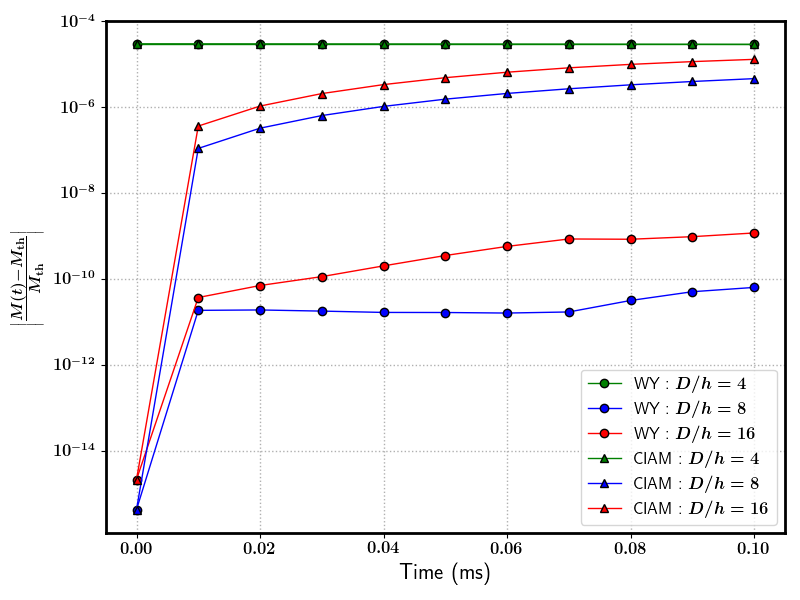} & 
\hspace{-0.2cm}%
\includegraphics[width=0.5\textwidth]{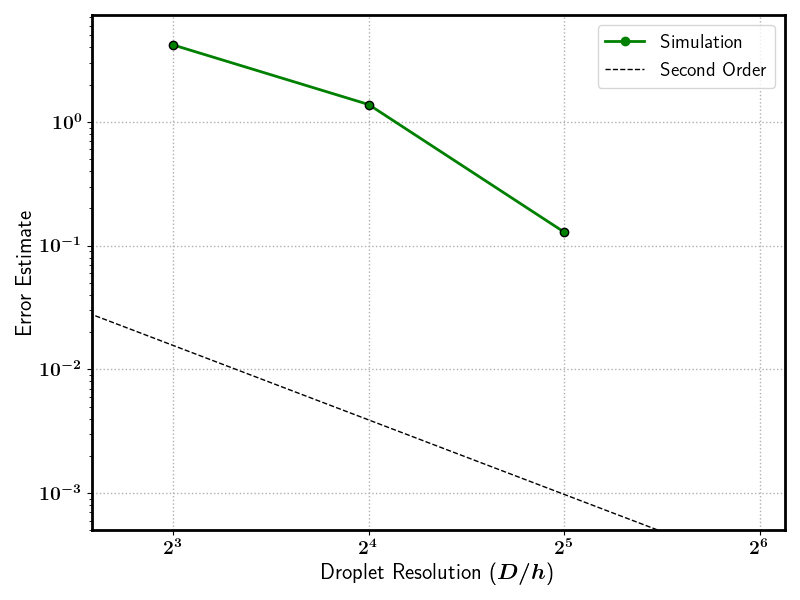} \\ 
\hspace{-0.2cm}%
(c) & (d)
\end{tabular}
\end{center}
\caption{Relative change in the droplet mass 
as a function of time in the first three plots. 
The simulations are carried out for a total time of $0.1$ milliseconds,
while the droplet resolution is $D/h = 16$ in the first two plots. 
The symbol ``WY'' in the legend refers to the combination of WY advection 
with the QUICK-UW velocity interpolation, and 
``CIAM'' to the CIAM advection with the Superbee slope limiter. 
Mass conservation properties of the two schemes:
(a) as a function of the Poisson solver tolerance; 
(b) as a function of the clipping parameter $\eps_c$; 
(c) as a function of the droplet resolution $D/h$.    
(d) Error estimate on the droplet acceleration, 
in the frame of reference of the static box. 
The corresponding droplet accelerations are plotted in the inset of Fig. \ref{drop_vel}.}
\label{mass_conv}
\end{figure}

We  use the WY advection scheme in combination with the QUICK-UW interpolation.
The quantities of interest while examining the robustness of the 
method are the temporal evolution of
(Fig. \ref{multi}a) and the moments of inertia (defined in Eq. \ref{immde}) of the droplet 
along the three coordinate directions (Fig. \ref{multi}b,c,d).  
These are used as a descriptor of the 'average' droplet shape.
The droplet kinetic energy is defined
relative to the droplet center-of-mass, that is
\be
E_k = \langle \rho_l C(x,t) || \U(x,t) - \U_{CM}(t) ||^2 \rangle
\nd
where $ \langle . \rangle$ is the spatial averaging operator of the entire domaine and $\U_{CM}(t)$ is the
droplet center of mass. 
 The kinetic energy of the droplet evolves in a relatively smooth manner, 
without the presence of sudden spikes and falls which are emblematic of 
the non-consistent version of our method (refer to Fig. \ref{pressure_2}). 
Such abrupt changes in kinetic energy of the droplet
typically occur when the droplet undergoes 
`artificial' atomization or breakup. 
We observe a strong decrease in the droplet 
kinetic energy as we increase resolution, an indication that
low resolutions add considerable spurious jitter in the
interfacial shape. Even at $D/h=64$ the simulation is not converged in
that respect. 
Finally, the moments of inertia of the droplet appear to evolve 
in a smooth manner for all droplet resolutions.

In Fig. \ref{drop_vel}, we show the velocity of 
the center of mass of the droplet (in the frame of reference 
of the box enclosing the droplet) as a function of time, 
and its behavior as we increase the droplet resolution. 
As one can observe it undergoes a near constant acceleration.
The observed acceleration is $\romandt U \simeq 5.8 \pm 0.1$ m/s
the error estimate being obtained from the difference
between the $2^5$ and the $2^6$ values of the droplet resolution $D/h$. 
This is consistent with equation \refeq{dtu} if one sets
$U=U_0$ and $U_t=6.6$m/s not far from the terminal velocity
of a solid sphere and of Section \ref{secdr1}.
One may also notice that on  Fig. \ref{drop_vel} the initial
center-of-mass velocity is not zero in the simulation frame
of reference but a small upward velocity of $10^{-2}$ m/s. 
This is due to the small velocity shift during the first time step projection
already discussed in Section \ref{sda} of order $rU_0 \simeq 6 \, 10^{-3}$ m/s. 

The temporal variation in the droplet velocity is fit
to a straight line in order to evaluate the droplet acceleration 
for each droplet resolution. The estimated acceleration is
shown in the inset of Fig. \ref{drop_vel}.
The decrease of the error, obtained
as above from the difference
between successive resolutions is plotted on Fig. \ref{mass_conv}d.
No clear convergence order emerges, although the second-order convergence
line is plotted to guide the eye. 

In Fig. \ref{mass_conv} we show the mass conservation properties
of the two most stable combinations, that is the WY advection with the QUICK-UW 
velocity interpolation (WY) and the CIAM advection with the Superbee slope 
limiter (CIAM) for the falling raindrop. We have pointed out after \eqref{sumfall2}
that the mass conservation of the WY advection is strongly dependent
on how accurately the divergence-free condition is enforced, which
in turns is determined by the Poisson's solver tolerance. A clipping
procedure, with no redistribution, affects as well mass conservation.
The WY combination is thus rather sensitive to these two parameters
but overall performs much better than the CIAM combination which is
inherently not mass-conserving (see \eqref{sumfall}) and not very
sensitive to the Poisson's solver tolerance and the parameter $\eps_c$.
Mass conservation is also plotted with various 
resolutions $D/h$ in Fig. \ref{mass_conv}c. It is seen that going from $D/h=32$ to $D/h=64$
the mass variation becomes larger for both methods, and grows over time at high resolution for WY, pointing to an accumulation of many machine precision errors.

}

\subsection{Atomizing air and water planar jets}

\begin{figure}
\begin{center}
\includegraphics[width=0.99\textwidth]{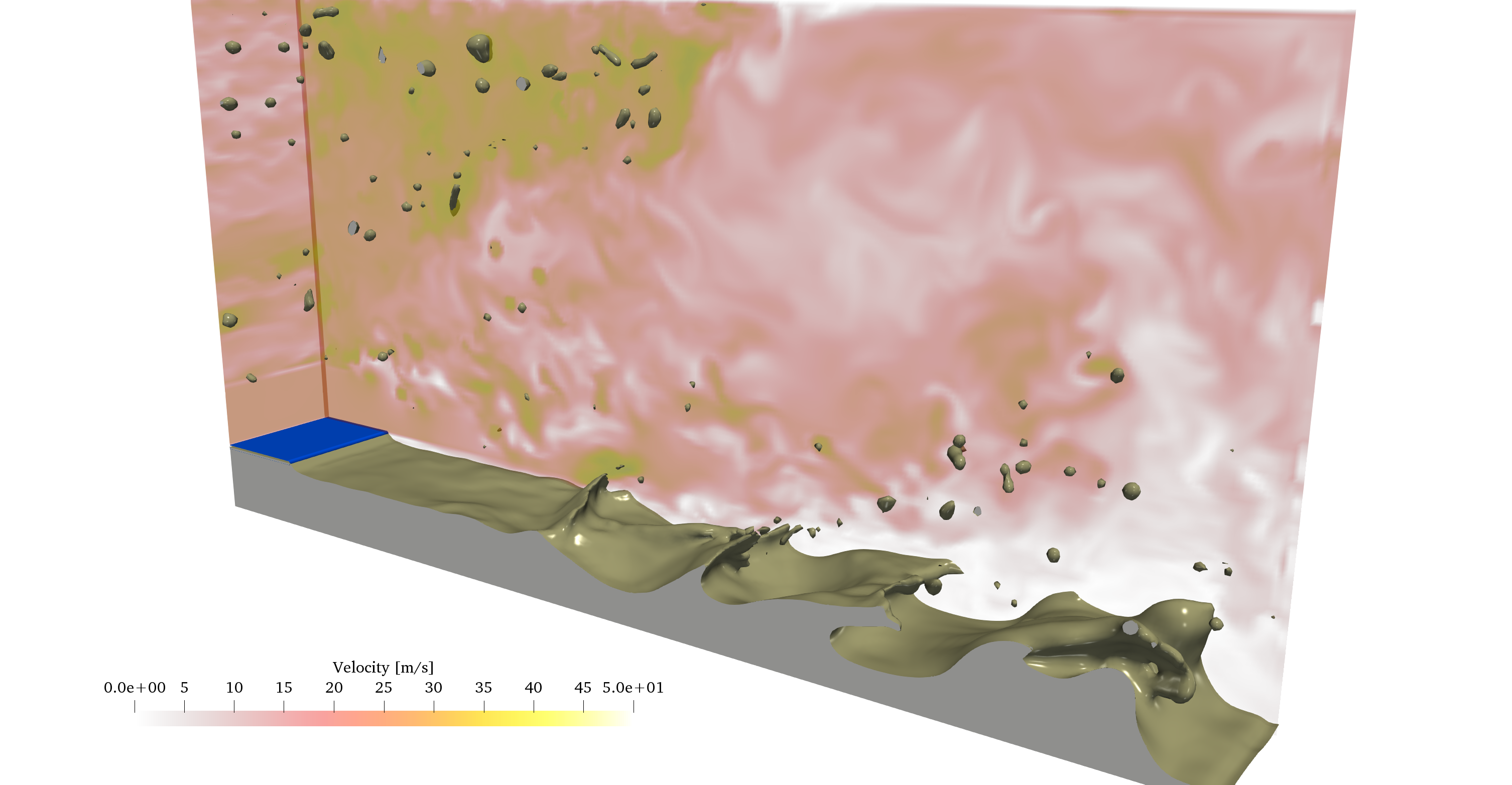}
\end{center}
\caption{Atomizing layer with air/water properties.}
\label{atom}
\end{figure}

We also test the capability of the mass-momentum-consistent scheme 
to simulate an atomizing air-water shear flow. For that 
purpose we repeat the setup of ref. \cite{Ling16}. Two jets of air and water 
are entering the computational domain from the left of Fig. \ref{atom}
at velocities comparable to those of experiments. However in order to 
save computer time the computational domain is smaller than the
experimental one. Physical 
properties of air and water are identical to those of the falling raindrop case 
given in Tab. \ref{raindropprop}. The flow and domain characteristics are 
given in Tab. \ref{PhysicalParam} including a gas boundary layer 
and separator-plate identical to those of ref. \cite{Ling16}. 
The notations are as in ref. \cite{Ling16}: $H_p$ is the thickness of the
jet of phase $p$, there is a separator plate of thickness $l_y$ and a 
gas boundary layer of thickness $\delta_g$.  The two streams have 
equal thickness $H_l=H_g$. The dimensionless parameters are given in 
Tab. \ref{dimensionlessParam}. The CIAM advection method has been used. 
The number of grid points in the layer $H_l/h = 16$ is relatively small,
when compared to $H_l/h = 32$ in the coarsest simulation
of ref. \cite{Ling16}. It is thus all the more remarkable that the 
simulation is stable since using a smaller number of grid points usually 
increases the trend towards instability. It is interesting to note 
that the VOF calculations accounts for 31.5\% of the total time, while the 
inversion of the Poisson operator for the pressure accounted for 51.5\%. 
The whole simulation runs overnight on a present-day workstation. 

\begin{table}
\begin{center}
\begin{tabular}{cccccc}
\hline
\hline
 $U_l$ & $U_g$ & $H_l$ & $h$ & $l_y$ & ${\delta_g}/{l_y}$ \\

 $(m/s)$ & $(m/s)$ & $(m)$ & $(m)$ & $(m)$ & $(-)$ \\
\hline
 $1$ & $25$ & $4\, 10^{-3}$ & $2.5\, 10^{-4}$ & $2.5\, 10^{-4}$ & $2$\\
\hline
\hline
\end{tabular}
\end{center}
\caption{Physical parameters (defined in the text) for the atomizing layer setup. The fluid 
properties are the same as in Table~\ref{raindropprop}.
}
\label{PhysicalParam}
\end{table}

\begin{table}
\begin{center}
\begin{tabular}{cccccc}
\hline
\hline
 $M$ & $r$ & $m$ & $\Re_{g,\delta}$ & $\We_{g,\delta}$  & $\Re_{g}$ \\
 $\rho_g U_g^2/\left(\rho_l U_l^2\right)$ & $\rho_l/\rho_g$ 
 & $\mu_l/\mu_g$ & $\rho_g U_g H_g/\mu_g$ & $\rho_l U_l H_l/\mu_l$ 
 & $\rho_g U_g^2 H_g/\sigma$ \\
\hline
 $0.75$ & $831.8$ & $45$ & $757.6$ & $5.151$ & $6061$  \\
\hline
\hline
\end{tabular}
\end{center}
\caption{Dimensionless parameters for the atomizing layer setup.}
\label{dimensionlessParam}
\end{table}

\section{Conclusion}
\red{
We have presented and tested a new simulation method for multiphase flow
that involves momentum advection that is consistent with VOF advection.
The method includes an implementation of the WY VOF advection method.

A simple and fast test of the method's stability and accuracy is offered by the analysis of the Kelvin-Helmholtz
instability. The method is converging when velocity profiles are continuous. However
in the case of vortex sheets spurious growth is often observed.
Progress on this front would probably provide benefits on other aspects of the
method's performance. 

The increased stability of the new method, on the other hand
is evidenced in several classical tests cases and especially for 
a $3$ mm droplet of water falling in air, a typical}
raindrop. It is a reflection on the challenging nature of multiphase
flow that such complex methods apparently need to be implemented to
resolve such an everyday and simple phenomenon.

The method comes with a significant saving of computer time, since for 
similar problems with raindrops, our attempts with a non-momentum-consistent
VOF approach led to catastrophic deformation of the drop or strong dimple 
formation. These problems have also been observed by us using other 
non-VOF-consistent and non-momentum-consistent methods such as the one of 
\cite{popinet09}. In that case whenever less than 200 grid points per 
diameter are used numerically stable air-water 
drops accelerated at moderate Weber number cannot be found. However for
higher resolutions they can be computed without difficulty as also found 
by the authors of ref. \cite{Jain15}.
Here, approximate solutions for the raindrop accurate within 15\% are found with only 15 
points per diameter. At lower resolution, down to 2 points per diameter
the simulations are often stable, but the new method is unable to stabilize them
in all conditions.

A particular advantage of the method is that it is conserving mass at the 
accuracy at which discrete incompressibility is enforced and opens a
perspective for similar momentum conservation using WY advection.
The method nevertheless is more complex and costly than a collocated method. 
This opens the perspective for systematic development of other methods with 
different grid arrangements. Another perspective is the potential
of stable methods with large density contrasts, exact mass and momentum conservation 
and small droplets, that could be smoothly merged into models that represent 
the small droplets as Lagrangian Point Particles \cite{Ling15}.

\section{Acknowledgements} 

This work has been supported by the ANR MODEMI project (ANR-11-MONU-0011) program, grant 
SU-17-R-PER-26-MULTIBRANCH from Sorbonne Universit\'e and the ERC Advanced Grant TRUFLOW.
%
%

This work was granted access to the HPC resources of TGCC-CURIE, TGCC-IRENE and CINES-Occigen under the allocations 
t20152b7325,  t20162b7760, 2017tgcc0080 and A0032B07760, made by GENCI and TGCC. 
The authors would also like to acknowledge the MESU 
computing facilities of Sorbonne Universit\'e. 

We would like to thank Dr.\ W.\ Aniszewski, Dr. S. Dabiri, Dr. Jiacai Lu and Dr. Ph. Yecko for their contribution to the development of the code \emph{PARIS-Simulator}, and we thank   Dr.\ W.\ Aniszewski, Dr. V. Le Chenadec, Mr. C. Pairetti, Dr. S. Popinet and Dr. S. Vincent for useful conversations on the topics of this paper.  

Finally, the simulation data are visualized by the software VisIt developed by the Lawrence Livermore National Laboratory. 

\appendix

\section{Interpolations for the advected velocity component $\phi = \bar u_q$}
\red{
\label{appinterp}
We want interpolate the value of $\phi$ on the face between the
two centered values $\phi_{-1/2}$ and $\phi_{1/2}$ of Fig. \ref{advect-ed-ing-fig},
where we have already interpolated with a centered scheme
the advecting velocity $u_f$ perpendicular to the face. We consider
the interpolating function \eqref{simpleinterp}
\be
\phi_0 = f \big( \phi_{-3/2}, \phi_{-1/2}, \phi_{1/2},\phi_{3/2},{\rm sign}(u_f)
\big) \,.
\nonumber
\nd
In the first scheme we use QUICK, a third order interpolant, away from the 
interface and a first-order upwind scheme near the interface. 
For positive advecting velocity, $u_f > 0$, and in the bulk we have
\be
\phi_0 = \frac 3 4   \phi_{-1/2} + \frac 3 8   \phi_{1/2} - \frac 1 8 \phi_{-3/2} \,,
\nd
while near the interface the upwind value is $\phi_0 = \phi_{-1/2}$. 
For negative advecting velocity, $u_f < 0$, and in the bulk we have
\be
\phi_0 = \frac 3 4   \phi_{1/2} + \frac 3 8   \phi_{-1/2} - \frac 1 8 \phi_{3/2} \,,
\nd
while near the interface $\phi_0 = \phi_{1/2}$.

In the second scheme we use a Superbee slope limiter away from the interface.
For a positive advecting velocity $u_f$, we consider the 
general family of interpolants 
\be
f(\phi_{-3/2}, \phi_{-1/2}, \phi_{1/2},\phi_{3/2},{\rm sign}(u_f)) = \phi_{-1/2} + S h/{2} \,, 
\nd
where the slope $S$ is given  by a slope-limiter function $g$,
$S = g(\phi_{-3/2}, \phi_{-1/2}, \phi_{1/2})$.
For a negative advecting velocity $u_f<0$ we consider
\be
f(\phi_{-3/2}, \phi_{-1/2}, \phi_{1/2},\phi_{3/2},{\rm sign}(u_f)) = \phi_{1/2} - S h/{2} \,,
\nd
where $S = g(\phi_{-1/2}, \phi_{1/2},\phi_{3/2})$. For $u_f > 0$, the two slopes
\be
\alpha^+ = \big( \phi_{1/2,j} -\phi_{-1/2,j} \big)/ h \,, \qquad
\alpha^- = 2 \big( \phi_{-1/2,j} -\phi_{-3/2,j} \big)/ h \,,
\nd
are first estimated, to compute
\be
\alpha = \min \big( \alpha^+, \alpha^- \big) \,,
\nd
then the other two slopes 
\be
\beta^+ = 2 \big( \phi_{1/2,j} -\phi_{-1/2,j} \big)/ h \,, \qquad
\beta^- = \big( \phi_{-1/2,j} -\phi_{-3/2,j} \big)/ h  \,,
\nd
to compute 
\be
\beta = \min \big( \beta^+, \beta^- \big) \,,
\nd
Finally the slope $S$ is given by the expression
\be
S = \max \big( 0, \alpha, \beta) \,.
\nd
A similar development can be done for $u_f<0$.

A slightly different estimate of the Superbee advected velocity is used near the interface. 
First we extend 
the definition of the interpolants as we shall predict $\bar u_q$ at a point $\hat x$ 
slightly upwind from $x_f$ using a new function $\hat f$ so that
\be
\phi_0 = \hat f \big( \hat x,\phi_{-3/2}, \phi_{-1/2}, \phi_{1/2},\phi_{3/2},
{\rm sign}(u_f) \big) \,. 
\label{mp}
\nd
We take  $\hat x = x_f - u_f \tau/2$ to be the midpoint of the fluxed region
$\Omega_D$ of Fig. \ref{advect-ed-ing-fig}(b).
The extended interpolant is defined for positive velocity $u_f$ as
\be
\hat f \big( \hat x,\phi_{-3/2}, \phi_{-1/2}, \phi_{1/2},\phi_{3/2},{\rm sign}(u_f)
\big) = \phi_{-1/2} +  S |\hat x - x_{-1/2}|
\nd
and for negative velocity $u_f$ as
\be
\hat f \big( \hat x,\phi_{-3/2}, \phi_{-1/2}, \phi_{1/2},\phi_{3/2},{\rm sign}(u_f)
\big) = \phi_{1/2} -  S |\hat x - x_{1/2}|
\nd
The rationale behind this choice is as follows. For a time-independent advecting velocity 
field $u(\X)$, the integrals in expression
(\ref{barudef}) can be simplified
\be
\phi_0 = \bar u_q = \frac{\int_{\Omega_D}  u_f u_q \, {\rm d}\X}{ \int_{\Omega_D}  
u_f  \,{\rm d}\X}.  \label{barphi3}
\nd
where ${\Omega_D} $ is the ``donating region'' of Fig. \ref{advect-ed-ing-fig}(b).
We approximate the advecting velocity $u_f$ by its midpoint value and the integral
can be further simplified as
\be 
\bar u_q = \frac 1{|\Omega_D|} \int_{\Omega_D}  u_q \, {\rm d}\X  \label{barphi4}
\nd
Since the ``midpoint'' at $\hat x$ is the center of mass of the donating region ${\Omega_D}$ 
the interpolation expression (\ref{mp}) follows. 

}

\section{Kelvin Helmholtz instability: numerical setup}
\red{
\label{appKH}
We consider the 2D base flow shown on Figure \ref{khappfig}. Coordinates are noted $(x,z)$ and vectors
$(u,w)$. The height of the interface is $h(x,t)$. The flow has density $\rho_1$ for $z<h(x,t)$ and 
$\rho_2$ for $z>h(x,t)$ and is incompressible. The base flow is a parallel shear flow.
The base flow is uniform with $u=-U$ for $z<0$
and $u=U$ for $z>2a$, with a linear (Couette flow) boundary layer in between.  When $\rho_1 \neq \rho_2$ the heavier fluid is the ``liquid'' and the lighter the ``gas''. For $\rho_1 > \rho_2$ the boundary layer is in the gas while otherwise the bounary layer is in the liquid. 
Similar flows have been studied
in \cite{Matas_2011a,Eggers08}.

\subsection{Dispersion relation}

The Euler and incompressibility 
equations are as in Eqs. (\ref{nse1}) with only
$
\LLL =  \LLL_{\rm conv} 
$, surface tension and viscosity are not included.
The interface height $h$ moves according to
\be
\dert h +  u \derx h =  w \label{heq}
\nd
We also use a stream function $\psi$
\be
w = - \derx \psi, \quad u = \derz \psi \label{psiu}
\nd
We consider a small perturbation of the base velocity 
in the form
\be
\U = \U_0  + \eps \U_1 + {\cal O}(\eps^2) \label{1}
\nd
the pressure expands as 
$p=p_0 + \eps p_1(x,z,t) + {\cal O}(\eps^2)$, the height as
$h= \eps h_1(x,z,t) + {\cal O}(\eps^2)$,
and similarly
the stream function. 
We assume the following form for the perturbation
\be
\left(
\ba{c} u_1 \\ w_1 \\ p_1 \\ \psi_1 \\ h_1 \ea \right)
=
\left(
\ba{c} U_1(z) \\ W_1(z) \\ P_1(z) \\ \Psi_1(z) \\ A_h \ea
\right)
e^{- \ii k x - \ii \om t}
\label{UVWdef} \nd
where $k$ is an  arbitrary wavenumber and $\om$ a frequency
to be determined.
Although the expressions on the rhs in (\ref{UVWdef}) are complex we understand the real part.

It is convenient to define
as Chandrasekhar \cite{Chandrasekhar} the reduced wavenumber  $\kappa = 2ka$
and the reduced frequency $\Omega=\om a/U$, which are related by 
\be
e^{-2\kappa} 
=  (1 - 2\Omega - \kappa)\frac{2 + (r+1)(2\Omega - \kappa) }{2+ ({r-1}) (2\Omega - \kappa)},
\label{ev}
\nd
see for example \cite{Eggers08}, equation (135).
The system is unstable whenever Eq. (\ref{ev}) has two complex conjugate non-real roots. Then the positive
imaginary part 
$\Omega_i$ is the growth rate, plotted on Figure \ref{grr} in the cases $r=1,10$ and $100$
where $r={\rm max}(\rho_1/\rho_2,\rho_2/\rho_1)$.
The case $\rho_1/\rho_2=100$ corresponds to the boundary layer in the gas
and is much less unstable than the case with the boundary layer in the liquid. As the ratio  $\rho_1/\rho_2=100$  is increased
the growth rate for the boundary layer in the gas decreases steadily while the growh rate for the boundary
layer in the liquid converges to the one for a free surface, with the gas replaced by a void. 

\subsection{Special case}

When $a=0$ the above is singular and a special computation, performed in Section \ref{appa0} below,
is needed. One has $\kappa=0$ and the frequencies are
\red{
  \be
\om =\left( \frac{r-1}{r+1} +\frac{2 \sqrt r}{r+1}  \ii  \right){ k U } \label{omkaz}
\nd
}

\subsection{Construction of the unstable mode}

We want to write the solution using the stream function $\psi$
in order to have a divergence free initial condition in the computations.
We introduce four arbitrary constants noted $A^{}_0$ and $B^{}_0$ in the region  $0<z<2a$, $A^{}_1$ in the region $z<0$ and $B^{}_2$ in the region  $z>2a$, so that $\psi$ is given by
\blue{
\bea
z<0 & \psi(x,z,t) & =  A_1 e^{kz} e^{-\ii k x - \ii \om t } \label{psi1} \\ 
0 <z< 2a & \psi(x,z,t) & = ( A_0 e^{kz} +  B_0 e^{-kz})  e^{-\ii k x - \ii \om t }  \label{psi2} \\
2a<z & \psi(x,z,t) &=  B_2 e^{-kz}  e^{-\ii k x - \ii \om t } \label{psi3}
\nda
}
We also introduce four additional constants
noted $A^\prime_0$ and $B^\prime_0$ in the region  $0<z<2a$, $A^\prime_1$ in the region $z<0$ and $B^\prime_2$ in the region  $z>2a$.
\blue{
\bea
z<0 & W_1(z) & = A^\prime_1 e^{kz} \label{w1} \\ 
0 <z< 2a & W_1(z) & = A^\prime_0 e^{kz} + B^\prime_0 e^{-kz} \label{w2} \\
2a<z & W_1(z) &= B^\prime_2 e^{-kz} \label{w3}
\nda
 }
 After substitution in the Euler equations one obtains the pressure perturbation in the three regions; for $z<0$ , 
\be
 P_1(z)  =  \frac{\ii \rho_1} k ( \om - k U ) ( A^\prime_0 + B^\prime_0)  e^{kz},
\nd
for $0<z<2a$ ,
\be
 P_1(z) = \frac{\ii \rho_2} k \left[  ( \om + k U\frac z a - k U) ( A^\prime_0 e^{kz} -  B^\prime_0 e^{-kz})
- (A^\prime_0 e^{kz} +  B^\prime_0 e^{-kz}) \frac U a \right] 
\nd
and for  $z> 2a$,
\be  
P_1(z) = - \frac{\ii \rho_2} k ( \om + k U ) ( B^\prime_0 + A^\prime_0 e^{4 ka} ) e^{-kz}.
\nd
From (\ref{heq}) 
\be
\dert h =  -U_0(0) \derx h + w \label{htw}
\nd
and from \refeq{psiu} and  \refeq{psi2} 
we get
$$
- \ii \om  A_h = -\ii k U  A_h +  \ii k (A_0 + B_0)
$$
hence
$$
(- \ii \om + \ii k U)  A_h =  \ii k (A_0 + B_0)
$$
and
$$
A_h = - \frac{k}{\om-kU} (A_0 + B_0) 
$$
Similar relations are obtained from the requirements of continuity of pressure
and normal velocity at $z=0$ and $z=2a$. 
We can then determine all the amplitudes of the constructed solution
after an amplitude for the interface has been chosen. Typically
the modulus $|A_h|$  and the argument $\phi$ are selected so that
$$A_h = | A_h | e^{\ii \phi}$$
then for $\kappa>0$ we have
\red{
\bea
C_0 & = & - (\frac{\om}k- U)A_h ( e^{-2\kappa}  +  2\Omega + \kappa - 1 )^{-1} \label{AhC0}\\
A_0 &=& C_0  e^{-2\kappa}\\
B_0 &=& C_0 (2\Omega + \kappa - 1 ) \label{AB0sol}\\
A_1 & = &  A_0 + B_0 \label{A1A0B0}  \\
B_2 & = &  A_0  e^{2 \kappa} + B_0  \label{B2A0B0}
\nda
}
These expressions for the amplitudes together with (\ref{psi1}-\ref{psi3}) are used to
initialize the stream function. The intermediate constant $C_0$ is used to simplify the expressions.

\subsection{Special cases: mode structure for $ka \rightarrow 0$}

\subsubsection{Mode structure for $a>0$ going to the limit $a \rightarrow  0$.}

For small or vanishing $a$ and $\kappa$ however the expression \refeq{AhC0} is singular.
Indeed in the limit $\kappa \rightarrow 0$ we also have $\Omega \rightarrow 0$ and
$$
e^{-2\kappa}  +  2\Omega + \kappa - 1 = - 2\kappa + 2\kappa^2 +   \kappa + \Order(\kappa^3) = - \kappa + \Order(\kappa^2)
$$
and then $A_0$ and $B_0$ become spurious as the region $(0,2a)$ vanishes.
From \refeq{AhC0}
$$
C_0 \simeq  - \frac{\om}k  \frac 1 \kappa A_h
$$
and from \refeq{AB0sol}  and  \refeq{B2A0B0} 
$$
B_2 = C_0 (2\Omega + \kappa) \simeq \kappa  C_0 
$$
Thus for $a=0$
\red{
$$
B_2 = - \frac{\om}k  A_h
$$
$$
A_1 =  \frac{\om}k A_h
$$
}
the latter being obtained directly from \refeq{A1A0B0}. Since $B_2 \neq A_1$
there is a $\Order(\eps)$ discontinuity of $\psi$ which results in a jump of $v(x,z,t)$
accross the interface at $z=0$ and a related thin jet $u_1(x,z,t) \simeq \epsilon f(x,t) \delta(z)$.
This is a consequence of placing the interface in the above calculations at $z=0$ instead of
$z=\eps h_1(x,t)$ and it conflicts with the solution obtained classically and also below with the
thin vortex sheet setup (that is, the setup in which $a=0$ is postulated at the beginning). 

\subsection{Mode structure in the $a=0$ case}
\label{appa0}

We now obtain the mode structure for the classic thin vortex sheet setup where one assumes $a=0$
from the start.
In that case we keep only the terms in \refeq{psi1} and \refeq{psi3}. The velocity continuity condition becomes
\be
h_t = - u h_x + w = - u_0 h_x + w_1 + \Order(\epsilon) \label{ht}
\nd
which replaces \refeq{htw}. Since $h_t$ must have the same expression above and below the interface
\be
[ w  -  h_x u_0 ]=0
\nd
thus $ [ w] = h_x [u_0] = 2 h_x U $ and from  \refeq{w1} and \refeq{w3}
\bea
z<0 & W_1 &= A^\prime_1 e^{kz}  \\ 
z > 0 & W_1 &= B^\prime_2 e^{-kz}
\nda
and from \refeq{ht}
\bea
A^\prime_1 &=&  -\ii (\om - kU) A_h \label{apht}\\
B^\prime_2 &=&  -\ii (\om + kU) A_h \label{bpht}\\
\nda
The pressure equality at $z=0$ leads to
\bea
z<0 &  P_1 &=    \frac{\ii \rho_1}{k^2} (\om - kU) A^\prime_1 k e^{kz}  \\ 
z>0 &  P_1 &=    \frac{\ii \rho_2}{k^2} (\om + kU) B^\prime_2 (- k e^{-kz}) 
\nda
hence introducing $\nu = \om/(kU)$
and from  \refeq{apht} and \refeq{bpht}
\bea
(\nu+1)A_1 - (\nu-1)B_2 & = &0\\
r (\nu-1)A_1 + (\nu + 1)B_2 & = &0\\
\nda
from which one obtains
  \be
\nu = \frac{r-1}{r+1} +\frac{2 \sqrt r}{r+1}  \ii 
\nd
identical to \refeq{omkaz}. Also from \refeq{apht} and \refeq{bpht}
\bea
A_1 &=&    (U - \om/k) A_h \label{aha}\\
B_2 &=&  - (U + \om/k)  A_h \label{ahb}
\nda
These expressions should be used whenever $a \ll A_h$ while the full expressions with the boundary layer
would be valid for  $A_h \ll a$. In both cases $\Delta x \ll {\rm min}(A_h,a)$ ma be required. 

}



\section*{References}
\bibliographystyle{elsarticle-num} 
\bibliography{multiphase,gretar,sagar}

\end{document}